\documentclass[%
 amsmath,amssymb,
 preprint,%
]{revtex4-1}
\usepackage{multirow}
\usepackage{color}
\usepackage{float}
\usepackage[caption=false]{subfig}
\usepackage{graphicx}
\usepackage{dcolumn}
\usepackage{bm}
\usepackage{adjustbox}

\usepackage{verbatim}
\usepackage[utf8]{inputenc}
\usepackage[T1]{fontenc}
\usepackage{mathptmx}
\usepackage[inline]{enumitem}

\begin{document}


\title[]{Hybrid Quantum-Classical Boson Sampling Algorithm for Molecular Vibrationally Resolved Electronic Spectroscopy with Duschinsky Rotation and Anharmonicity}
\author{Yuanheng Wang}
\affiliation{MOE Key Laboratory of Organic OptoElectronics and Molecular Engineering, Department of Chemistry, Tsinghua University, Beijing 100084,
 People's Republic of China }
\author{Jiajun Ren}
\affiliation{MOE Key Laboratory of Organic OptoElectronics and Molecular Engineering, Department of Chemistry, Tsinghua University, Beijing 100084,
 People's Republic of China }
\affiliation{Current Address: Key Laboratory of Theoretical and Computational Photochemistry, Ministry of Education, College of Chemistry, Beijing Normal University, Beijing 100875, People’s Republic of China}
\author{Weitang Li}
\affiliation{MOE Key Laboratory of Organic OptoElectronics and Molecular Engineering, Department of Chemistry, Tsinghua University, Beijing 100084,
 People's Republic of China }
\author{Zhigang Shuai}%
 \email{zgshuai@tsinghua.edu.cn}
\affiliation{MOE Key Laboratory of Organic OptoElectronics and Molecular Engineering, Department of Chemistry, Tsinghua University, Beijing 100084,
 People's Republic of China }

\begin{abstract}
Using a photonic quantum computer for boson sampling has been demonstrated a tremendous advantage over classical supercomputers. It is highly desirable to develop boson sampling algorithms for realistic scientific problems. In this work, we propose a hybrid quantum-classical sampling (HQCS) algorithm  to calculate the optical  spectrum for complex molecules considering anharmonicity and Duschinsky rotation (DR) effects. 
The classical sum-over-state method for this problem has a computational complexity that exponentially increases with system size. 
In the HQCS algorithm, an intermediate harmonic potential energy surface (PES) is created, bridging the initial and final PESs. The magnitude and sign ($-1$ or $+1$) of the overlap between the initial state and the intermediate state are estimated by quantum boson sampling and by classical algorithms respectively, achieving an exponential speed-up. Additionally, the overlap between the intermediate state and the final state is efficiently evaluated by classical algorithms. 
The feasibility of HQCS is demonstrated in calculations of the emission spectrum of a Morse model as well as pyridine molecule by comparison with the nearly exact time-dependent density matrix renormalization group solutions.
A near-term quantum advantage for realistic molecular spectroscopy simulation is proposed.
\end{abstract}

\maketitle

\begin{figure}
    \centering
    \subfloat[]{\includegraphics[width=0.45 \textwidth]{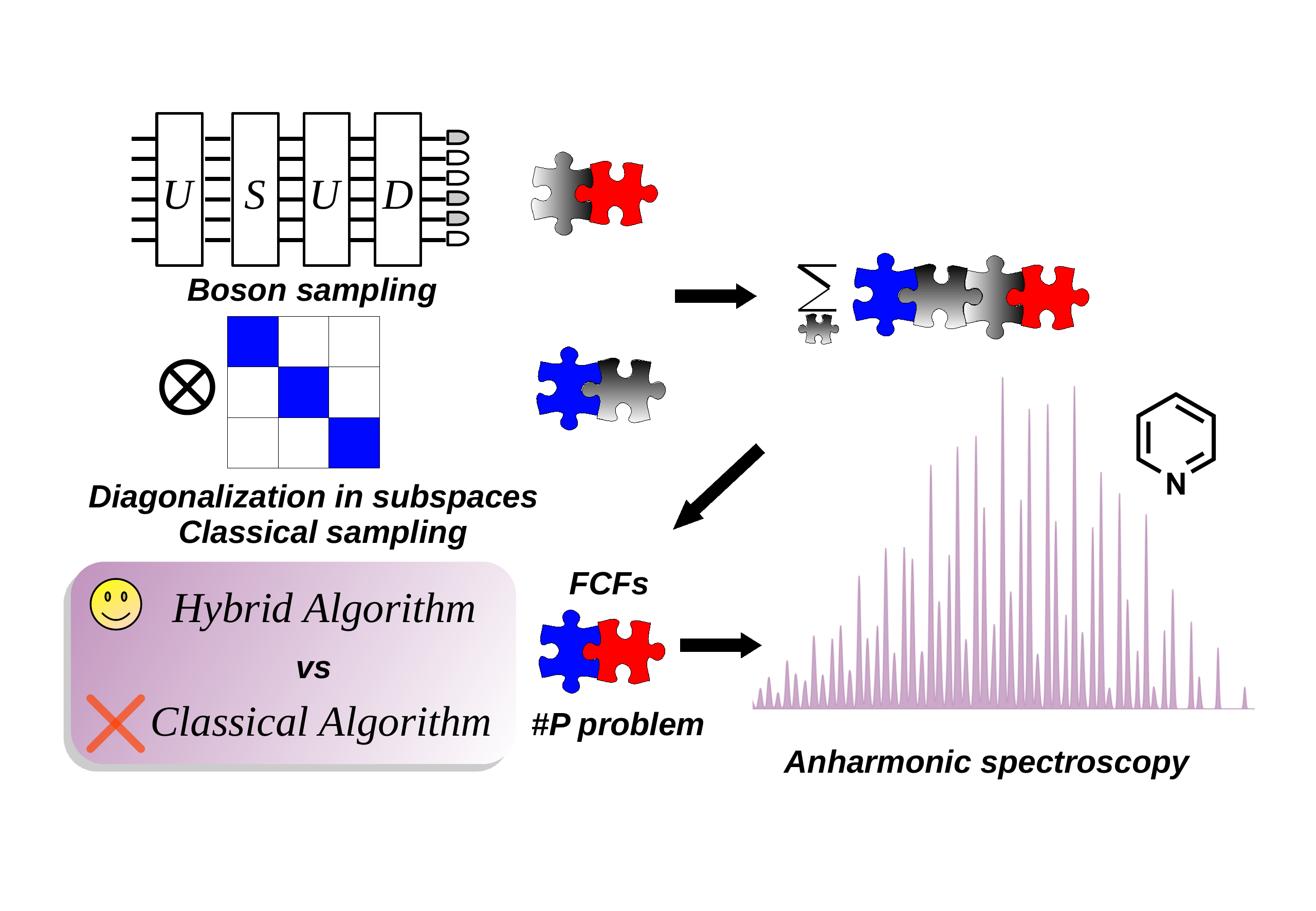}}
    \caption{TOC graphic }
    \label{fig:toc}
\end{figure}

\section{Introduction}
In the early 1980s, Feynman proposed quantum computation as a natural solution for many-body problems hard for classical computers~\cite{Feynman1982Simulating}. Since then, many quantum algorithms have been devised to solve classically hard problems. However, the presently known methods such as Shor's algorithm~\cite{Shor1994algorithms} for factorization and quantum phase estimation (QPE)\cite{Abrams1999Quantum} for the eigenvalue problems  
require an error-tolerant quantum computer which is not available in the near future. Variational quantum algorithms ~\cite{Peruzzo2014variational}, hybridizing classical and quantum computation, is regarded as a practical choice for quantum computation applications with the current noisy intermediate-scale quantum (NISQ) device. On the other hand, quantum advantage has been reported through quantum sampling algorithms such as random circuit sampling~\cite{arute2019quantum}, boson sampling~\cite{Aaronson2011the,wang2019boson} and its variants Gaussian boson sampling (GBS)~\cite{hamilton2017gaussian, Zhong2020quantum, Zhong2021Phase}. 
Recently, quantum computer "Jiu Zhang 2.0" performed GBS on a $ \sim 10^{43}$ large Hilbert space which was shown to be $ \sim 10^{24}$ faster than a brute-force simulation on the state-of-the-art classical computers ~\cite{Zhong2020quantum,Zhong2021Phase}. 
Currently, it is valuable to develop algorithms that profit from the quantum advantage of the quantum sampling algorithms and devices to solve practical problems. 

In this work,
we focus on the computation of molecular vibrationally resolved electronic  spectrum taking both Duschinsky rotation (DR) and anharmonicity into account.
DR refers to the mode-mixing between the normal modes of the initial and final states in complex molecular systems. A number of experimental ~\cite{Shen2018quantum,Clements2018approximating,Wang2020Efficient} and theoretical ~\cite{Huh2015boson,Jnane2021Analog} works have focused on the simulation of molecular spectrum with DR effect under harmonic approximation using the boson sampling algorithms. 
It should be noted that such problem has been solved exactly and analytically on classical computers by Peng \textit{et al.} by the thermal vibration correlation function (TVCF) method back in 2007 ~\cite{Peng2007Excited,niu2008promoting}. Hence, there is no point to demonstrate quantum advantages for such a problem.  Besides the DR effect, the anharmonic effect on molecular spectrum has also been found to be very important in general for complex molecules, especially, for flexible molecules ~\cite{Petrenko2017General,liang2006electronic,heimel2005breakdown}. To calculate the spectrum with both anharmonicity and DR is indeed a hard problem for classical algorithms because there is no analytical solution and the numerical approaches have an exponential scaling with the number of modes. Therefore, it is necessary to develop quantum algorithms for such a problem, while no near-term solutions have been proposed yet.
The previously suggested boson sampling algorithms can only simulate Franck-Condon factors (FCFs) between harmonic states, so they are not suitable for anharmonic spectrum.  
Vibrational relaxation of anharmonic Hamiltonian using vibrational pertubation theory to consider influence from 
all third derivatives and the semi-diagonal quartic derivatives had been simulated using nonlinear optics in analogue quantum simulation \citep{sparrow2018simulating}, but this method cannot deal with more general anharmonic potential energy surface (PES).  
QPE based solutions have been reported to calculate eigenstates on anharmonic potential~\cite{McArdle2019Digital} and anharmonic spectrum~\cite{Sawaya2019Quantum}, but unable to realize with NISQ device. 

In this work, we propose a hybrid quantum-classical sampling (HQCS) algorithm to calculate the molecular vibrationally resolved electronic spectrum including both the anharmonic effect and the DR effect.
We first present an analysis of the computational complexity for this classically hard problem.
Second, we will describe the HQCS algorithm in detail, which includes three sub-components. 
The effectiveness of this HQCS algorithm is then demonstrated by calculating the emission spectrum of a two-mode model as well as a pyridine molecule using a simulator for photonic quantum computer. We believe that the real quantum device can accelerate the calculations in the near term.

\section{Results}
\subsection{Analysis of the complexity of computing molecular vibrationally resolved electronic spectrum}
\label{sec:analysis}

The molecular vibrationally resolved electronic spectrum is commonly calculated under the Born-Oppenheimer approximation~\cite{Lin2002Ultrafast}, in which two adiabatic electronic states are considered, denoted with subscript ``i/f'' for the initial/final electronic state (Symbols without subscript refer to a general state). With mass-weighted rectilinear coordinates, the Hamiltonian can be expressed as 

\begin{equation}
  \hat{H}_0 = \sum_{n=1}^{N}  -\frac{1}{2}\frac{\partial^2}{\partial q_n^2} +  
         \left[ \begin{matrix}
         V_\mathrm{f}( \mathbf{q}) & 0  \\
          0 &  V_\mathrm{i}( \mathbf{q}) \\
         \end{matrix} \right],    \label{eq:ham1}
\end{equation}
where $N$ is the total number of modes. $\mathbf{q} \equiv \{q_1, q_2,\cdots, q_N\}$ indicates that the PES $V$ is a multi-dimensional function. For semi-rigid molecules, it is preferred to express PES in the normal coordinates, because the mode coupling in this coordinate system is minimized. The two sets of normal coordinates of the initial and final PESs are related by the DR matrix $\mathbf{S}$ and the normal-mode-projected displacement $\Delta \mathbf{q}$.
\begin{equation}
\begin{aligned}
\mathbf{q}_\mathrm{f} &=\mathbf{S}\mathbf{q}_{\mathrm{i}} + \Delta \mathbf{q}_{\mathrm{f}}
\label{eq:mass_r}
\end{aligned}
\end{equation}

The lowest order approximation to PES is the harmonic approximation,
\begin{gather}
        V = \sum_{n} \frac{1}{2}\omega_{n}^2 q_{n}^2 + V_\textrm{eq} \label{eq:ha}
\end{gather}
Beyond that, the PES can be hierarchically expanded as 1-mode terms, 2-mode terms, 3-mode terms, \textit{etc.,} which is known as the $n$-mode representation ($n$-MR)~\cite{li2001high,bowman2003multimode} . In this work, we only consider 1-MR PES, in which the intra-mode anharmonicity is included and the modes are still independent of each other.
\begin{gather}
    V = \sum_{n} V_{n} (q_n) + V_\textrm{eq} \label{eq:1mr}
\end{gather}

With Fermi's golden rule and Condon's approximation, the optical transition rate at zero temperature can be expressed as
\begin{gather}
\sigma_{\textrm{abs/emi}}(E) 
= |\mu_{\textrm{if}} |^2 \sum_{ v_\mathrm{f}}  |\langle \phi_\mathrm{i}^\mathbf{0} (\mathbf{q}_\mathrm{i})|  \phi_\mathrm{f}^{\mathbf{v}_\mathrm{f}} (\mathbf{q}_\mathrm{f})\rangle|^2 \delta(E_{\mathrm{f},{\mathbf{v}_\mathrm{f}}}+E-E_{\mathrm{i},\mathbf{0}})
\label{eq:FGR}
\end{gather}
$|\phi_\mathrm{i/f}^{\mathbf{v}_\mathrm{i/f}} \rangle$ is the initial/final vibrational state with configuration $\mathbf{v}_\mathrm{i/f}$. $\mu_{\textrm{if}}$ is the transition dipole moment.
At zero temperature, the harmonic approximation is reasonable for $V_\textrm{i}$, but is inappropriate for $V_\textrm{f}$, because the optical transition starts from the equilibrium geometry of the initial state, which however could be far from the equilibrium geometry of the final state if the electron-vibration coupling is large\cite{wang2009anharmonic}. Therefore, in this work, $V_\mathrm{i}$ is described by a harmonic PES (Eq.~\eqref{eq:ha}) and $V_\mathrm{f}$ is described by a 1-MR PES (Eq.~\eqref{eq:1mr}). 

Unlike the harmonic spectrum with analytical solution, no accurate and efficient classical algorithm for the anharmonic spectrum has been proposed yet to calculate Eq.~\eqref{eq:FGR}. The computational cost of the sum-over-state algorithm will exponentially increase with the system size for the following two reasons.
First, the multi-dimensional integration of  FCFs $|\langle \phi_\mathrm{i}(\mathbf{q}_\mathrm{i})|\phi_\mathrm{f}(\mathbf{q}_\mathrm{f}) \rangle|^2$ with DR has been recognized as a $\#P$ problem\cite{Huh2015boson} with exponential computational scaling on classical computers. 
Second, the summation over $\mathbf{v}_\mathrm{f}$ is also an exponentially hard problem for classical computers, because the dimension of the Hilbert space increases exponentially with the system size (if $d$ vibrational states are considered in each mode, the number of $\mathbf{v}_\textrm{f}$ is $d^N$). 
To overcome these two exponentially hard problems, we propose a hybrid quantum-classical sampling algorithm. The details are described in the next section.

\subsection{hybrid quantum-classical sampling algorithm}

\begin{figure}[htbp]
\includegraphics[width=0.45 \textwidth]{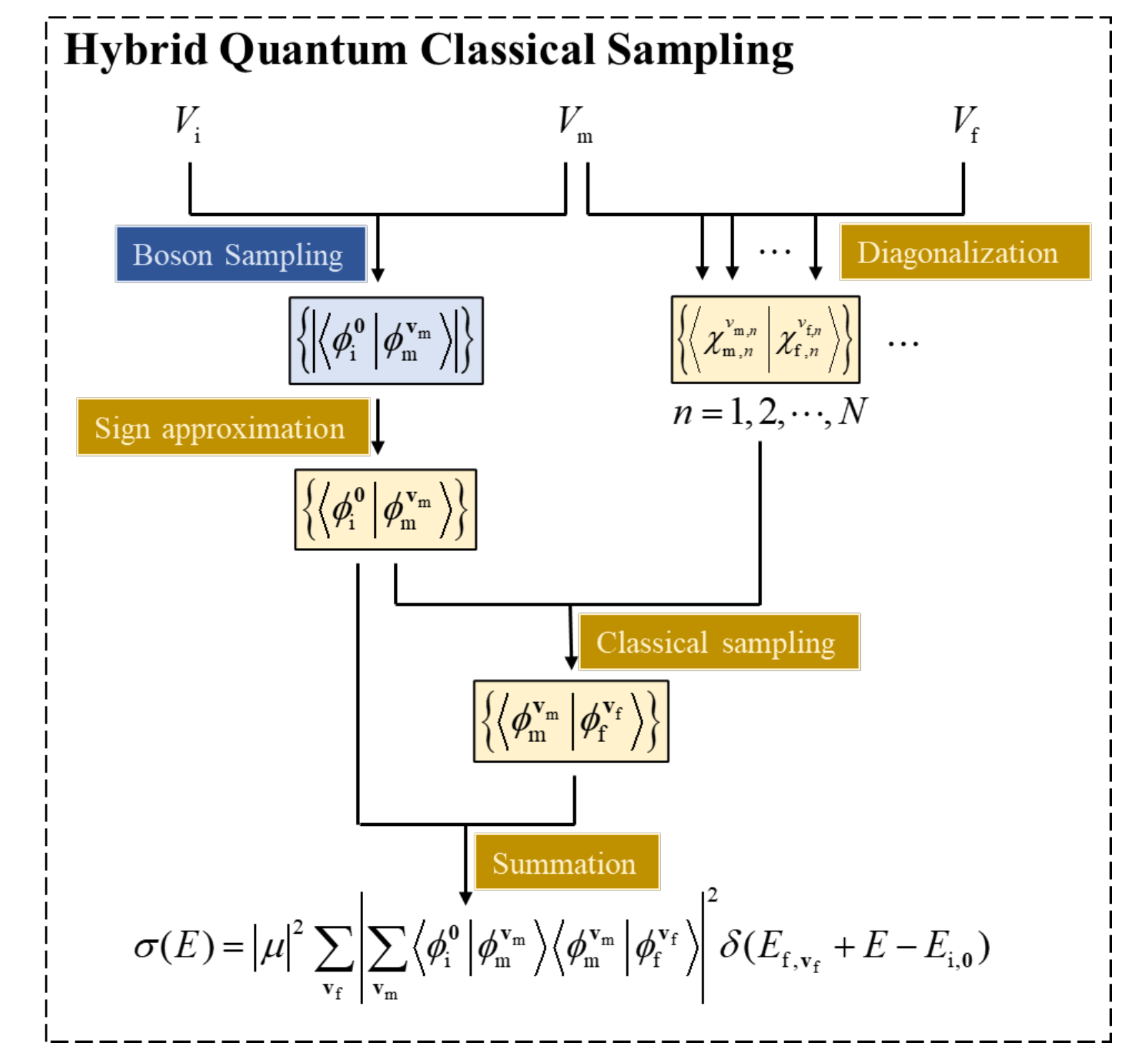}%
\caption{\label{fig:process} A schematic flowchart of Hybrid Quantum-Classical Sampling algorithm calculating the optical spectroscopy for transition from harmonic PES $V_\mathrm{i}$ to anharmonic and Duschinsky-rotated PES $V_\mathrm{f}$ by introducing a harmonic and Duschinsky-rotated PES $V_\mathrm{m}$
. Blue steps are performed on the quantum device, while yellow steps are performed on the classical computer. }
\end{figure}

The key to the HQCS algorithm is to build an intermediate harmonic PES, $V_{\mathrm{m}} = \sum_{n} \frac{1}{2}\omega_{\mathrm{m},n}^2 q_{\mathrm{m},n}^2$ and insert its complete eigenbasis set into Eq.~\eqref{eq:FGR}.
\begin{equation}
\begin{aligned}
     \sigma_\mathrm{abs/emi}(E)= |\mu_{\textrm{if}} |^2 \sum_{\mathbf{v}_\mathrm{f}}|\sum_{\mathbf{v}_{\mathrm{m}}}\langle \phi_{\mathrm{i}}^\mathbf{0}(\mathbf{q}_\mathrm{i})|\phi_{\mathrm{m}}^{\mathbf{v}_{\mathrm{m}}} (\mathbf{q}_\mathrm{m})\rangle\langle \phi_{\mathrm{m}}^{\mathbf{v}_{\mathrm{m}}}(\mathbf{q}_\mathrm{m})|\phi_{\mathrm{f}}^{\mathbf{v}_\mathrm{f}} (\mathbf{q}_\mathrm{f})\rangle|^2  \\
 \delta(E_{\mathrm{f},{\mathbf{v}_\mathrm{f}}}+E-E_{\mathrm{i},\mathbf{0}})
 \label{eq:FGR-lr}
\end{aligned}
\end{equation}
It seems that the evaluation of Eq.~\eqref{eq:FGR-lr} is even more complicated compared to Eq.~\eqref{eq:FGR}. But, with a smart choice of the parameters of the intermediate PES and taking advantage of both quantum and classical sampling algorithms, Eq.\eqref{eq:FGR-lr} can be efficiently calculated. The two requirements of $V_{\mathrm{m}}$ is that 
\begin{enumerate*}[label=\roman*)]
    \item $\omega_{\mathrm{m},n} = \mathrm{max}(\omega_{\mathrm{f},n},\omega_{\mathrm{i},n})$;
    \item $\mathbf{q}_{\mathrm{m}} = \mathbf{q}_\mathrm{f}$. 
\end{enumerate*}
With this intermediate PES, there are three steps to evaluate Eq.\eqref{eq:FGR-lr}, which are
\begin{enumerate}
    \item estimation of $\langle \phi_{\mathrm{i}}^\mathbf{0} (\mathbf{q}_\mathrm{i})|\phi_{\mathrm{m}}^{\mathbf{v}_{\mathrm{m}}}(\mathbf{q}_\mathrm{m})\rangle$;
    \item estimation of $\langle \phi_{\mathrm{m}}^{\mathbf{v}_{\mathrm{m}}}(\mathbf{q}_\mathrm{m})| \phi_{\mathrm{f}}^{\mathbf{v}_\mathrm{f}}(\mathbf{q}_\mathrm{f}) \rangle$;
    \item summation over $\mathbf{v}_\mathrm{m}$ and $\mathbf{v}_\mathrm{f}$.
\end{enumerate}
The schematic flowchart is shown in FIG.~\ref{fig:process}.
The first requirement is key to approximating the sign of $\langle \phi_{\mathrm{i}}^\mathbf{0} (\mathbf{q}_\mathrm{i})|\phi_{\mathrm{m}}^{\mathbf{v}_{\mathrm{m}}}(\mathbf{q}_\mathrm{m})\rangle$. The second requirement is to make the estimation of $\langle \phi_{\mathrm{m}}^{\mathbf{v}_{\mathrm{m}}}(\mathbf{q}_\mathrm{m})| \phi_{\mathrm{f}}^{\mathbf{v}_\mathrm{f}}(\mathbf{q}_\mathrm{f}) \rangle$ much easier. In the following, we replace $\mathbf{q}_\mathrm{m}$ with $\mathbf{q}_\mathrm{f}$ for simplicity.
The Dirac delta function in Eq.~\eqref{eq:FGR-lr} is broadened with Gaussian function  $\delta(E_{\mathrm{f},{\mathbf{v}_\mathrm{f}}}+E-E_{\mathrm{i},\mathbf{0}}) \to e^{-\frac{(E-E_{\mathrm{i},\mathbf{0}}+E_{\mathrm{f},{\mathbf{v}_\mathrm{f}}})^2}{2\sigma^2}}$. 

\subsubsection{estimation of $ \langle \phi_{\mathrm{i}}^\mathbf{0} (\mathbf{q}_\mathrm{i})|\phi_{\mathrm{m}}^{\mathbf{v}_{\mathrm{m}}}(\mathbf{q}_\mathrm{f}) \rangle $}

The calculation of the overlap $\langle \phi_{\mathrm{i}}^\mathbf{0} (\mathbf{q}_\mathrm{i})|\phi_{\mathrm{m}}^{\mathbf{v}_{\mathrm{m}}}(\mathbf{q}_\mathrm{f})\rangle$ between two harmonic wavefunctions with DR is a classically hard problem as we discussed in Section ~\ref{sec:analysis}. 
Here, we propose to calculate the norm and the sign ($-1$ or $+1$) separately. The former can be obtained through boson sampling on a quantum device. The latter can be approximated by a classical algorithm.
    
The Doktorov unitary operator $\mathbf{\hat{U}}_\mathrm{Dok}$ can connect two harmonic wavefunctions with rotation, distortion, and displacement ~\cite{doktorov1975dynamical,doktorov1977dynamical}. The expression 
of $\mathbf{\hat{U}}_\mathrm{Dok}$ can be found in the literature\cite{Quesada2019franck,barnett2002methods} (See the Supplementary Information for more details).
Hence, the FCFs can be expressed as
\begin{align}
|\langle \phi_\mathrm{i}^\mathbf{0}(\mathbf{q}_\mathrm{i})|  \phi_\mathrm{m}^{\mathbf{v}_\mathrm{m}}(\mathbf{q}_\mathrm{f}) \rangle|^2 = & |\langle \varphi_\mathrm{i}^\mathbf{0}(\mathbf{R}_\mathrm{i})| \varphi_\mathrm{m}^{\mathbf{v}_\mathrm{m}}(\mathbf{R}_\mathrm{f}) \rangle|^2 \nonumber \\ 
= &|\langle \varphi_\mathrm{i}^\mathbf{0}(\mathbf{R}_\mathrm{f})| \mathbf{\hat{U}}_\mathrm{Dok}| \varphi_\mathrm{m}^{\mathbf{v}_\mathrm{m}}(\mathbf{R}_\mathrm{f}) \rangle|^2.  \label{eq:dimensionless}
\end{align}
where $\mathbf{R}$ is the dimensionless normal coordinates as $\mathbf{R}=\mathbf{\omega}^{\frac{1}{2}} \mathbf{q}$ and $\phi(\mathbf{q}) = \varphi(\mathbf{R})$.  
The right-hand side of Eq.~\eqref{eq:dimensionless} is the same as the expression describing the sampling probability $P$ of one specific output harmonic state $|\phi_\mathrm{m}^{\mathbf{v}_\mathrm{m}} \rangle$ for an input harmonic state $|\phi_\mathrm{i}^\mathbf{0} \rangle $ passing through an interferometer accounted by unitary transformation $\mathbf{\hat{U}}_\mathrm{Dok}$  during the boson sampling process. 
The sampling probability has been confirmed as a $\#P$ problem hard on classical computers \cite{Aaronson2011the} but can be efficiently simulated with the boson sampling algorithm using a quantum device.
Because of the equivalence, the boson sampling algorithm~\cite{Huh2015boson,Jnane2021Analog} has been proposed to simulate FCFs between two harmonic states, in which the sampling probability $P_{\mathbf{v}_\mathrm{i} \rightarrow \mathbf{v}_{\mathrm{m}}}$ corresponds to the magnitude of FCFs. Thus, the norm can be obtained, that is,  
\begin{gather}
    |\langle \phi_{\mathrm{i}}^\mathbf{0}(\mathbf{q}_\mathrm{i})|\phi_{\mathrm{m}}^{\mathbf{v}_{\mathrm{m}}}  (\mathbf{q}_\mathrm{f})\rangle| = \sqrt{P_{\mathbf{0} \rightarrow \mathbf{v}_{\mathrm{m}}}}.
\end{gather}
The next step is to approximate the sign. For a one-dimensional problem, previous works~\cite{lin2007Symmetric,lin2010theoretical} have derived an analytic expression of the overlap between two one-mode harmonic wavefunctions $|\chi \rangle$.
The detailed expression at zero temperature is given in the method section. When $\omega_{\mathrm{m},n} = \mathrm{max}(\omega_{\mathrm{f},n}, \omega_{\mathrm{i},n}) \ge \omega_{\mathrm{i},n} $, the sign of the overlap can be obtained through Eq.~\eqref{eq:one-mode-sign}. 
\begin{equation}
    \mathrm{sgn}(\langle \chi_{\mathrm{i},n}^{0}(q_{\mathrm{f},n}) | \chi_{\mathrm{m},n}^{v_{\mathrm{m},n}}(q_{\mathrm{i},n}) \rangle) = (\mathrm{sgn}(\Delta q_{\mathrm{f},n}))^{v_{\mathrm{m},n}}
    \label{eq:one-mode-sign}
\end{equation} 
Thus, if without DR,
\begin{equation}
     \mathrm{sgn}(\langle \phi_{\mathrm{i}}^\mathbf{0} (\mathbf{q}_\mathrm{i}) | \phi_{\mathrm{m}}^{\mathbf{v}_{\mathrm{m}}}(\mathbf{q}_\mathrm{f}) \rangle) = \prod_n^N (\mathrm{sgn}(\Delta q_{\mathrm{f},n}))^{v_{\mathrm{m},n}}
    \label{eq:overlap-sign}
\end{equation}
With DR, the  factorized formula above is not exact anymore.
However, considering that the amplitude of the initial ground vibrational state is positive and nodeless, if DR only occurs between modes with the same frequency, DR will not change the sign of the overlap. 
Fortunately, in molecular systems, DR commonly occurs within groups of modes having close frequencies. 
Therefore, it can be speculated that DR could hardly flip the sign of the overlap and thus Eq.~\eqref{eq:overlap-sign} is still a good approximation. 
To verify the assumption, we calculate the sign of a two-mode and a four-mode model numerically in the Supplementary Information. The results showed that the assumption is quite reliable.
However, it should be noted that this approximation to the sign of the overlap is only reliable for the ground vibrational state as initial state, which limits our algorithm to the zero temperature case currently.

\subsubsection{estimation of $\langle \phi_{\mathrm{m}}^{\mathbf{v}_{\mathrm{m}}}(\mathbf{q}_\mathrm{f})| \phi_{\mathrm{f}}^{\mathbf{v}_\mathrm{f}}(\mathbf{q}_\mathrm{f}) \rangle$}
\label{sec:step2}

As the coordinates of $\phi_{\mathrm{m}}^{\mathbf{v}_{\mathrm{m}}}$ and $\phi_{\mathrm{f}}^{\mathbf{v}_{\mathrm{f}}}$ are the same and the modes are assumed to be independent, 
\begin{equation}
\langle \phi_{\mathrm{m}}^{\mathbf{v}_{\mathrm{m}}}(\mathbf{q}_\textrm{f})| \phi_{\textrm{f}}^{\mathbf{v}_\mathrm{f}}(\mathbf{q}_\textrm{f}) \rangle = \prod_n\langle \chi_{\mathrm{m}}^{v_{{\mathrm{m}},n}}(q_{\textrm{f},n})| \chi_{\mathrm{f}}^{v_{\mathrm{f},n}}(q_{\textrm{f},n})\rangle,
\label{eq:ed}
\end{equation}
where $\langle \chi_{\mathrm{m}}^{v_{{\mathrm{m}},n}}(q_{\textrm{f},n})| \chi_{\mathrm{f}}^{v_{\mathrm{f},n}}(q_{\textrm{f},n})\rangle$ is an one-dimensional problem and thus is quite easy to calculate on classical computers. Using  the harmonic states $\{ | \chi_{\mathrm{m}}^{v_{{\mathrm{m}},n}} \rangle \}$ with a cutoff on $v_{\mathrm{m},n}$ as the primitive basis set, we can obtain the corresponding discrete variable representation (DVR) basis set~\cite{light1985generalized} and the transformation matrix between the primitive harmonic basis and the DVR basis. The matrix elements of $\hat{H}_{\mathrm{f},n}= -\frac{1}{2}\frac{\partial^2}{\partial q_{\mathrm{f},n}^2} + V_{\mathrm{f},n}(q_{\mathrm{f},n})$ can be first expanded in DVR basis and then transformed to the primitive harmonic basis. Through exact diagonalizaiton of this Hamiltonian, the eigenstates $\{ | \chi_{\mathrm{f}}^{v_{\mathrm{f},n}} \rangle \}$ are obtained, so are their overlaps with the primitive basis.
For an $N$-mode system with $d$ basis functions for each mode,  the cost of the current sub-step has a linear scaling with the system size, \textit{i.e.}, $O(Nd^3)$. 
Finally, it should be noted that although the number of the overlap matrix elements $\langle \phi_{\mathrm{m}}^{\mathbf{v}_{\mathrm{m}}}(\mathbf{q}_\mathrm{f})| \phi_{\mathrm{f}}^{\mathbf{v}_\mathrm{f}}(\mathbf{q}_\mathrm{f}) \rangle$ is exponentially large, only the dominant elements sampled out in the next section actually need to be calculated.

\subsubsection{summation over $\mathbf{v}_\mathrm{m}$ and $\mathbf{v}_\mathrm{f}$}

\begin{figure*}[htbp]
\includegraphics[width=0.95 \textwidth]{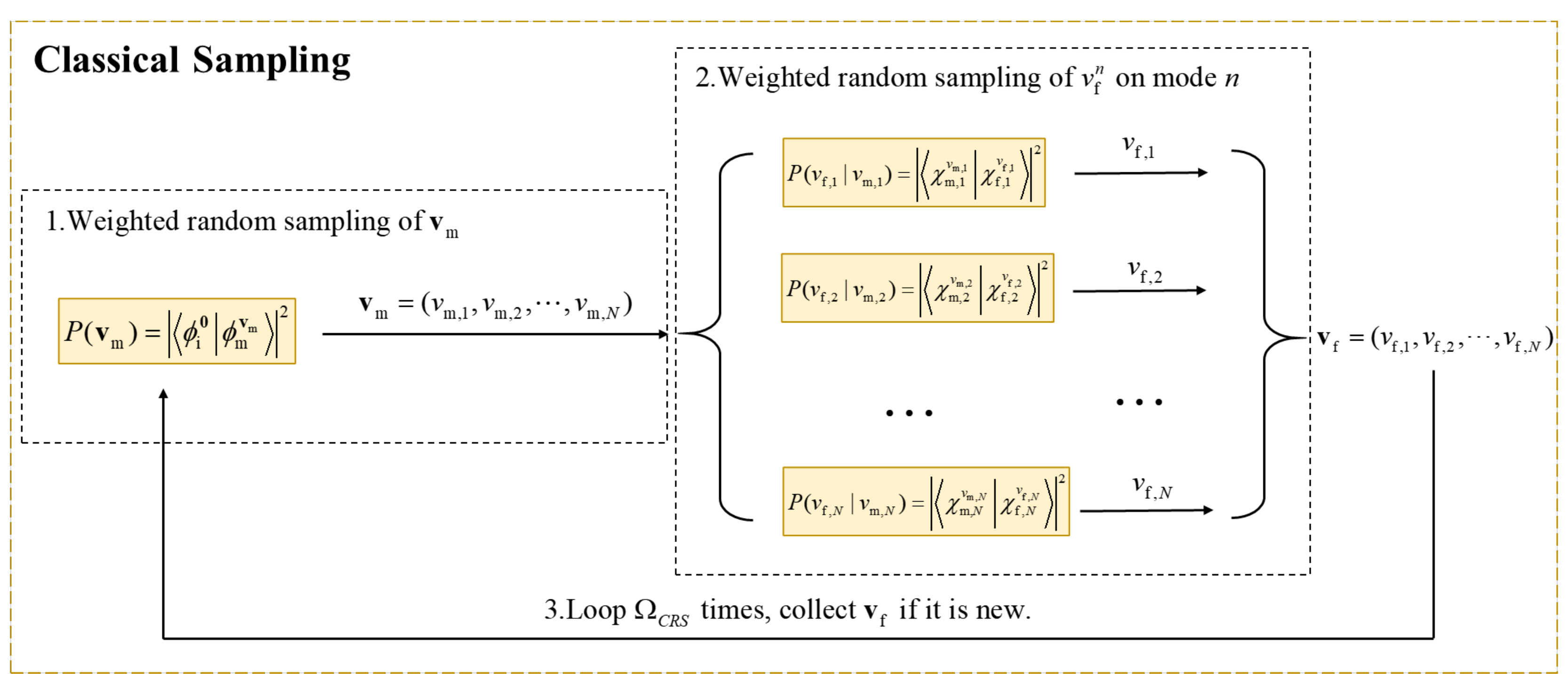}%
\caption{\label{fig:CS} A schematic flowchart of the classical sampling step used in HQCS.}
\end{figure*}

The last step is to make the summation over $ \mathbf{v}_\mathrm{m} $ and $ \mathbf{v}_\mathrm{f} $ in Eq.~\eqref{eq:FGR-lr}. Although the total number of $\mathbf{v}_\mathrm{m}$ and $ \mathbf{v}_\mathrm{f} $ both grow exponentially with the system size, in realistic systems only a small portion of the vibrational states with large FCFs determines the vibrational structure in electronic spectrum. 
Because the boson sampling will naturally screen out $\mathbf{v}_{\mathrm{m}}$ with dominant $|\langle \phi_\mathrm{i}^\mathbf{0}|\phi_{\mathrm{m}}^{\mathbf{v}_{\mathrm{m}}}\rangle|^2$, the summation over $\mathbf{v}_{\mathrm{m}}$ in Eq.~\eqref{eq:FGR-lr} will be classically efficient~\cite{Dierksen2005efficient,Santoro2007effective}.
Given a specific $\mathbf{v}_\mathrm{m}$, the dominant $\mathbf{v}_\mathrm{f}$ can also be obtained  according to the weight $ |\langle \chi_{\mathrm{m},n}^{v_{\mathrm{m},n}}| \chi_{\mathrm{f},n}^{v_{\mathrm{f},n}} \rangle|^2 $ by a  classical sampling (CS) algorithm.
A schematic flowchart of CS is shown in FIG.~\ref{fig:CS}. It consists of $\Omega_\mathrm{CS}$ sampling loops, among each of which a weighted random sampling is first performed for the intermediate vibrational  state configuration $ \mathbf{v}_{\mathrm{m}}=(\cdots,v_{\mathrm{m},n},\cdots) $ with the weight $ |\langle \phi_\mathrm{i}^\mathbf{0} |  \phi_{\mathrm{m}}^{\mathbf{v}_{\mathrm{m}}} \rangle|^2 $ collected from boson sampling result. 
Then, given one $\mathbf{v}_{\mathrm{m}}$, another weighted random sampling is carried out for  $v_{\mathrm{f},n}$ with the weight $ |\langle \chi_{\mathrm{m},n}^{v_{\mathrm{m},n}}| \chi_{\mathrm{f},n}^{v_{\mathrm{f},n}} \rangle|^2 $ calculated in Section~\ref{sec:step2}. 
After each loop, the newly appeared $\mathbf{v}_\mathrm{f} = (\cdots,v_{\mathrm{f},n},\cdots)$ will be collected. The eventual summations over $\mathbf{v}_\mathrm{f}$ is done only for those sampled out.
The upper bound of the number of summations is $O(M_\mathrm{m} M_\mathrm{f})$, where $M_\mathrm{m/f}$ are the number of intermediate/final state configurations sampled by boson sampling/CS. From CS, we can also obtain a set of $\{ (\mathbf{v}_\mathrm{m},\mathbf{v}_\mathrm{f}) \}$ pairs. The size of this set $M_{\mathrm{pair}}$ should be smaller than $M_\mathrm{m} M_\mathrm{f} $. Hence, for higher efficiency, we can just sum over this paired set $\{ (\mathbf{v}_\mathrm{m},\mathbf{v}_\mathrm{f}) \}$, giving cost $O(M_\mathrm{pair})$.    

For molecules with complex spectrum, we need to sample some small probability events to achieve high resolution. For such a purpose, a tunable bias $k$ can be added on the weight for CS such that $P(v_{\mathrm{f},n} |v_{\mathrm{m},n}) = \frac{|\langle \chi_{\mathrm{m},n}^{v_{\mathrm{m},n}}| \chi_{\mathrm{f},n}^{v_{\mathrm{f},n}} \rangle|^{2/k}}{\sum_{v_{\mathrm{f},n}}{|\langle \chi_{\mathrm{m},n}^{v_{\mathrm{m},n}}| \chi_{\mathrm{f},n}^{v_{\mathrm{f},n}} \rangle|^{2/k}}} $ to enhance the probability for sampling less probable events. 

\subsection{Results}
\begin{figure}[htbp]
    \centering
    \includegraphics[width=0.5 \textwidth]{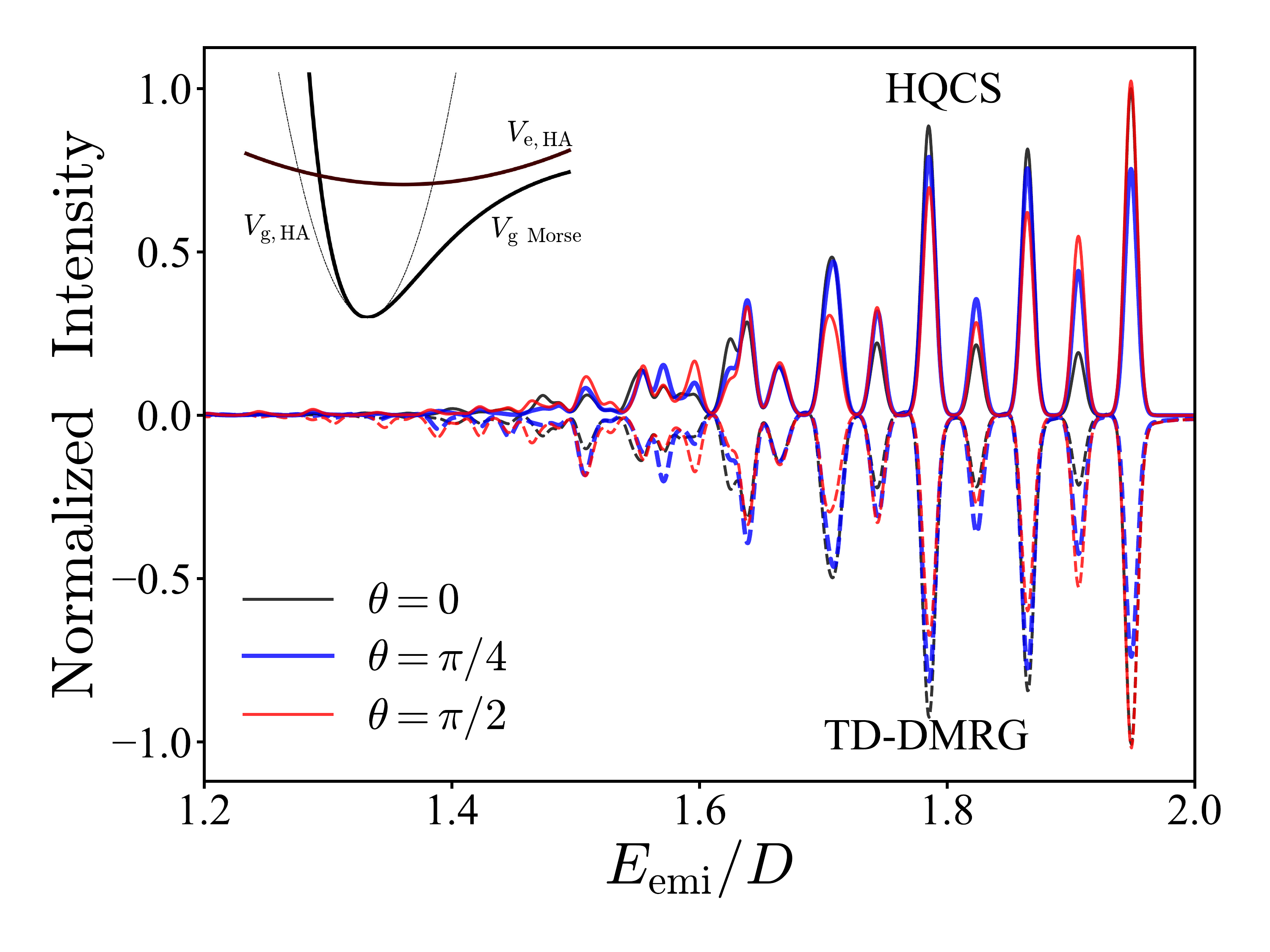}
    \caption{The emission spectrum of the two-mode Morse model. The reference spectrum produced by TD-DMRG are shown on the bottom panel (dashed lines). On the top (solid lines) are results produced by HQCS. Different $\theta$ values are used to manipulate the Duschinsky rotation effects. The intensity is normalized by the highest peak of the spectrum. Inset is the schematic PES along one mode of the two-mode Morse model. The initial excited state PES $V_{\mathrm{e},\mathrm{HA}}$ is harmonic (top thick line) and the final ground state PES $V_{\mathrm{g},\mathrm{Morse}}$is Morse (bottom 
   thick line). Performing harmonic approximation on $V_{\mathrm{g},\mathrm{Morse}}$ at equilibrium point, we obtain $V_{\mathrm{g},\mathrm{HA}}$  (bottom thin line), which is used as the intermediate PES. }
    \label{fig:2mode_model_compare}
\end{figure}

In order to validate the HQCS algorithm, we use the python package ``strawberry fields''\cite{Killoran2019strawberryfields} and ``Walrus''~\cite{Gupt2019the} to simulate the quantum boson sampling on classical computers.
The results are compared with reference spectrum calculated by the nearly-exact time-dependent density matrix renormalization group (TD-DMRG) method with the package Renormalizer~\cite{Renormalizer} developed by our group.
We calculate the emission spectrum of a two-mode model with initial excited state PES $V_\mathrm{i}$ as harmonic potential and the final ground state PES $V_\mathrm{f}$ as Morse potential. 
The intermediate PES $V_{\mathrm{m}}$ is obtained by performing harmonic expansion at equilibrium point on the Morse potential $V_\mathrm{f}$. 
This model has been used previously to investigate the anharmonic effect on the internal conversion process\cite{wang2021evaluating,Ianconescu2011Semiclassical}. Harmonic frequency and Morse parameters on the different between modes here and more details and parameters can be seen in the Supplementary Information. The DR matrix $\mathbf{S}$ between the PESs is parameterized by an angle $\theta$. 
\begin{equation}
\mathbf{S} =  \left[ \begin{matrix}
         \mathrm{cos} \theta & -\mathrm{sin} \theta  \\
          \mathrm{sin} \theta &  \mathrm{cos} \theta \\
         \end{matrix} \right] \label{eq:s-two-mode}
\end{equation}
The spectrum with different $\theta$ are shown in FIG.~\ref{fig:2mode_model_compare}. No matter $\theta$ is 0, $\pi/4$, or $\pi/2$, the HQCS algorithm reproduces the results of TD-DMRG quite well. 

\begin{figure}[htbp]
    \centering
    \includegraphics[width=0.5\textwidth]{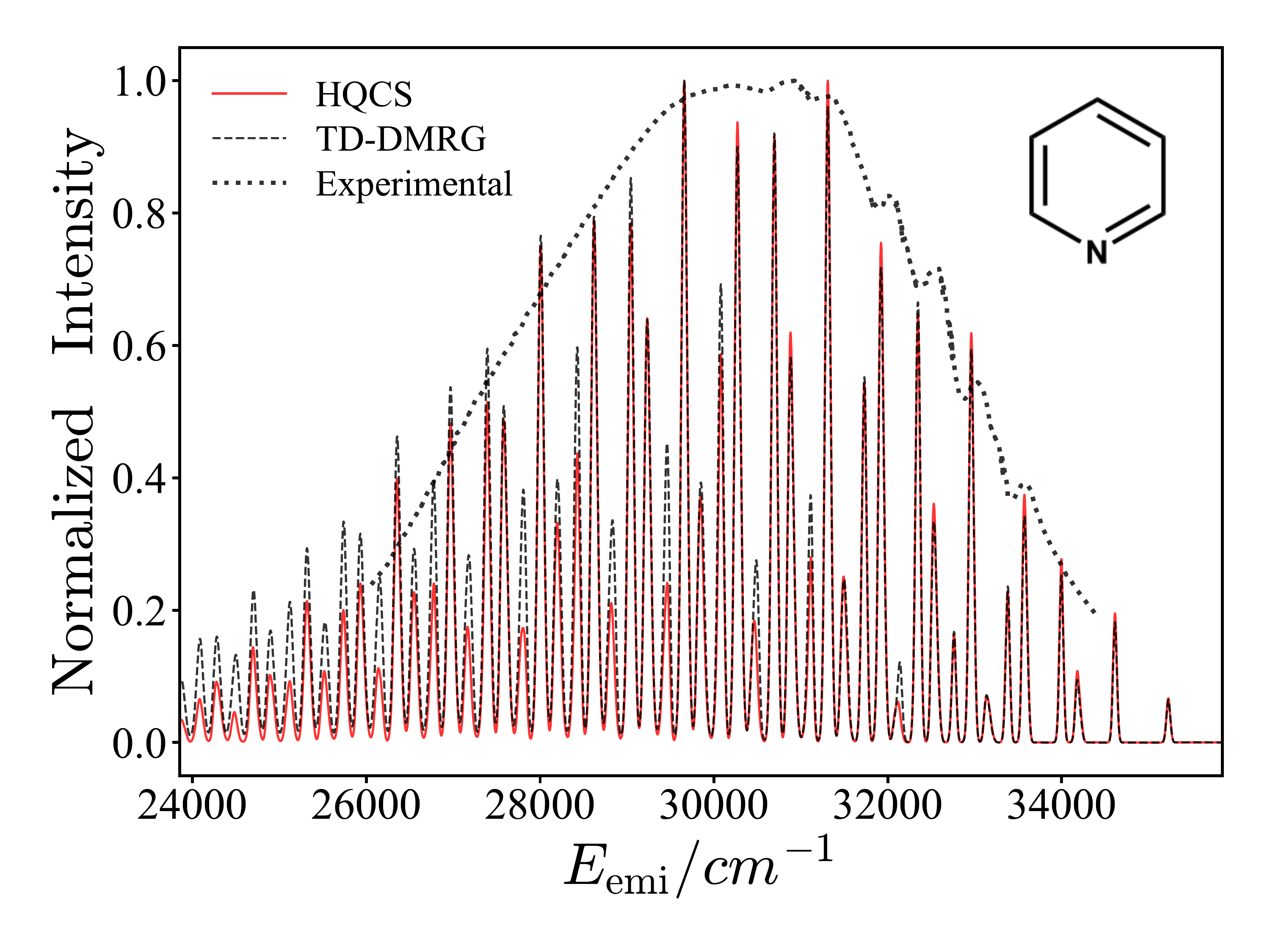}%
    \caption{The $S_1$ to $S_0$ emission spectrum of pyridine calculated with HQCS (red solid line) and reference result obtained by TD-DMRG (black dashed line). The anharmonic $S_0$ is approximated using 1-MR PES. The final states configurations $\mathbf{v}_\mathrm{f}$ are sampled by CS with $10^4$ loop times, altogether $M_\mathrm{f} = 936$ different configurations of $\mathbf{v}_\mathrm{f}$ are sampled out based on $M_{\mathrm{m}} = 1806$ configurations of $\mathbf{v}_{\mathrm{m}}$ sampled by boson sampling. The experimental results ('Experimental', black dotted curves) is also showed. spectrum are normalized with their highest peak. The theoretical (HQCS \& TD-DMRG) results have been blue-shifted by 1350 $\textrm{cm}^{-1}$ for better correspondence.}
    \label{fig:py_compare}
\end{figure}

Furthermore, we move to a real molecular system: the $S_1$ to $S_0$ emission spectrum of pyridine. Previous studies have used the perturbation method to handle the third-order force constants\cite{wang2009anharmonic} of $S_0$ PES to investigate the anharmonic effect on the emission spectrum. 
In our calculations, the electronic structure of pyridine and normal mode analysis are calculated at the B3LYP/6-31g(d) level with Gaussian 16 package \cite{g16}. Following that, the DR matrix $\mathbf{S}$, the displacement $\Delta \mathbf{q}$ are calculated by the program MOMAP\cite{niu2018molecular} developed by our group. For the anharmonicity, 1-MR $S_0$ PES fitted with polynomial functions up to 12th order is constructed by the adaptive density-guided approach implemented in MidasCpp ~\cite{midascpp} developed by Christiansen \textit{et al}.~\cite{sparta2009adaptive}. 
From a total of 27 normal modes, we select 7 important modes involving large displacements or DR effect for the HQCS simulation. 
For the rest 20 modes, only the difference in their zero-point energies of $S_0$ and $S_1$ PESs are considered, which shifts the spectrum. 
Detailed information of electronic structure calculation on pyridine and the way to select important modes are given in Supplementary Information. There, the reduced 7-mode model is demonstrated to be sufficient for the emission spectrum of pyridine.
Limited by the capability of the quantum simulator, we allow a maximum of 5 quanta on one mode for boson sampling and sampling $\Omega_\mathrm{BS}=10^5$ times in total.
The HQCS simulated emission spectrum are shown in FIG.~\ref{fig:py_compare} in comparison with TD-DMRG (The experimental fluorescence spectrum~\cite{Yamazaki1977Observation} is also shown). It shows that HQCS can well reproduce the features in the spectrum. The deviations at the weak-intensity peak originate from the limited sampling times and the 
cutoff on quanta. These deviations are expected to reduce with a real boson sampling device, which can simulate sufficiently large quanta and sampling times.

\begin{table*}[htbp]
\caption{\label{tab:CS-loop}Comparison of the number of configurations sampled with different $\Omega_\mathrm{CS}$.}
\begin{ruledtabular}
\begin{tabular}{c@{\hspace{2cm}}c@{\hspace{2cm}}c@{\hspace{2cm}}c}
 $\Omega_\mathrm{CS}$ & Number of sampled configuration $v_\mathrm{f}$  & $\sum_{\mathbf{v}_\mathrm{f}}|\sum_{\mathbf{v}_{\mathrm{m}}}\langle \phi_{\mathrm{i}}^\mathbf{0}|\phi_\mathrm{m}^{\mathbf{v}_{\mathrm{m}}} \rangle\langle \phi_{\mathrm{m}}^{\mathbf{v}_{\mathrm{m}}}|\phi_{\mathrm{f}}^{\mathbf{v}_\mathrm{f}} \rangle|^2$ \\
 \hline
 $10^1$&10&0.023  \\
  $10^2$&78&0.362  \\
  $10^3$&359&0.824  \\
  $10^4$&936&0.969  \\
  $10^5$&1797&0.996  \\
\end{tabular}
\end{ruledtabular}
\end{table*}

The deviation of $\sum_{\mathbf{v}_\mathrm{f}}|\sum_{\mathbf{v}_{\mathrm{m}}}\langle \phi_{\mathrm{i}}^\mathbf{0}|\phi_{\mathrm{m}}^{\mathbf{v}_{\mathrm{m}}} \rangle\langle \phi_{\mathrm{m}}^{\mathbf{v}_{\mathrm{m}}}|\phi_{\mathrm{f}}^{\mathbf{v}_\mathrm{f}} \rangle|^2$ from 1 can be used as a measure of  whether the classical sampling process is converged or not.
Table.~\ref{tab:CS-loop} shows the results with different $\Omega_\mathrm{CS}$. The results indicate that the summation over 936 final configurations $\mathbf{v}_\mathrm{f}$ from CS after $10^4$ loops and with 1806 intermediate configurations $\mathbf{v}_{\mathrm{m}}$ from the boson sampling can already give a quantitatively correct spectrum (the deviation is less than 5\%). It should be noted that the whole size of the final configuration space is $10^7$ (10 quanta for each of the 7 modes). Namely, the HQCS algorithm can vastly reduce the computational cost for the summation. 

\section{Discussion}

To perform the HQCS algorithm on an actual photonic quantum device, more error sources including photon loss, noise, and distinguishability\cite{Clements2018approximating} should be considered.
The HQCS algorithm described in this work can be further improved for higher efficiency in specific cases. For example, considering that the real experimental devices have limitations on the maximum number of quanta (basis cutoff) simulated on one optical mode\cite{Wang2020Efficient}, the intermediate PES $V_{\mathrm{m}}(\mathbf{q}_{\mathrm{m}})$ in HQCS should be fine-tuned to reduce the basis size needed on each optical mode to simulate the important configurations with large FCFs. One prescription is to decease the displacement between $\mathbf{q}_\mathrm{i}$ and $\mathbf{q}_{\mathrm{m}}$ because larger displacement often leads to larger overlap between $| \phi_{\mathrm{i}}^\mathbf{0} \rangle$ and high level
$|\phi_{\mathrm{m}}^{\mathbf{v}_{\mathrm{m}}} \rangle$ .

In conclusion, we have proposed a hybrid quantum-classical sampling algorithm, combining the boson sampling on quantum devices and the classical sampling on classical computers, for simulating molecular vibrationally resolved electronic spectroscopy including both Duschinsky rotation effect and anharmonic effect. 
Although the current algorithm is only for the zero temperature case and independent modes, the simplified problem is still classically hard.
The effectiveness of the HQCS algorithm is demonstrated in a two-mode Morse potential model and pyridine molecule by comparing with the nearly exact TD-DMRG method. 
Moreover, as we have intentionally designed the HQCS algorithm to contain only one step of boson sampling that have already been realized in many platforms~\cite{Clements2018approximating,Shen2018quantum,Wang2020Efficient} and other three steps on robust classical computers, 
we suggest HQCS is a practical near-term quantum simulation scheme to achieve quantum advantage for molecular spectroscopy and boost relating photophysical and photochemical research.

\section{Methods}
\subsection{Time correlation function formalism of spectrum }
spectrum from TD-DMRG for benchmark are obtained from the time correlation function (TCF) formalism. Here we will give a brief introduction. Detailed derivations can be seen from our former works\cite{Ren2018Time,Li2020Numerical}. Performing Fourier transformation of Delta function in  Eq.\eqref{eq:FGR} and then removing the completeness relation $\mathbf{I} = \sum_{\mathbf{v}_\mathrm{f}}| \phi_\mathrm{f}^{\mathbf{v}_\mathrm{f}} (\mathbf{q}_\mathrm{f})\rangle \langle \phi_\mathrm{f}^{\mathbf{v}_\mathrm{f}} (\mathbf{q}_\mathrm{f}) | $, we can obtain the TCF formalism as Eq.\eqref{eq:TCF}.
By TD-DMRG, the TCF $C(t)$  is first calculated within a finite time period and then is Fourier transformed to obtain the spectrum.
\begin{align}
\sigma_{\textrm{abs/emi}}(E) 
&= |\mu_{\textrm{if}} |^2 \int_{-\infty}^{\infty} dt \, e^{-iEt} C(t) \nonumber \\
C(t) &= e^{iE_{\mathrm{i},\mathbf{0}}t} 
 \langle \phi_\mathrm{i}^\mathbf{0} (\mathbf{q}_\mathrm{i}) |
e^{-i \hat{H}_\textrm{f} t} |   \phi_\mathrm{i}^\mathbf{0} (\mathbf{q}_\mathrm{i})\rangle
\label{eq:TCF}
\end{align}
\subsection{Parameters for computation}
\subsubsection{The spectrum of the 2-mode Morse model}
For HQCS, the quanta cutoff is 15 for each mode to perform boson sampling, which are operated $10^4$ times. Simple harmonic basis with frequency $\omega_{\mathrm{m},n}$ and size as 60 are used for each mode $n$ as primitive basis set in the exact diagonalization. Only 15 lowest eigenstates of each mode are used in the summation step. The whole Hilbert space is very small and thus CS is not needed here.  
For the reference spectrum produced by TD-DMRG, simple harmonic oscillator basis with frequency $\omega_{\mathrm{f},n}$ and size as 60 are used for each mode $n$ as primitive basis set. Bond dimension $M_S$ for matrix product state (MPS) is 10 for the initial state optimization and 4 for time evolution. 400 steps are propagated with time-step size $\Delta t = 8 \, \mathrm{a.u.}$ to calculate the TCF. The spectrum is broadened by a Gaussian function with $\sigma = 10^{-3} \, \mathrm{a.u.}$.
Corresponding DVR basis are used to expand the Morse form potential operator and then transformed to the primitive basis for both methods.
\subsubsection{The spectrum of pyridine}
For HQCS, the quanta cutoff is 5 for each mode to perform boson sampling because of the heavy computational cost of the classical simulation of the boson sampling algorithm. $10^5$ times of boson sampling operation are simulated. Simple harmonic oscillator basis with frequency $\omega_{\mathrm{m},n}$ and size as 60 on each mode $n$ are used in the exact diagonalization and only the lowest 10 eigenstates on each mode are used in CS and then the summation step. 
For the reference spectrum produced by TD-DMRG, simple harmonic oscillator basis with frequency as $\omega_{\mathrm{f},n}$ and size as 60 are used for each mode $n$. Bond dimension $M_S$ for MPS is 40 for the initial state optimization and 10 for time evolution. 1200 steps are propagated with time-step size $\Delta t = 32 \, \mathrm{a.u.}$ to calculate TCF. The spectrum is broadened by a Gaussian function  with $\sigma = 10^{-4} \, \mathrm{a.u.}$.
DVR basis is not used here because pyridine 1-MR PES in this work is described by polynomial functions so the matrix elements of potential operator on simple harmonic oscillator basis used is analytical.  
\subsection{Analytical expression of the overlap between one-dimensional harmonic states}
\begin{equation}
\langle\chi_{\mathrm{i}}^0 | \chi_{\mathrm{m}}^{v_\mathrm{m}} \rangle =\Gamma (\Delta q_\mathrm{f})^{v_{\mathrm{m}}} \times (\sum_l C_l )
\label{eq:one-mode-ana}
\end{equation}
\begin{equation}
\begin{aligned}
C_l = & \frac{(-1)^{v_{\mathrm{m}}-l}+1}{2} (\Delta q_\mathrm{f})^{-(v_{\mathrm{m}}-l)}\beta_{\mathrm{m}}^{l/2}   \\
& \frac{2^l}{l!} (\frac{\beta_\mathrm{i}}{\beta_\mathrm{i} +\beta_{\mathrm{m}}})^{l} (\frac{\beta_{\mathrm{m}}- \beta_\mathrm{i}}{\beta_\mathrm{i} +\beta_{\mathrm{m}}})^{\frac{v_{\mathrm{m}}-l}{2}}  \frac{1}{(\frac{v_{\mathrm{m}}-l}{2})!}  \\
\Gamma= &(\frac{v_{\mathrm{m}}!}{2^{v_{\mathrm{m}}}} \times \frac{2\sqrt{\beta_\mathrm{i} \beta_{\mathrm{m}} }}{\beta_\mathrm{i} +\beta_{\mathrm{m}}})^\frac{1}{2} \mathrm{exp}(-\frac{\beta_\mathrm{i} \beta_{\mathrm{m}} (\Delta q_\mathrm{f})^2}{2(\beta_\mathrm{i} +\beta_{\mathrm{m}})})
\label{eq:cm}
\end{aligned}
\end{equation}
where $\beta_\mathrm{i} = \omega_\mathrm{i} / \hbar$, $l$ are non-negative integers meeting the condition $l \le v_{\mathrm{m}}$. As $\Gamma$-function is always non-negative and $C_l$ is non-negative when $\omega_{\mathrm{m},n}$ $\ge$ $\omega_{\mathrm{i},n}$, the sign of the overlap only depends on the sign of the displacement $\Delta q_\mathrm{f}$ and the quanta $v_\mathrm{m}$ of the intermediate state.

\section{references}

\section{Acknowledgment}
This work is supported by the National Natural Science Foundation of China Grant Nos. 21788102 and by the Ministry of Science and Technology of China through the National Key R\&D Plan Grant No. 2017YFA0204501

\section{Author contributions}
Z.S. supervised the study. Y.W. performed the numerical calculations,
analyzed the data, and wrote the manuscript with the help of J.R and W.L
\section{Competing interests}
The authors declare no competing interests.

\section{Materials \& Correspondence}
Correspondence and requests for materials should be addressed to Z.S.
\end{document}


\title[]{Supplementary information for 'Hybrid Quantum-Classical Boson Sampling Algorithm for Molecular Vibrationally Resolved Electronic Spectroscopy with Duschinsky Rotation and Anharmonicity'}
\author{Yuanheng Wang}
\affiliation{MOE Key Laboratory of Organic OptoElectronics and Molecular Engineering, Department of Chemistry, Tsinghua University, Beijing 100084,
 People's Republic of China }
\author{Jiajun Ren}
\affiliation{MOE Key Laboratory of Organic OptoElectronics and Molecular Engineering, Department of Chemistry, Tsinghua University, Beijing 100084,
 People's Republic of China }
\affiliation{Current Address: Key Laboratory of Theoretical and Computational Photochemistry, Ministry of Education, College of Chemistry, Beijing Normal University, Beijing 100875, People’s Republic of China}
\author{Weitang Li}
\affiliation{MOE Key Laboratory of Organic OptoElectronics and Molecular Engineering, Department of Chemistry, Tsinghua University, Beijing 100084,
 People's Republic of China }
\author{Zhigang Shuai}%
 \email{zgshuai@tsinghua.edu.cn}
\affiliation{MOE Key Laboratory of Organic OptoElectronics and Molecular Engineering, Department of Chemistry, Tsinghua University, Beijing 100084,
 People's Republic of China }

\maketitle

\section{Benchmark of sign approximation method}
Here, We compare the sign: $\mathrm{sgn}(\langle \phi_\mathrm{i}^\mathbf{0} | \phi_\mathrm{f}^{\mathbf{v}_\mathrm{f}} \rangle)$ calculated through numerical methods (exact diagonalization) and sign approximation method introduced in the main article as Eq.~\ref{eq:sign-approximate} with a two-mode model and a four-mode model. Studied model Hamiltonian can be written as Eq.~\ref{eq:ham1}. $V_\mathrm{i}$ is set more smooth than $V_\mathrm{f}$ as in Eq.~\ref{eq:smooth} .

\begin{equation}
  \hat{H}_0 = \sum_{n=1}^{N} -\frac{1}{2}\frac{\partial^2}{\partial q_n^2} + \left[ \begin{matrix}
         V_\mathrm{f}( \mathbf{q}) & 0  \\
          0 &  V_\mathrm{i}( \mathbf{q}) \\
         \end{matrix} \right] \label{eq:ham1}
\end{equation}

\begin{gather}
V_\mathrm{i} = \sum_{n=1}^N \frac{ \omega_{\mathrm{i},n}^2 q_{\mathrm{i},n}^2}{2} \\
V_\mathrm{f} = \sum_{n=1}^N \frac{\omega_{\mathrm{f},n}^2 q_{\mathrm{f},n}^2}{2}  \\
\omega_{\mathrm{i},n}=\omega_{\mathrm{f},n}/5 \label{eq:smooth}
\end{gather}
\begin{gather}
q_{\mathrm{f},n} =\sum_{l}S_{nl} q_{\mathrm{i},l} + \Delta q_{\mathrm{f},n}
\end{gather}

As $\omega_\mathrm{f} \ge \omega_\mathrm{i}$ , the requirement for reliable sign approximation as Eq.~\ref{eq:sign-approximate} is met. Here we set $\Delta q_{\mathrm{f},n} = -\frac{1}{\omega_0}\sqrt{\frac{D}{2}}$ for all modes unless special noticed , where $D=5.52 \mathrm{eV}$ and $\omega_0 = 3868 \mathrm{cm^{-1}}$. In fact, any $\Delta q_{\mathrm{f},n}$ to ensure $\mathrm{sgn}(\Delta q_{\mathrm{f},n})= -1$  can meet the condition for this benchmark to make the sign (featured by color) flips when quanta $v_{\mathrm{f},n}$ on any one mode $n$ increases or decreases by one. We select this number only for more obvious displaying. For simplicity, in the graphics below we directly use $v_n$ to represent $v_{\mathrm{f},n}$ .
\begin{gather}
\mathrm{sgn}(\langle \phi_\mathrm{i}^\mathbf{0} | \phi_\mathrm{f}^{\mathbf{v}_\mathrm{f}} \rangle) = \prod_{n=1}^N (\mathrm{sgn}(\Delta q_{\mathrm{f},n}))^{v_{\mathrm{f},n}}
\label{eq:sign-approximate}
\end{gather}

\subsection{Two-mode model}
\begin{figure*}[h]
\subfloat[]{
\includegraphics[width=0.3\textwidth]{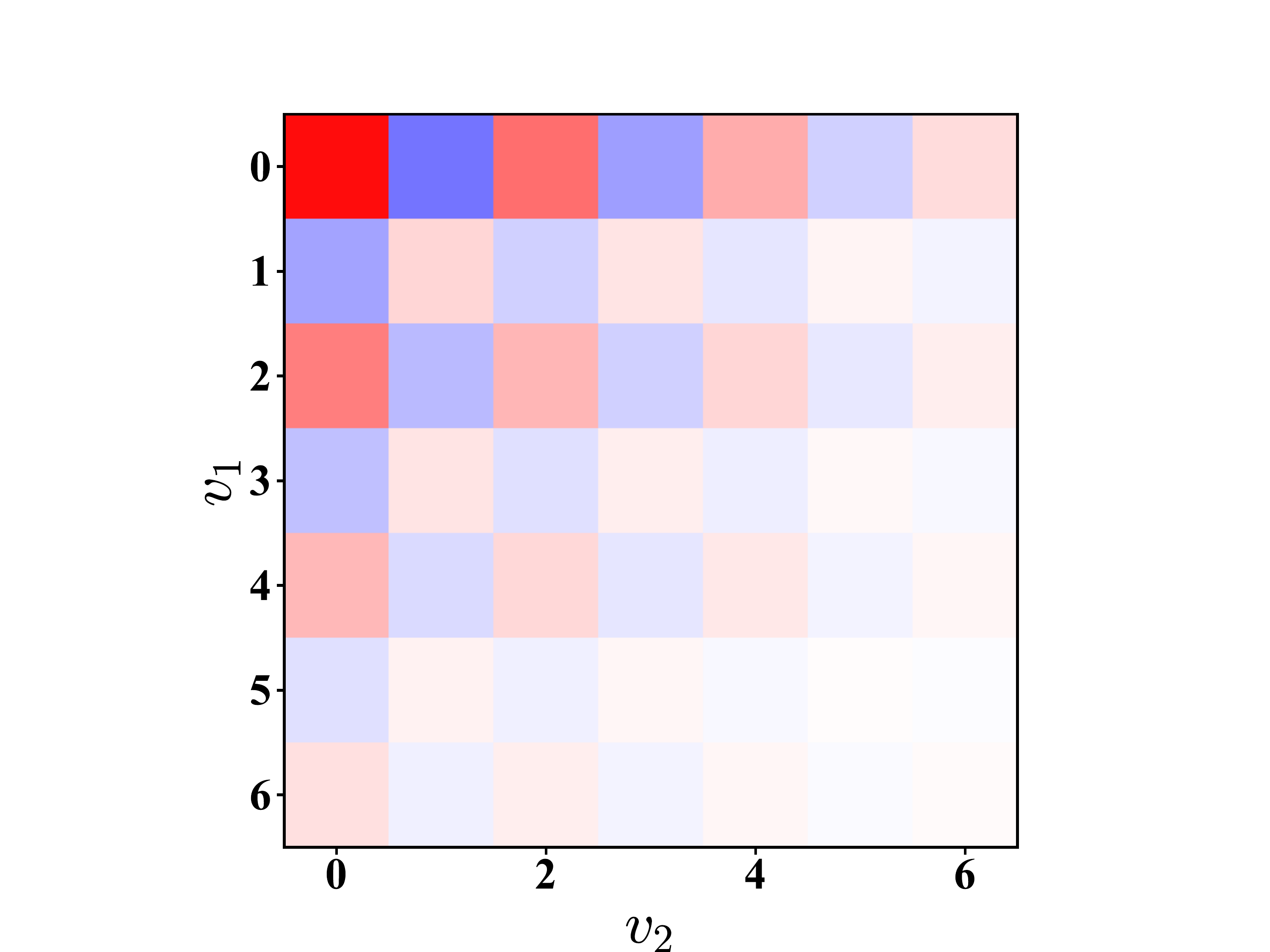}%
    }
\subfloat[]{
\includegraphics[width=0.3\textwidth]{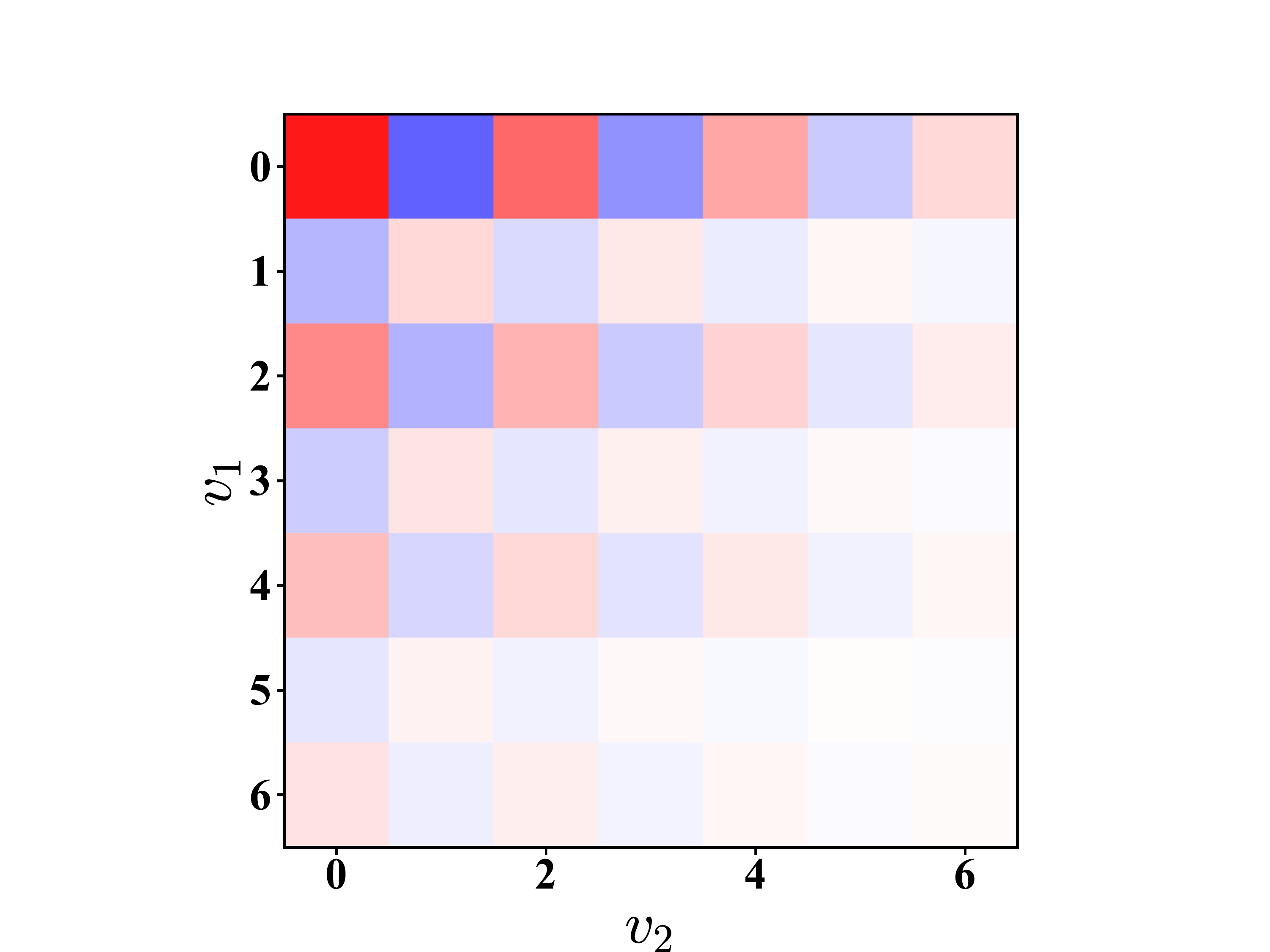}%
    }
\subfloat[]{
\includegraphics[width=0.3\textwidth]{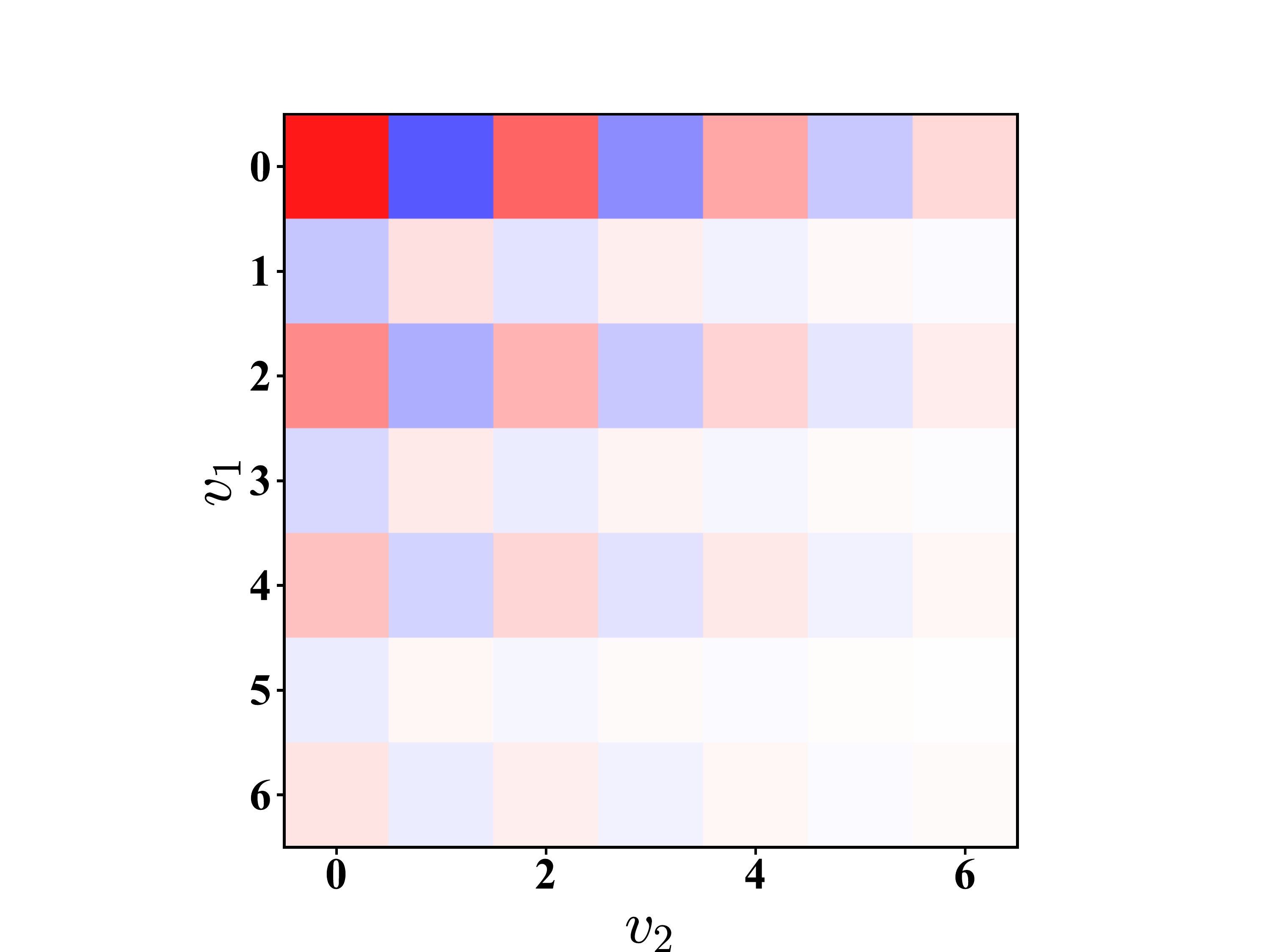}%
    }\\
\subfloat[]{
\includegraphics[width=0.3\textwidth]{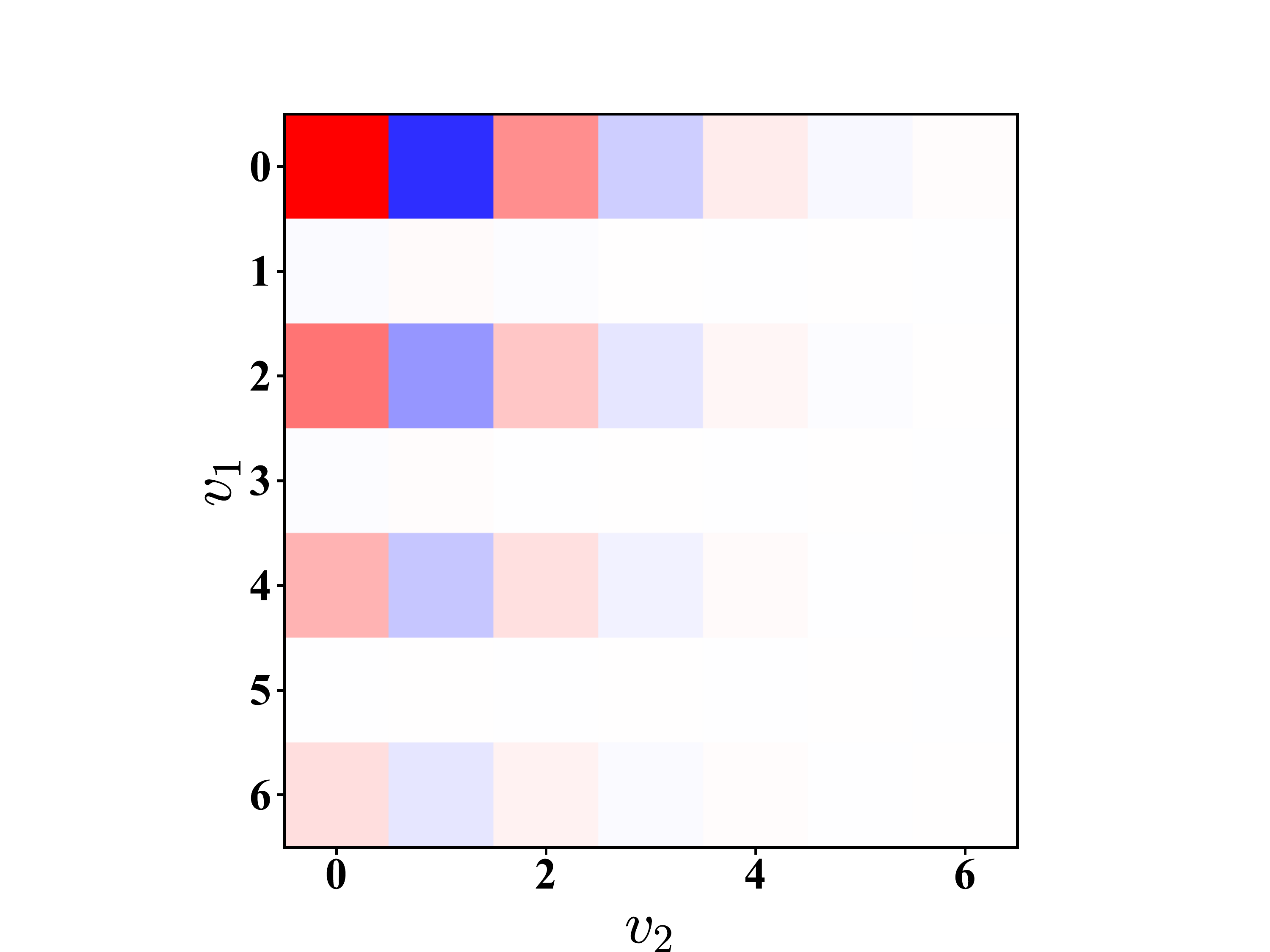}%
    }
\subfloat[]{
\includegraphics[width=0.3\textwidth]{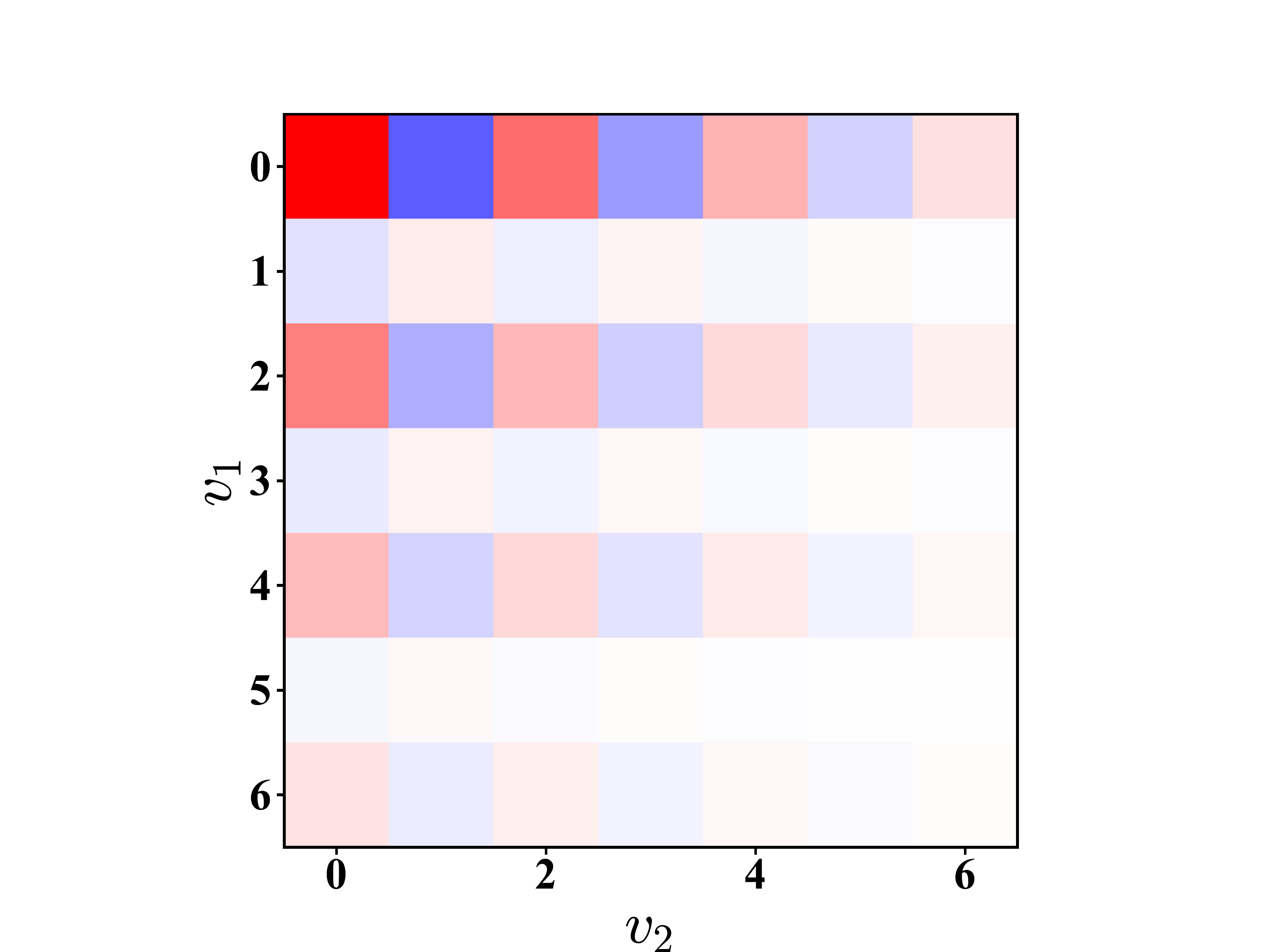}%
    }
\subfloat[]{
\includegraphics[width=0.3\textwidth]{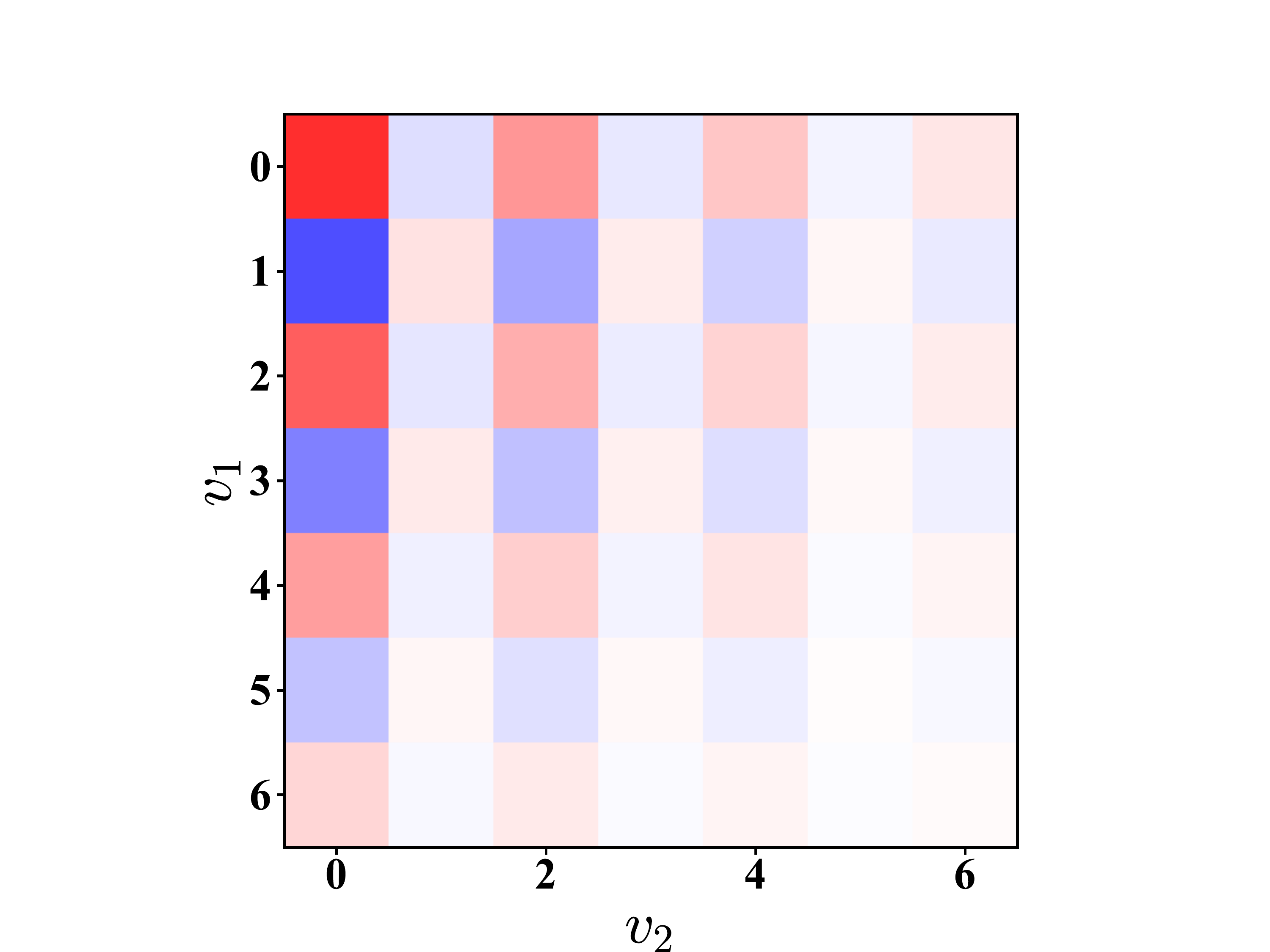}%
    }
    
\caption{\label{fig:two mode}Benchmark of the sign of overlap $\langle \phi_\mathrm{i}^\mathbf{0} | \phi_\mathrm{f}^{(v_1,v_2)} \rangle$ in different two-mode models through exact diagonalization. Red means positive overlap and blue means negative overlap. Darker color means larger norm of the overlap. $\omega_{\mathrm{f},1}= \omega_{0}$ for all models, (a): $\omega_{\mathrm{f},2} = 0.5\omega_{\mathrm{f},1}$, $\theta =\pi/6$; (b): $\omega_{\mathrm{f},2} = 0.5\omega_{\mathrm{f},1}$, $\theta =\pi/4$; (c): $\omega_{\mathrm{f},2} = 0.5\omega_{\mathrm{f},1}$, $\theta =\pi/3$; (d): $\omega_{\mathrm{f},2} = 0.2\omega_{\mathrm{f},1}$, $\theta =\pi/2$; (e): $\omega_{\mathrm{f},2} = 0.5\omega_{\mathrm{f},1}$, $\theta =\pi/2$ and (f): $\omega_{\mathrm{f},2} = 2\omega_{\mathrm{f},1}$, $\theta =\pi/2$. The basis for mode $n$ is simple harmonic oscillator with frequency $\omega_{\mathrm{f},n}$ and size as 7.}
\end{figure*}

First, we perform benchmark in a two-mode model where $N=2$. We carefully set the frequency of two modes and manipulate the duschinsky rotation matrix $\mathbf{S}$ through a variable $\theta$ as in Eq.~\ref{eq:s-two-mode} to perform benchmark. The results in FIG.~\ref{fig:two mode} show our sign approximation method is reliable within wide situations studied.

\begin{equation}
\mathbf{S} =  \left[ \begin{matrix}
         \mathrm{cos}\theta & -\mathrm{sin}\theta  \\
          \mathrm{sin}\theta &  \mathrm{cos}\theta \\
         \end{matrix} \right] \label{eq:s-two-mode}
\end{equation}

\subsection{Four-mode model}
\begin{figure*}[h]
\subfloat[]{
\includegraphics[width=0.2\textwidth]{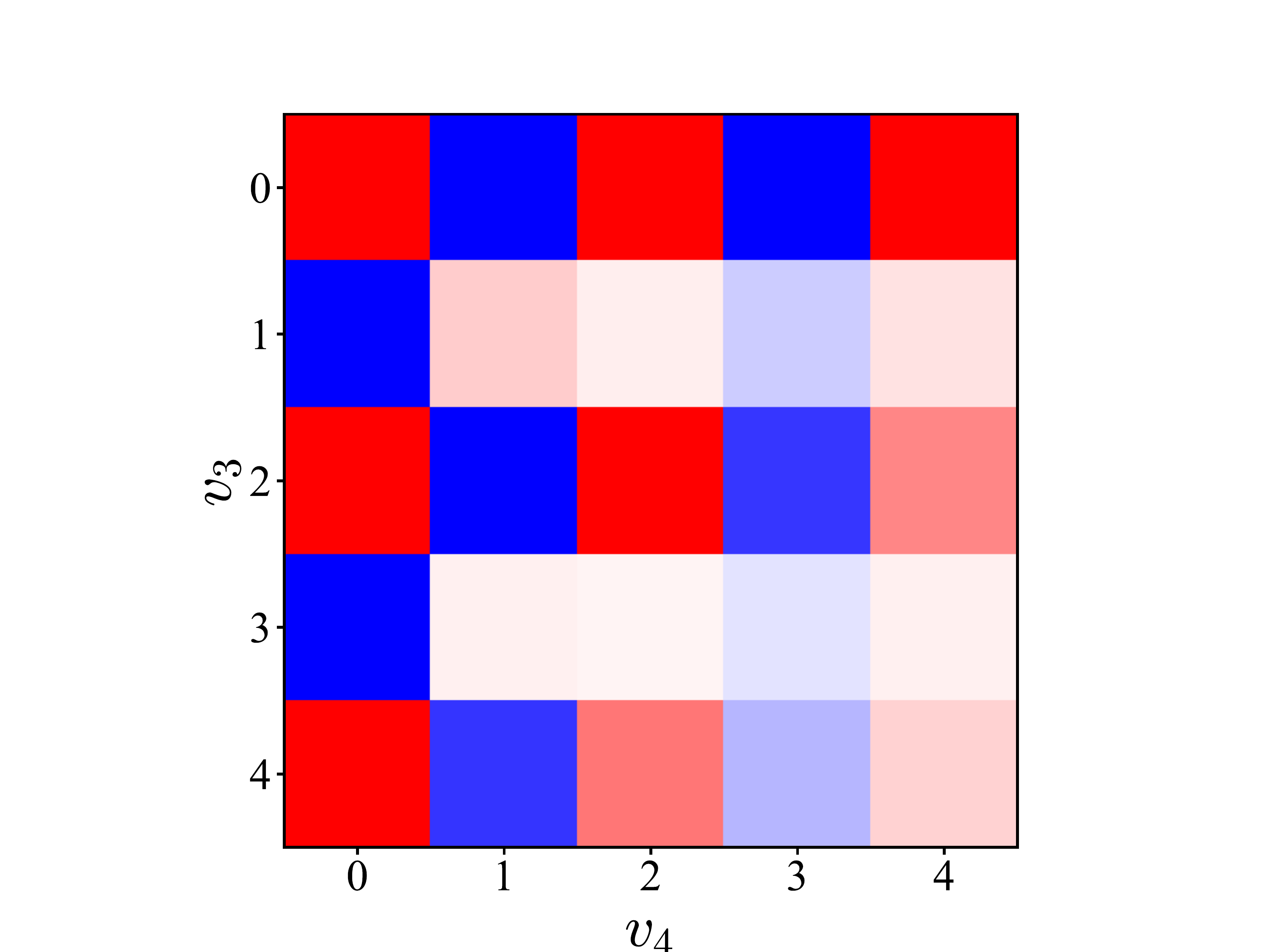}%
    }
\subfloat[]{
\includegraphics[width=0.2\textwidth]{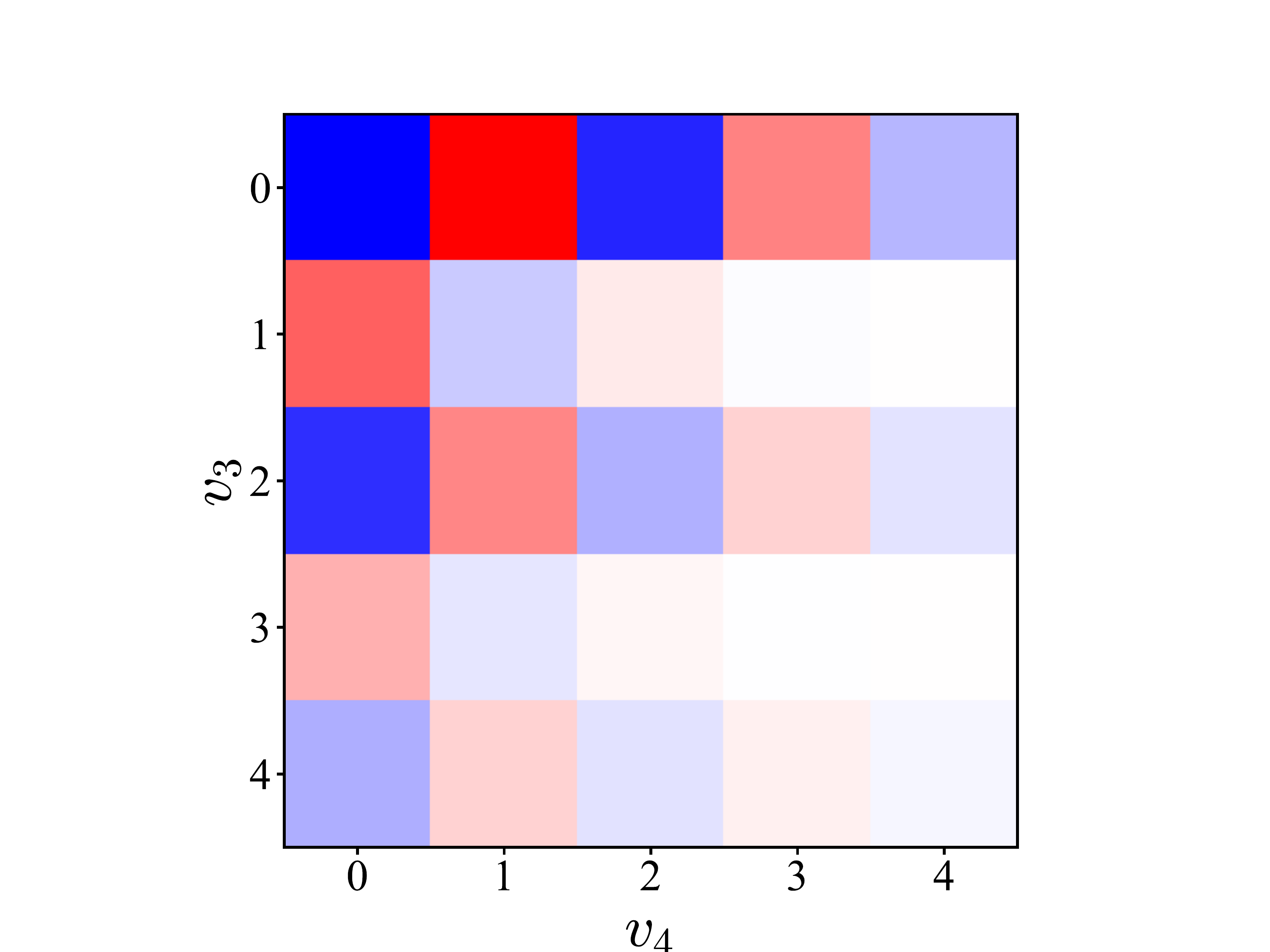}%
    }
\subfloat[]{
\includegraphics[width=0.2\textwidth]{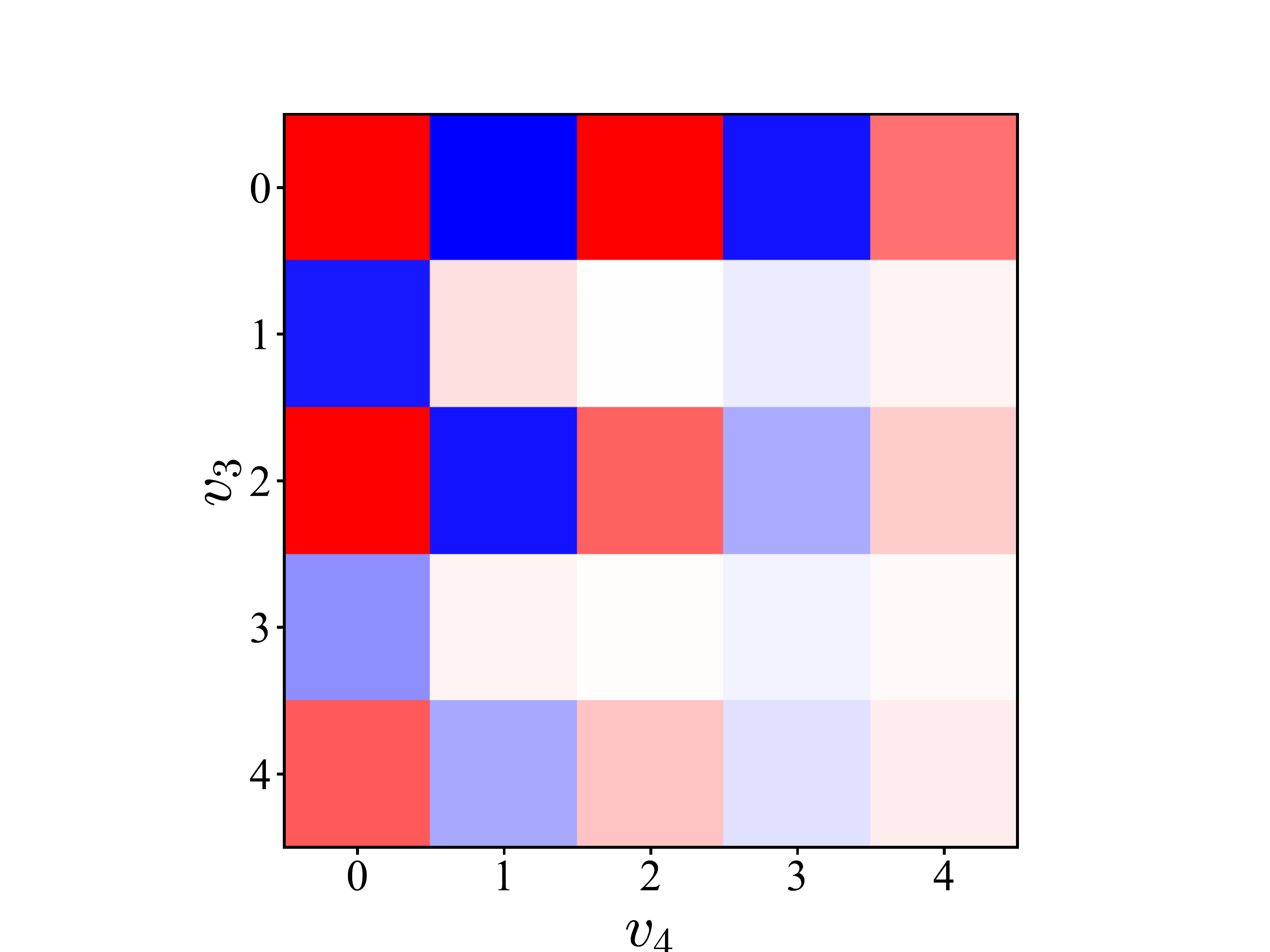}%
    }
\subfloat[]{
\includegraphics[width=0.2\textwidth]{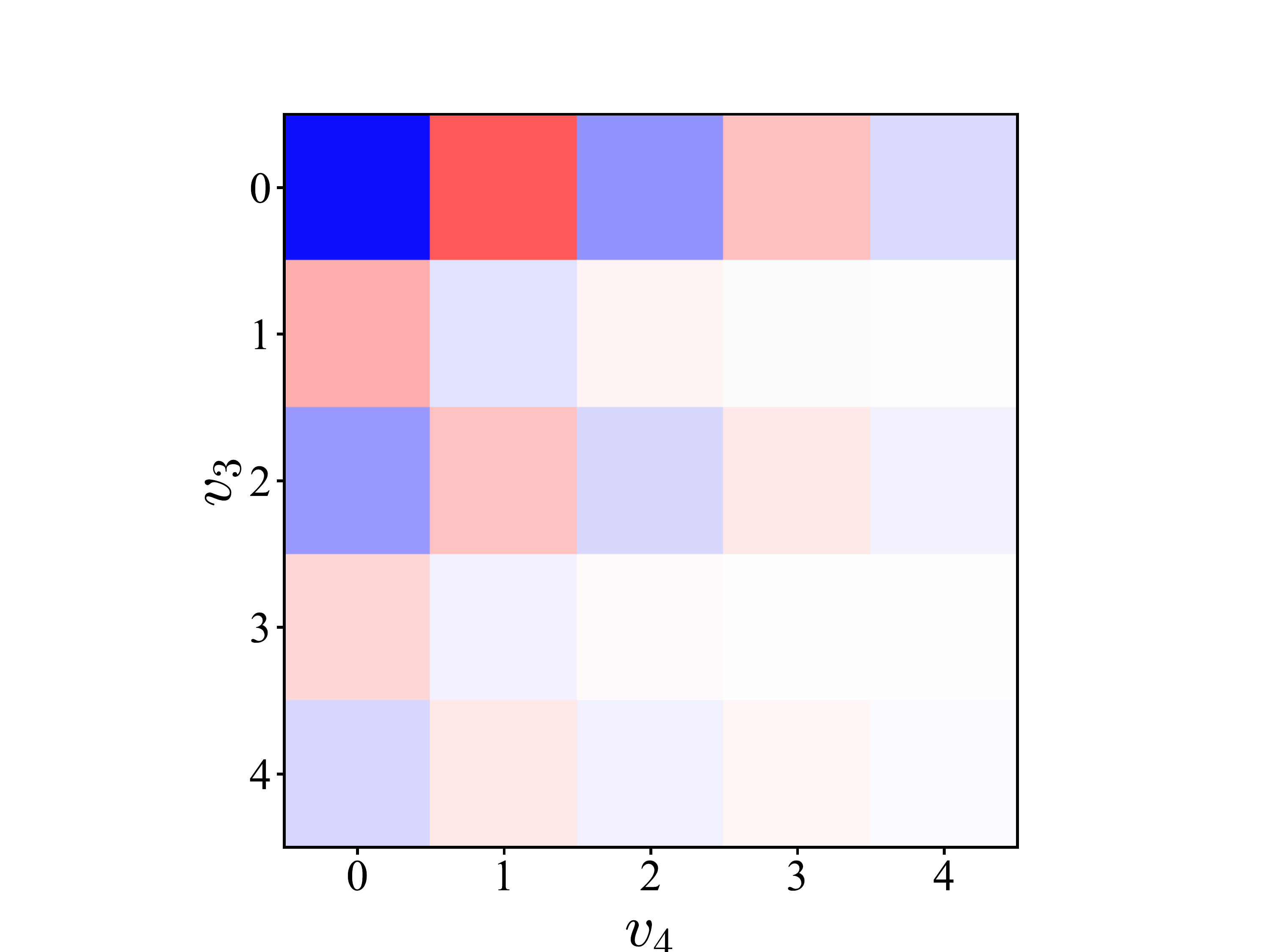}%
    }
\subfloat[]{
\includegraphics[width=0.2\textwidth]{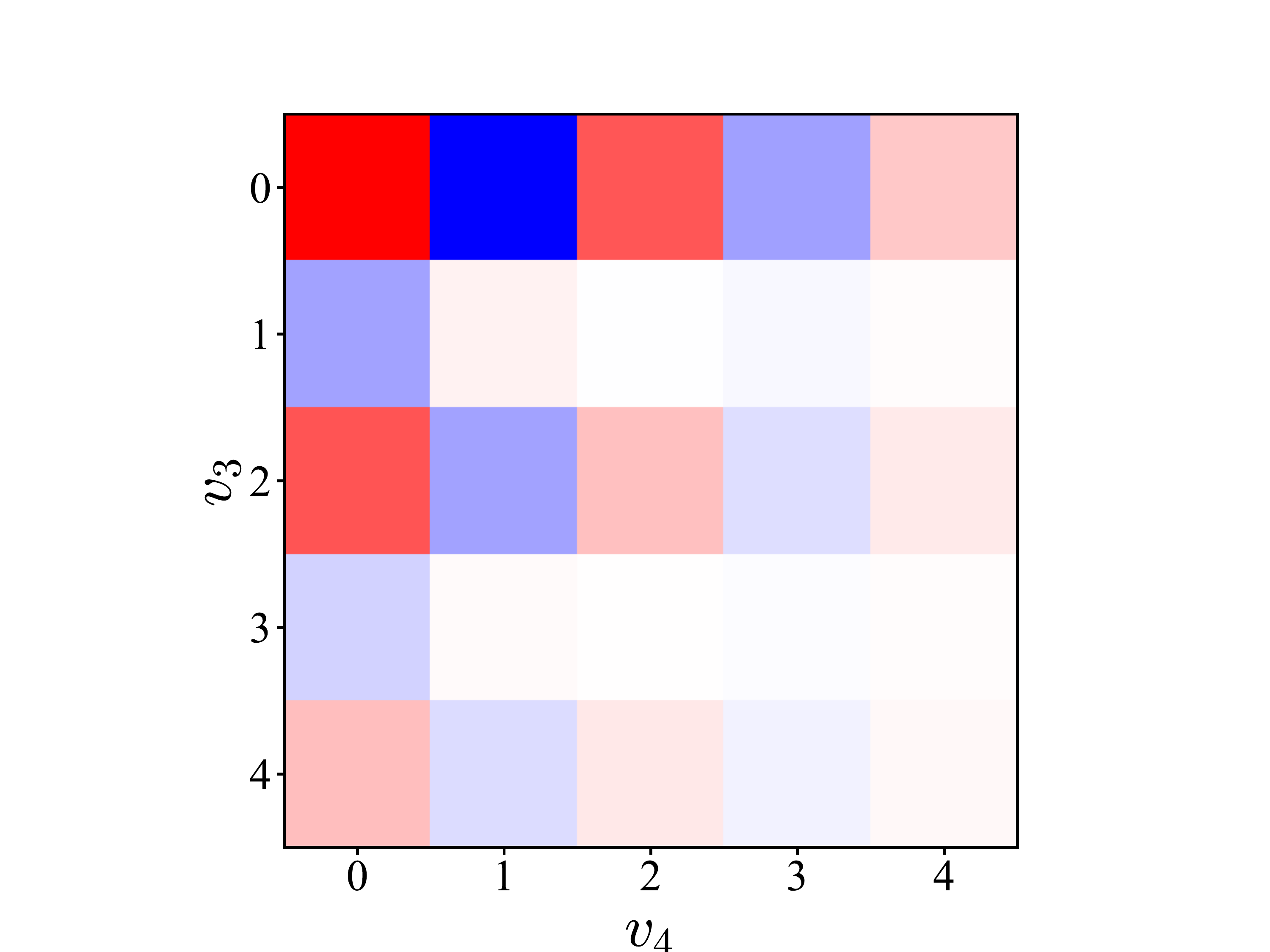}%
    }\\
\subfloat[]{   
\includegraphics[width=0.2\textwidth]{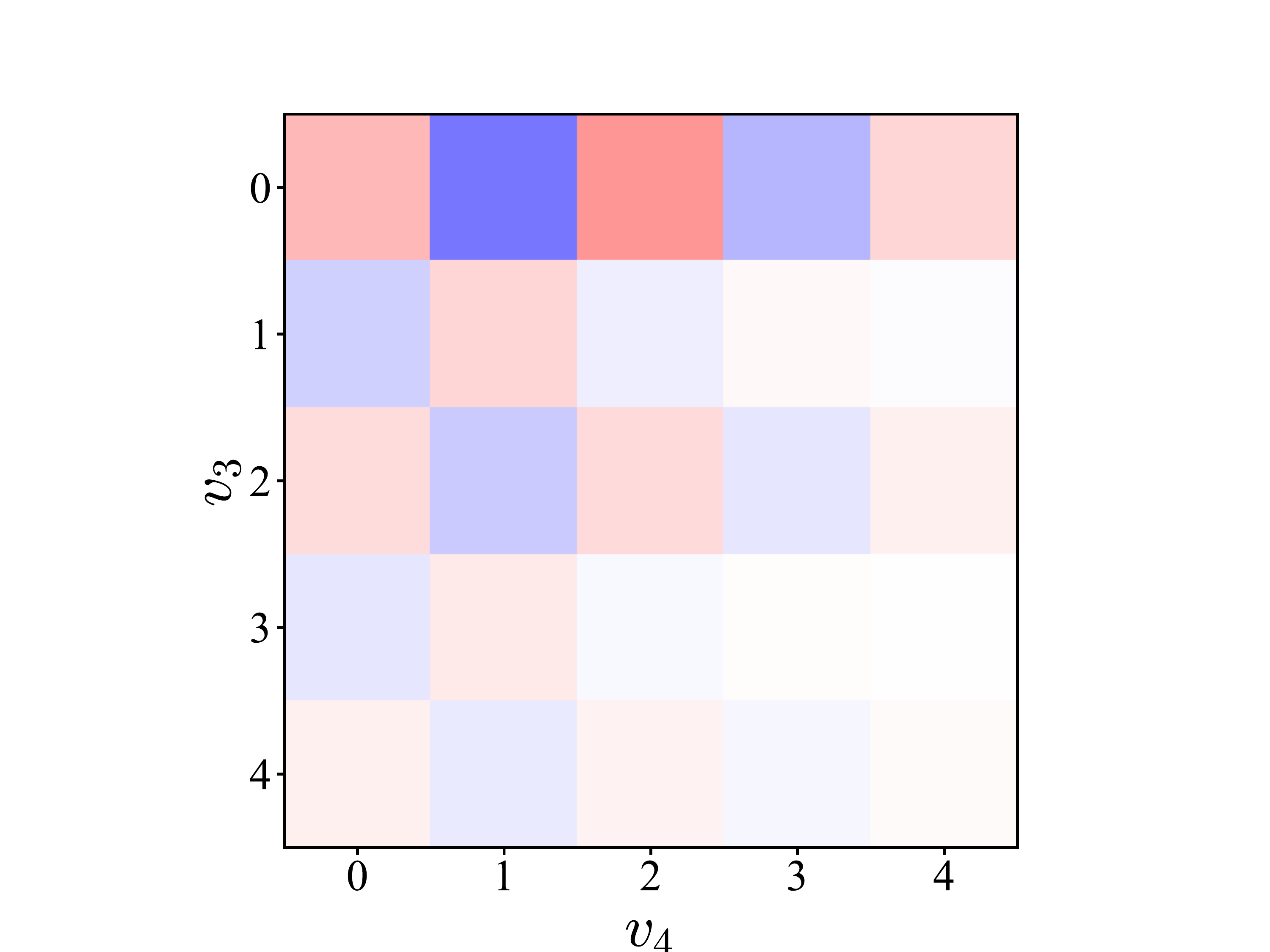}%
    }
\subfloat[]{
\includegraphics[width=0.2\textwidth]{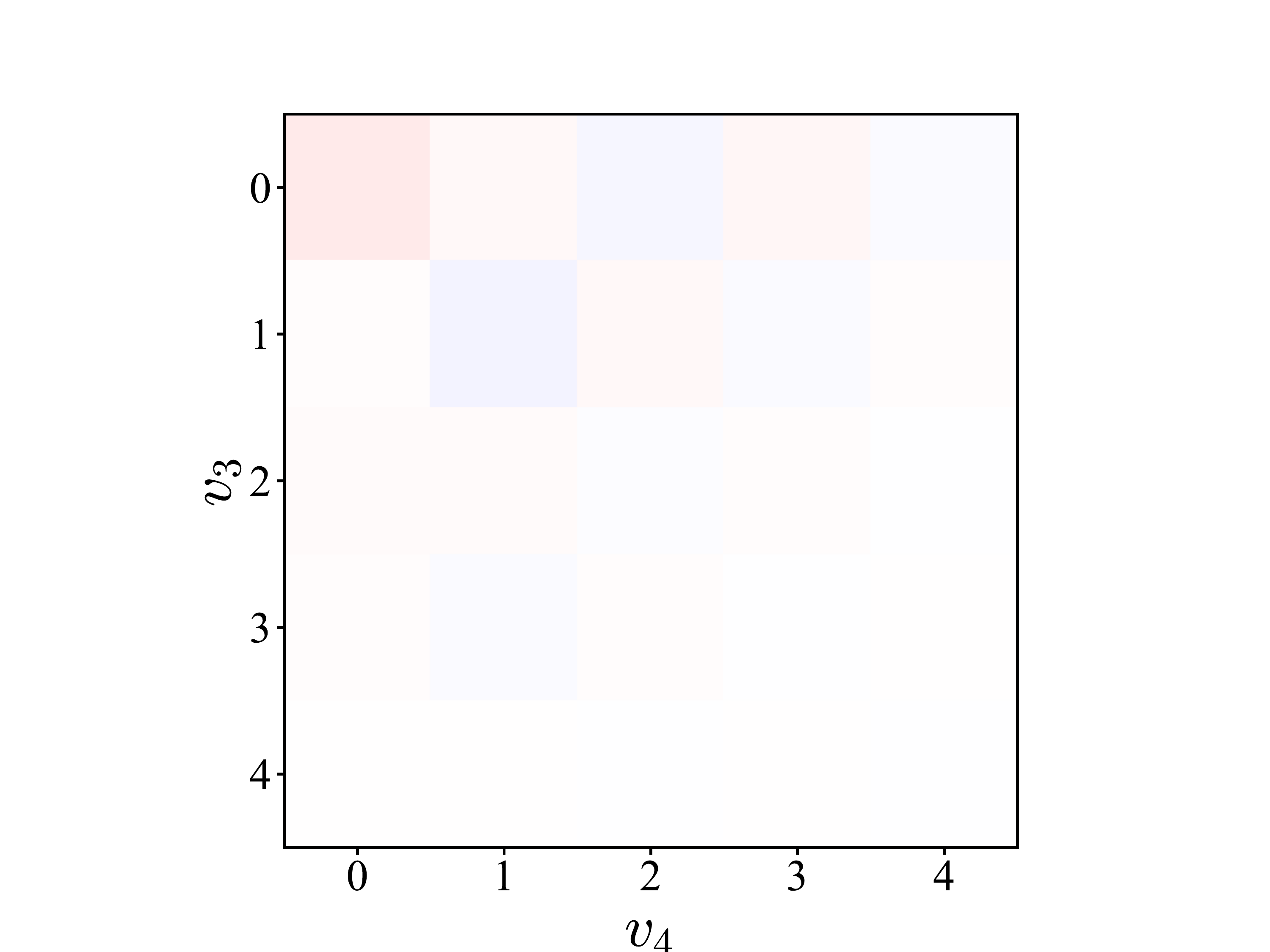}%
    }
\subfloat[]{
\includegraphics[width=0.2\textwidth]{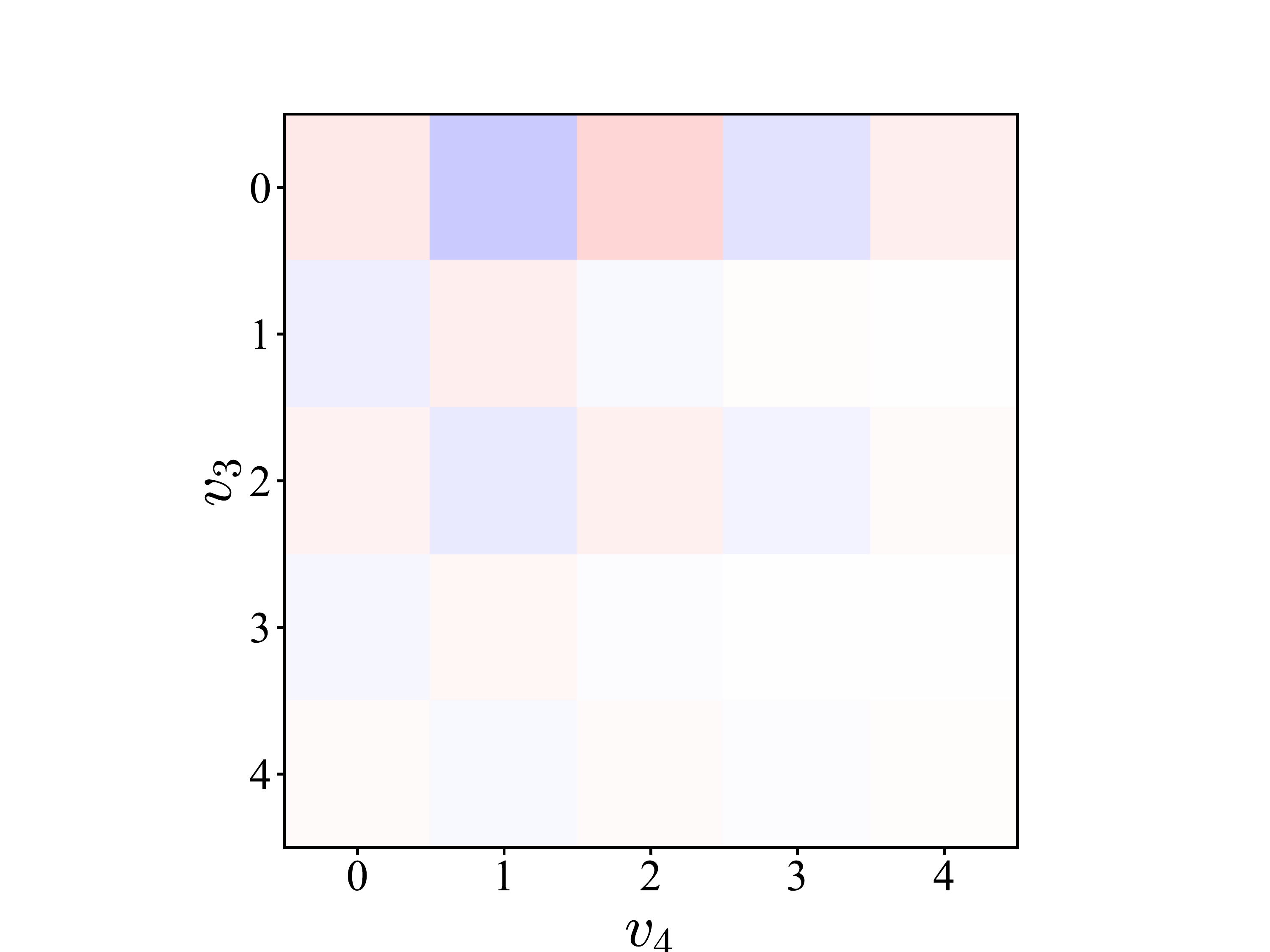}%
    }
\subfloat[]{
\includegraphics[width=0.2\textwidth]{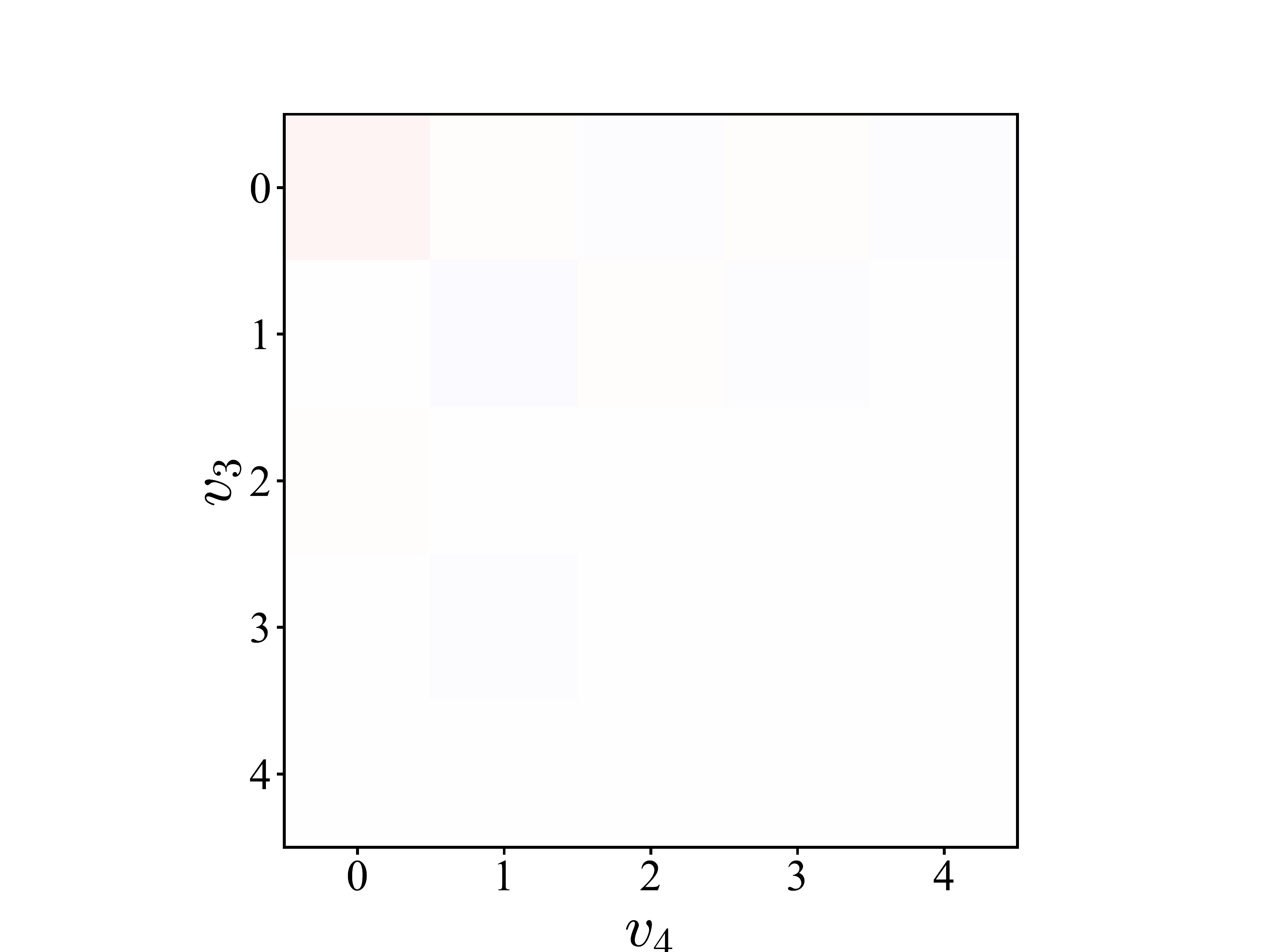}%
    }
\subfloat[]{
\includegraphics[width=0.2\textwidth]{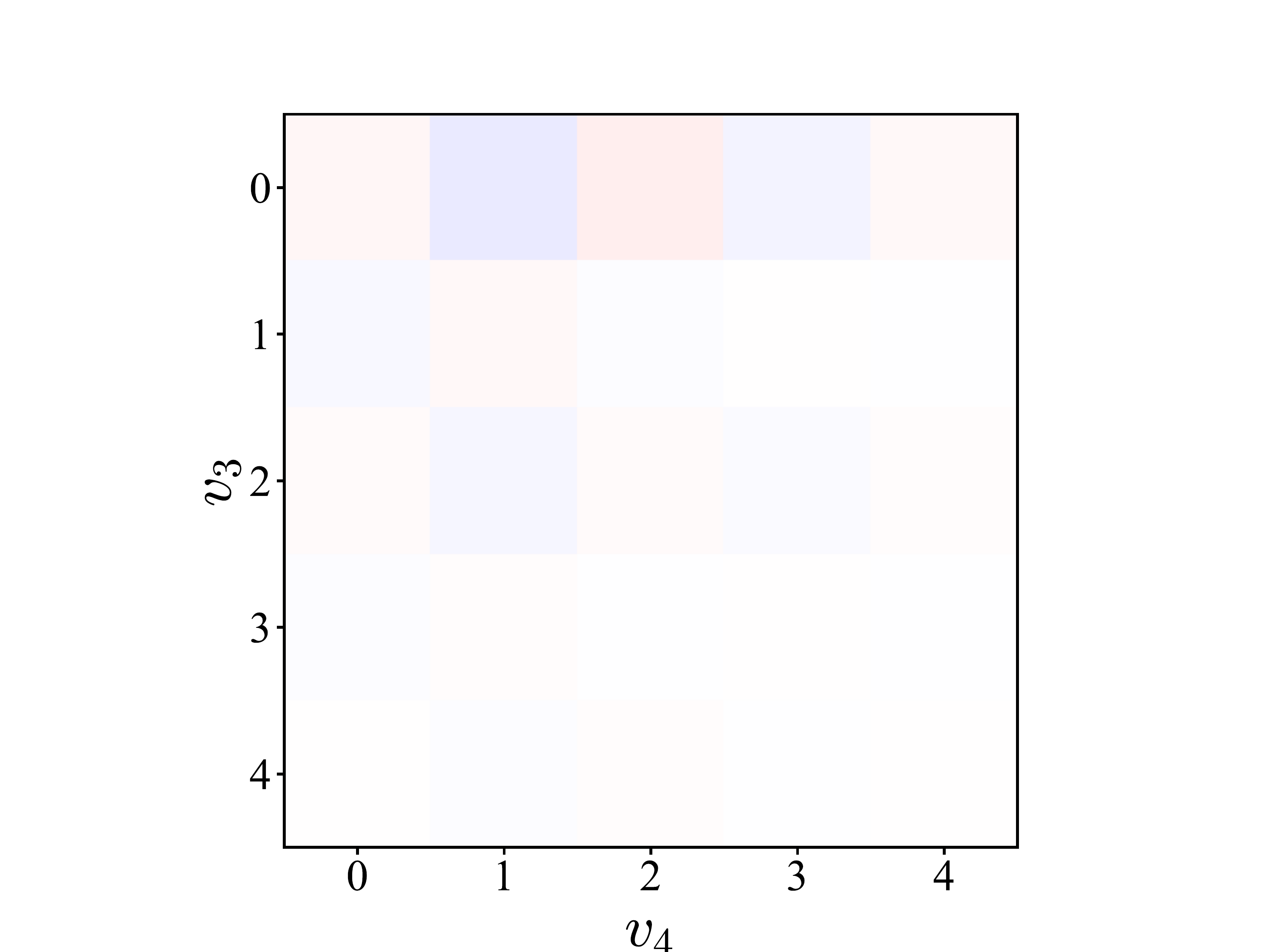}%
    }\\    
\subfloat[]{   
\includegraphics[width=0.2\textwidth]{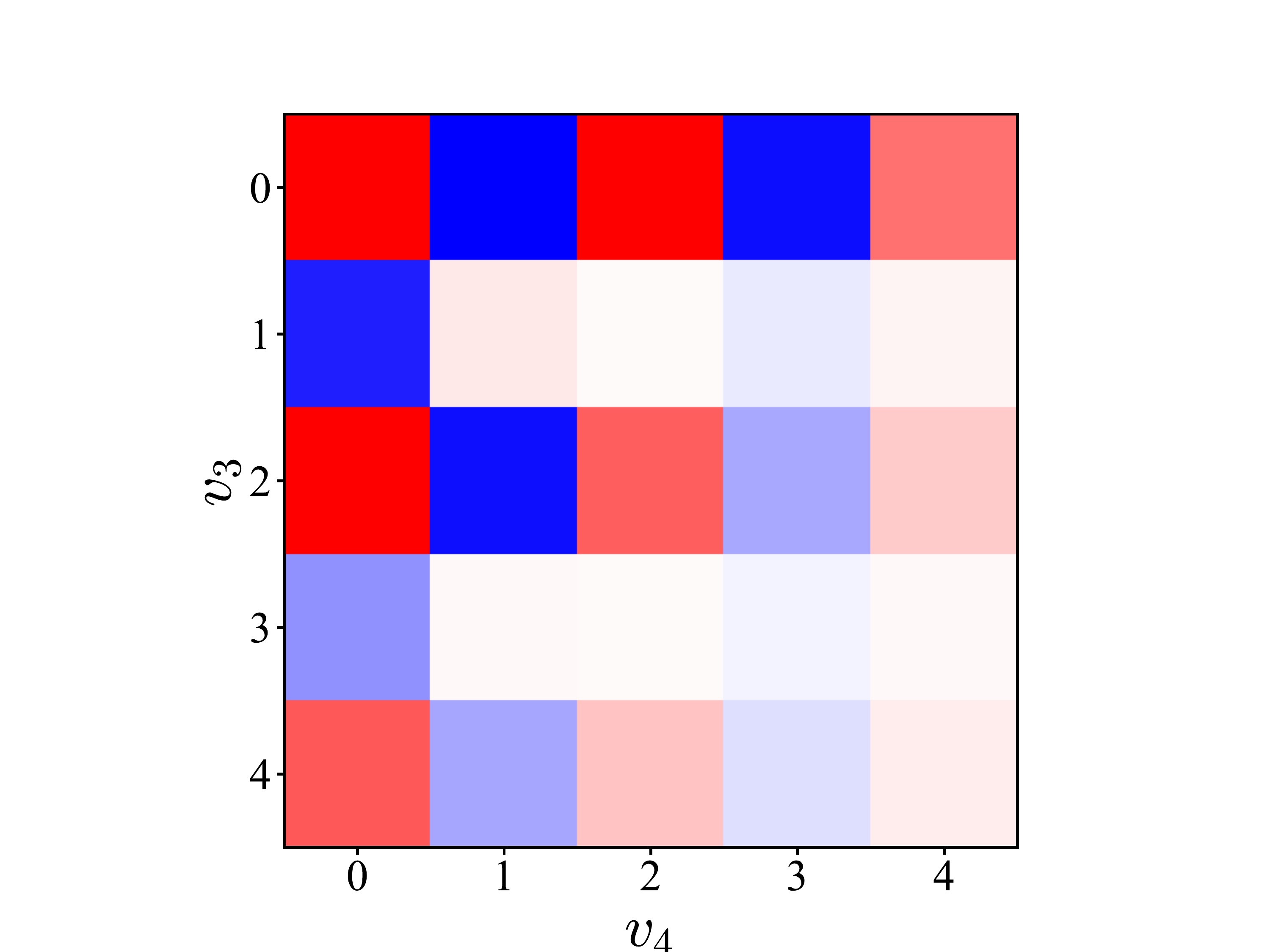}%
    }
\subfloat[]{
\includegraphics[width=0.2\textwidth]{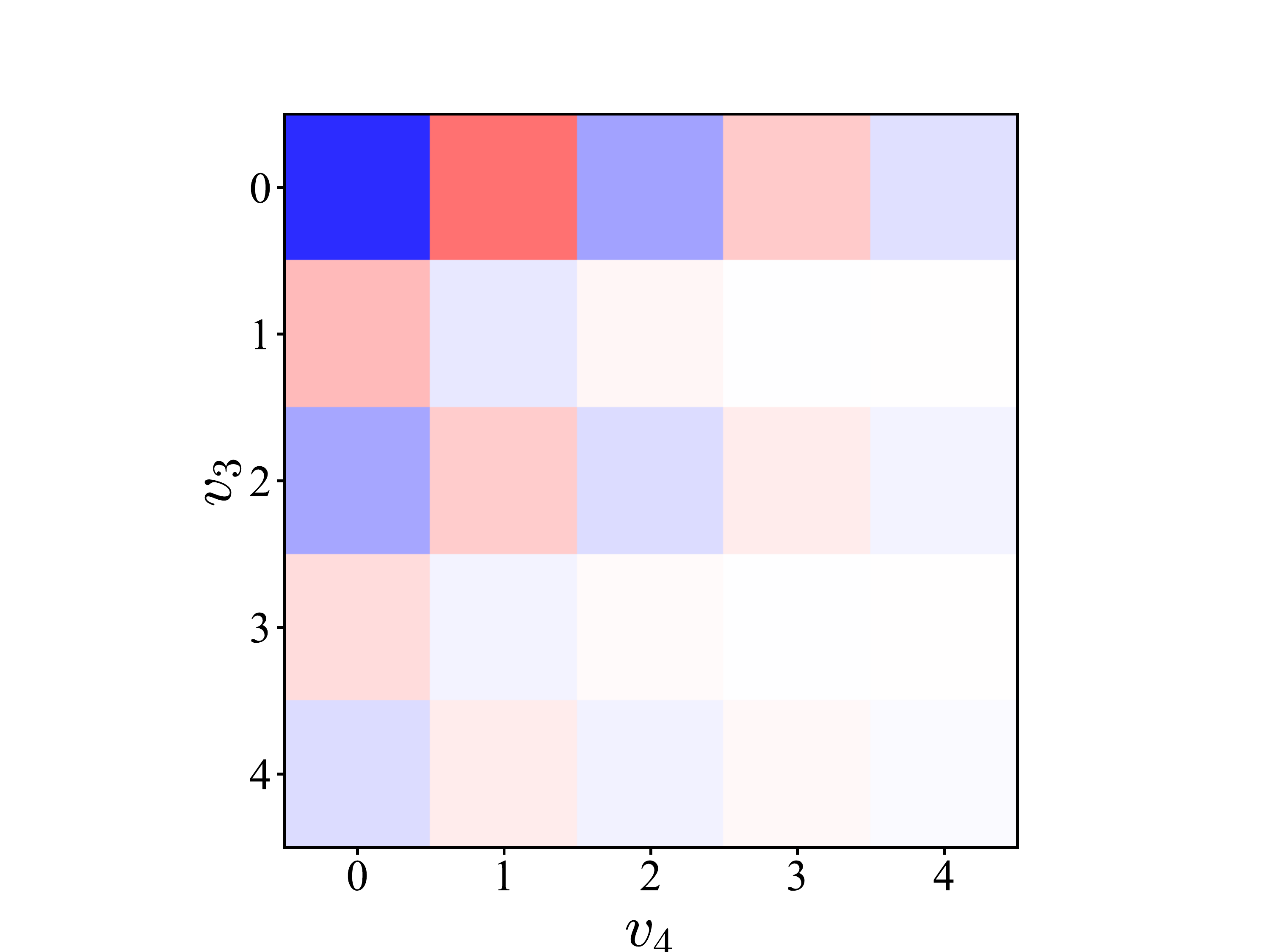}%
    }
\subfloat[]{
\includegraphics[width=0.2\textwidth]{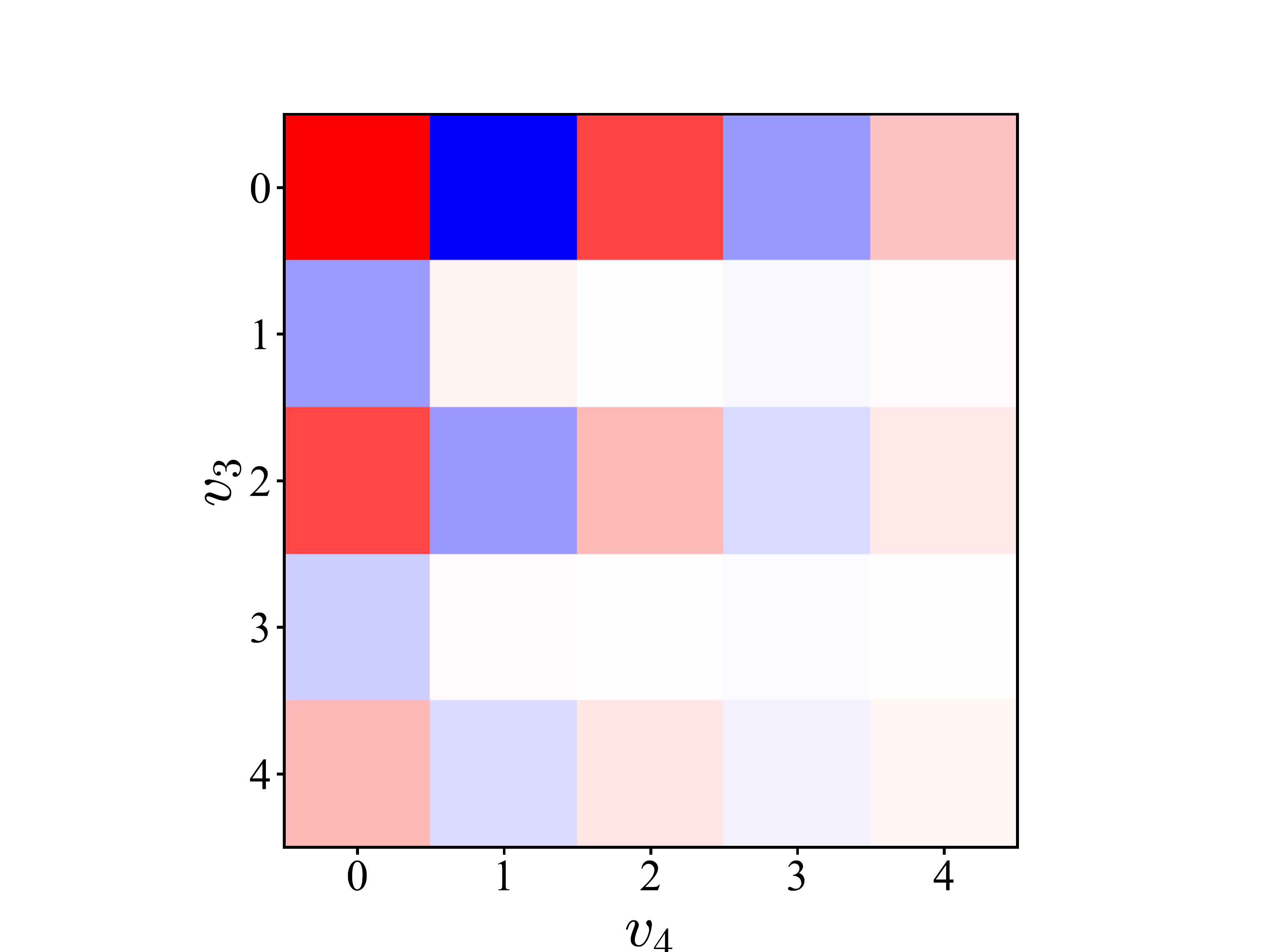}%
    }
\subfloat[]{
\includegraphics[width=0.2\textwidth]{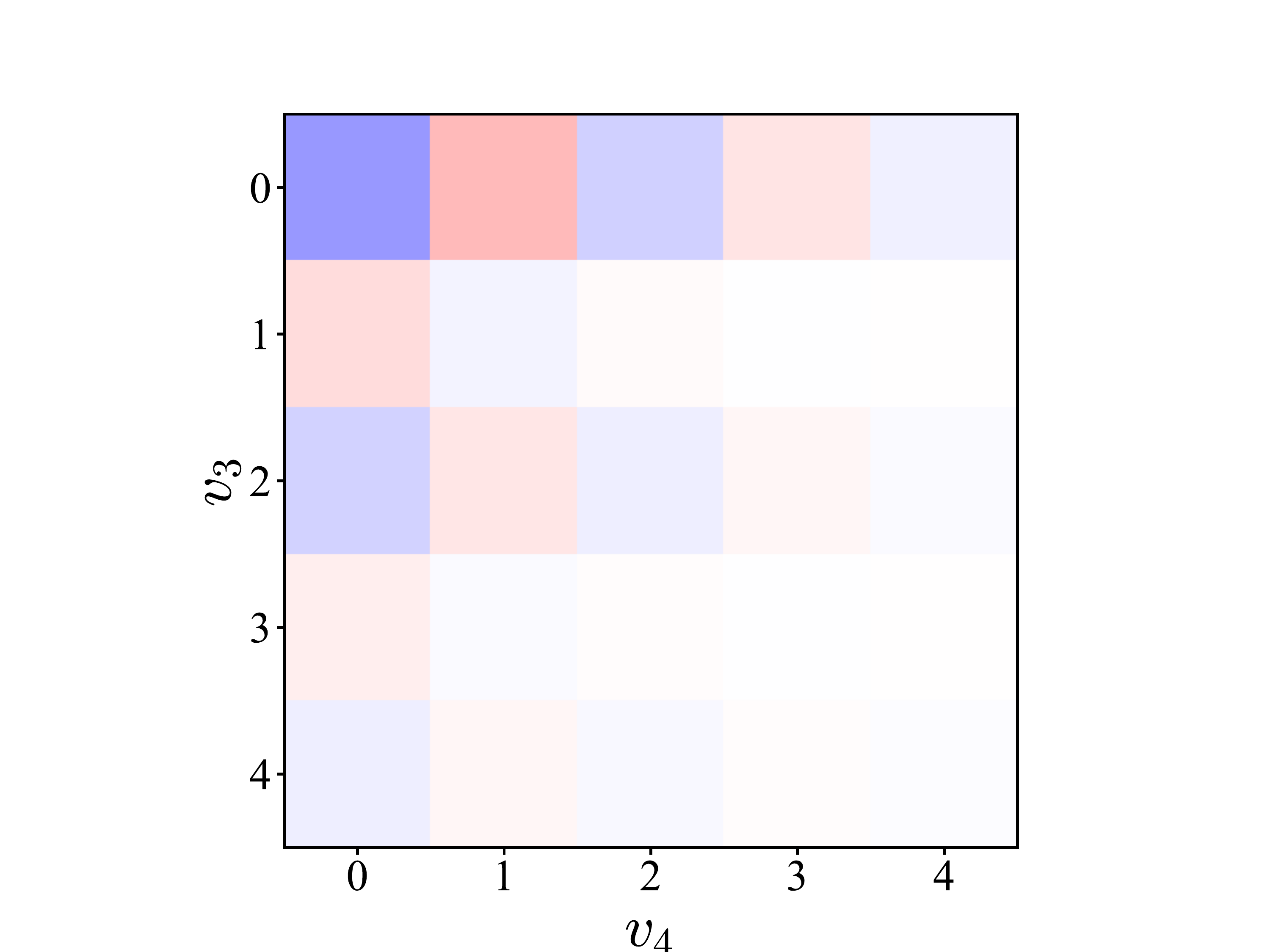}%
    }
\subfloat[]{
\includegraphics[width=0.2\textwidth]{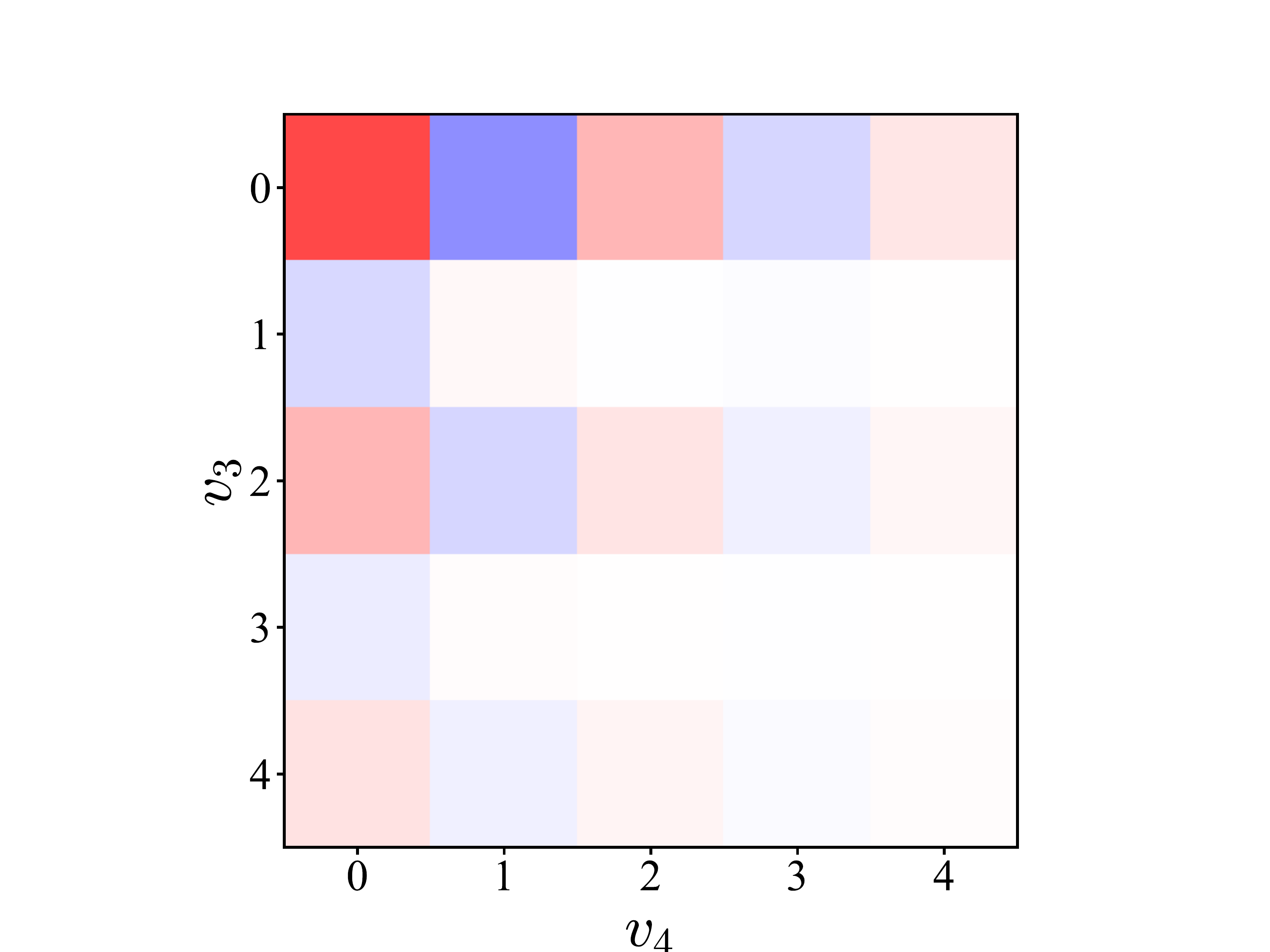}%
    }\\
\subfloat[]{   
\includegraphics[width=0.2\textwidth]{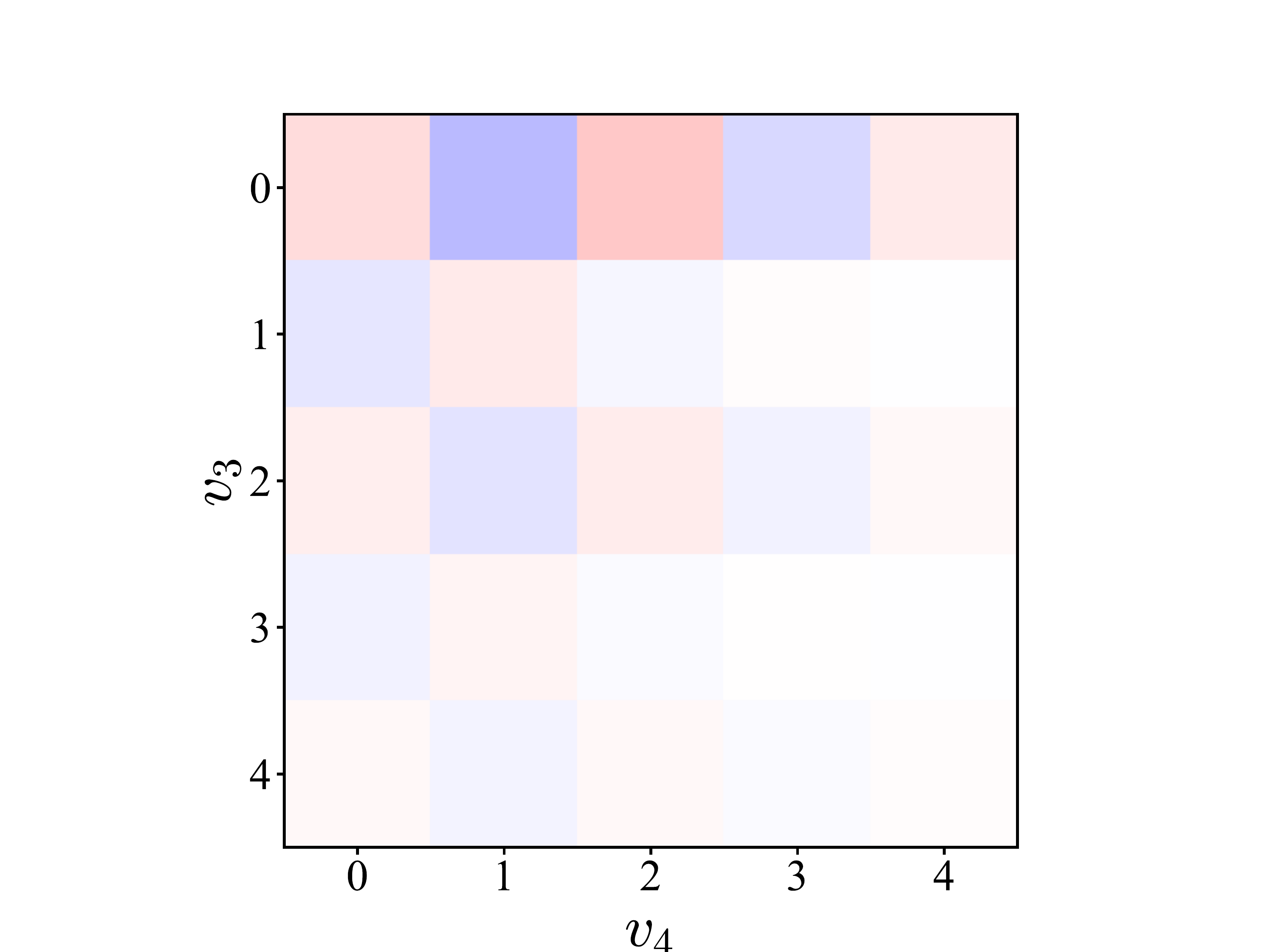}%
    }
\subfloat[]{
\includegraphics[width=0.2\textwidth]{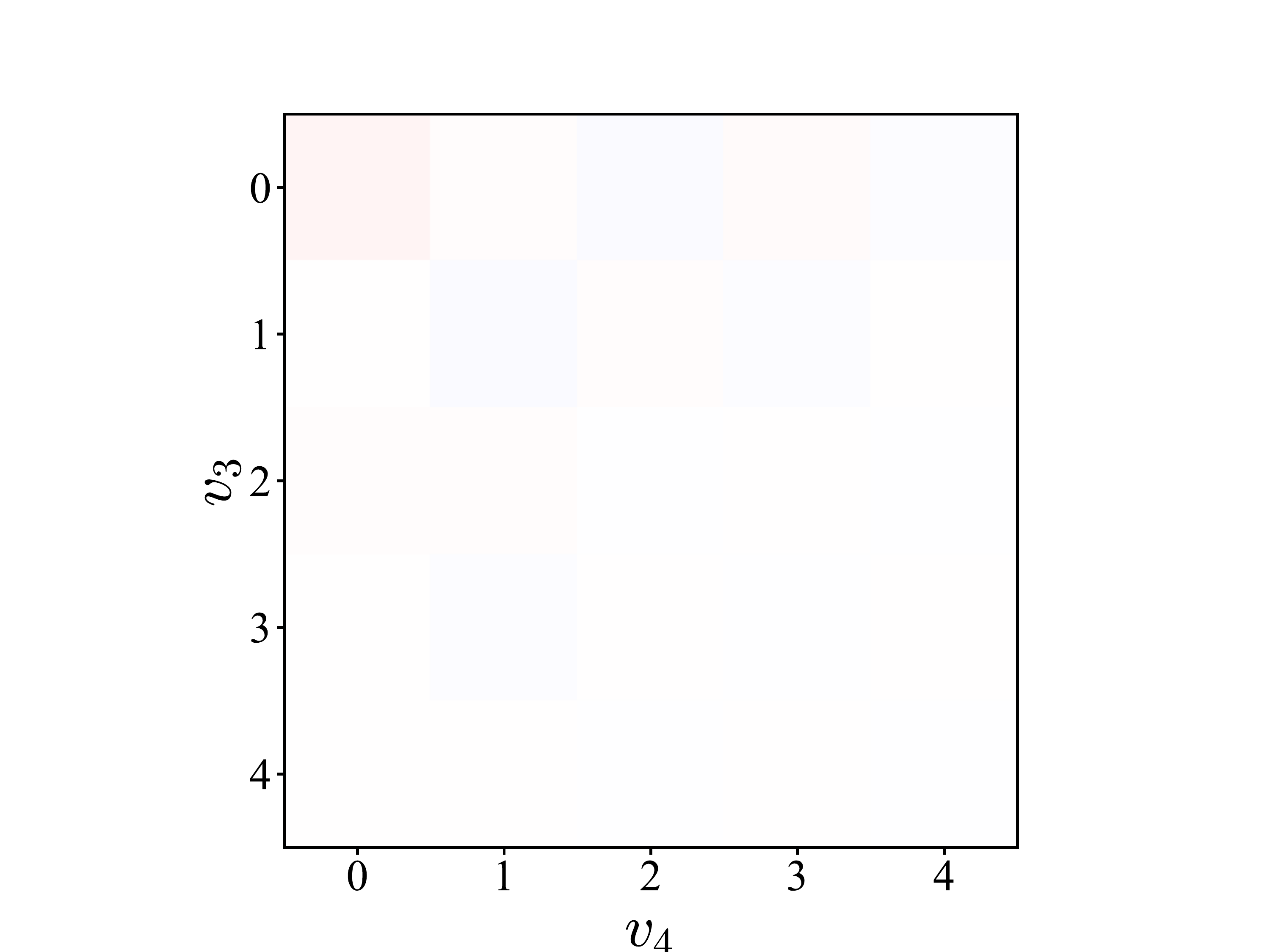}%
    }
\subfloat[]{
\includegraphics[width=0.2\textwidth]{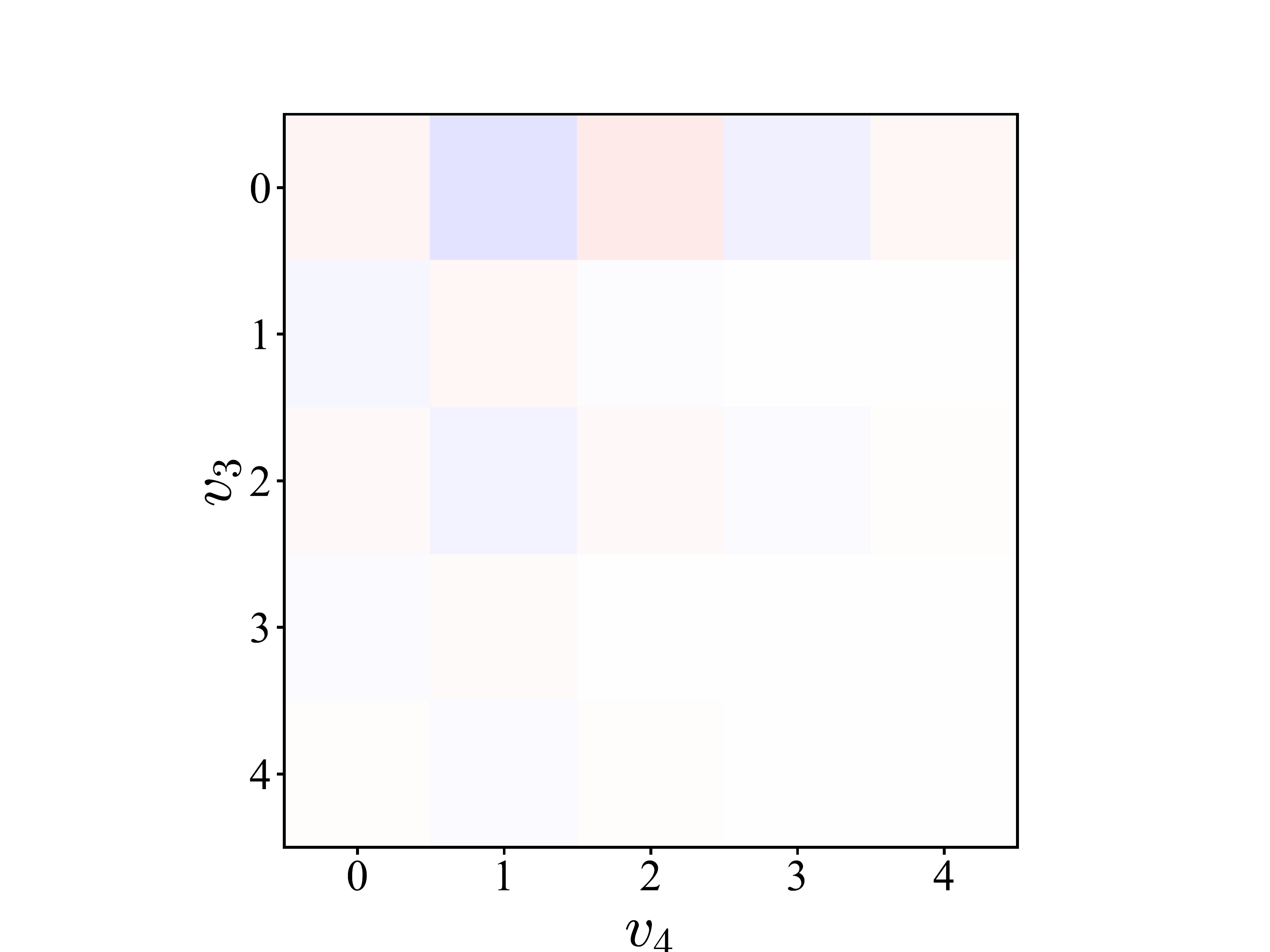}%
    }
\subfloat[]{
\includegraphics[width=0.2\textwidth]{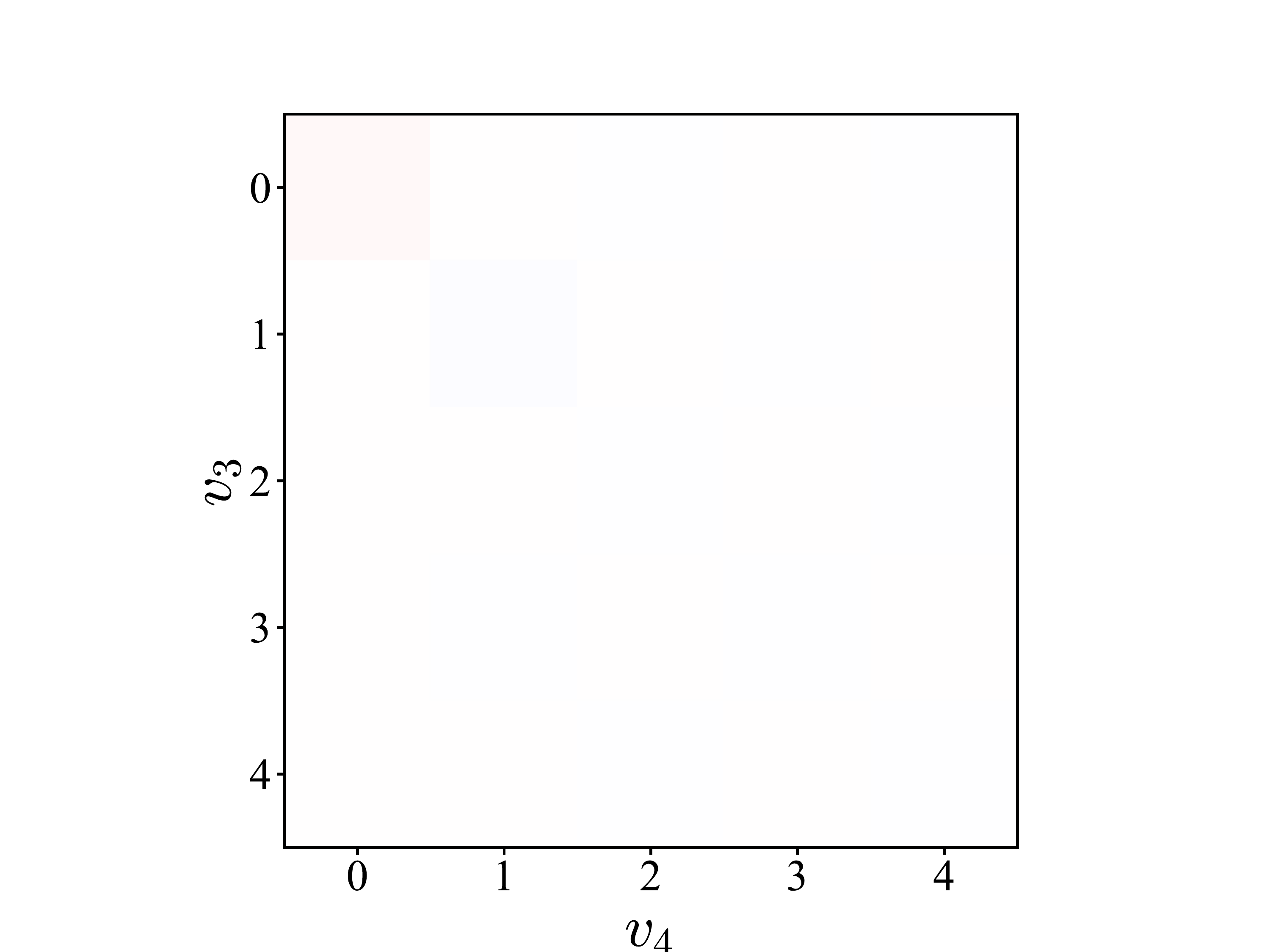}%
    }
\subfloat[]{
\includegraphics[width=0.2\textwidth]{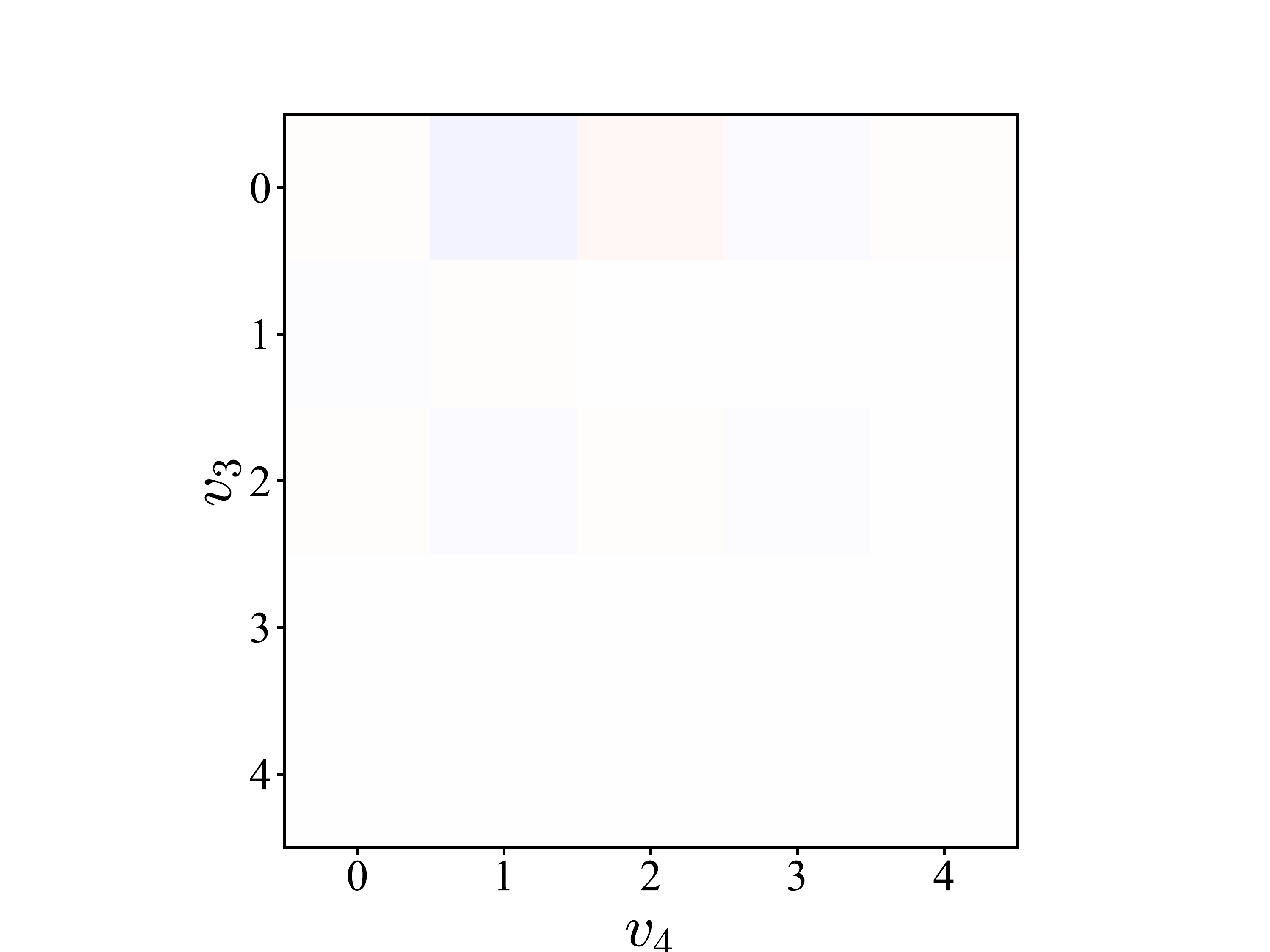}%
    }\\   
\subfloat[]{   
\includegraphics[width=0.2\textwidth]{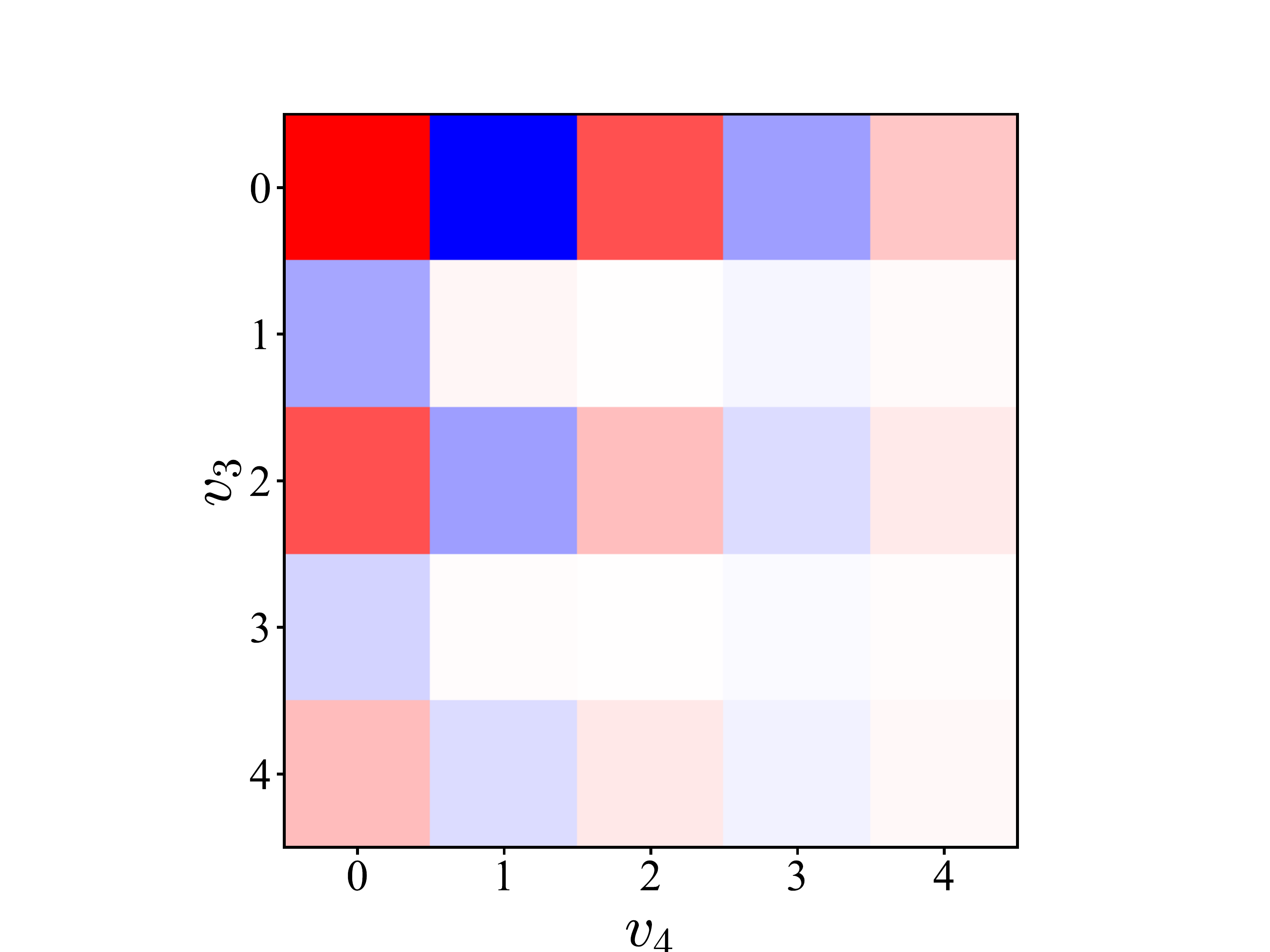}%
    }
\subfloat[]{
\includegraphics[width=0.2\textwidth]{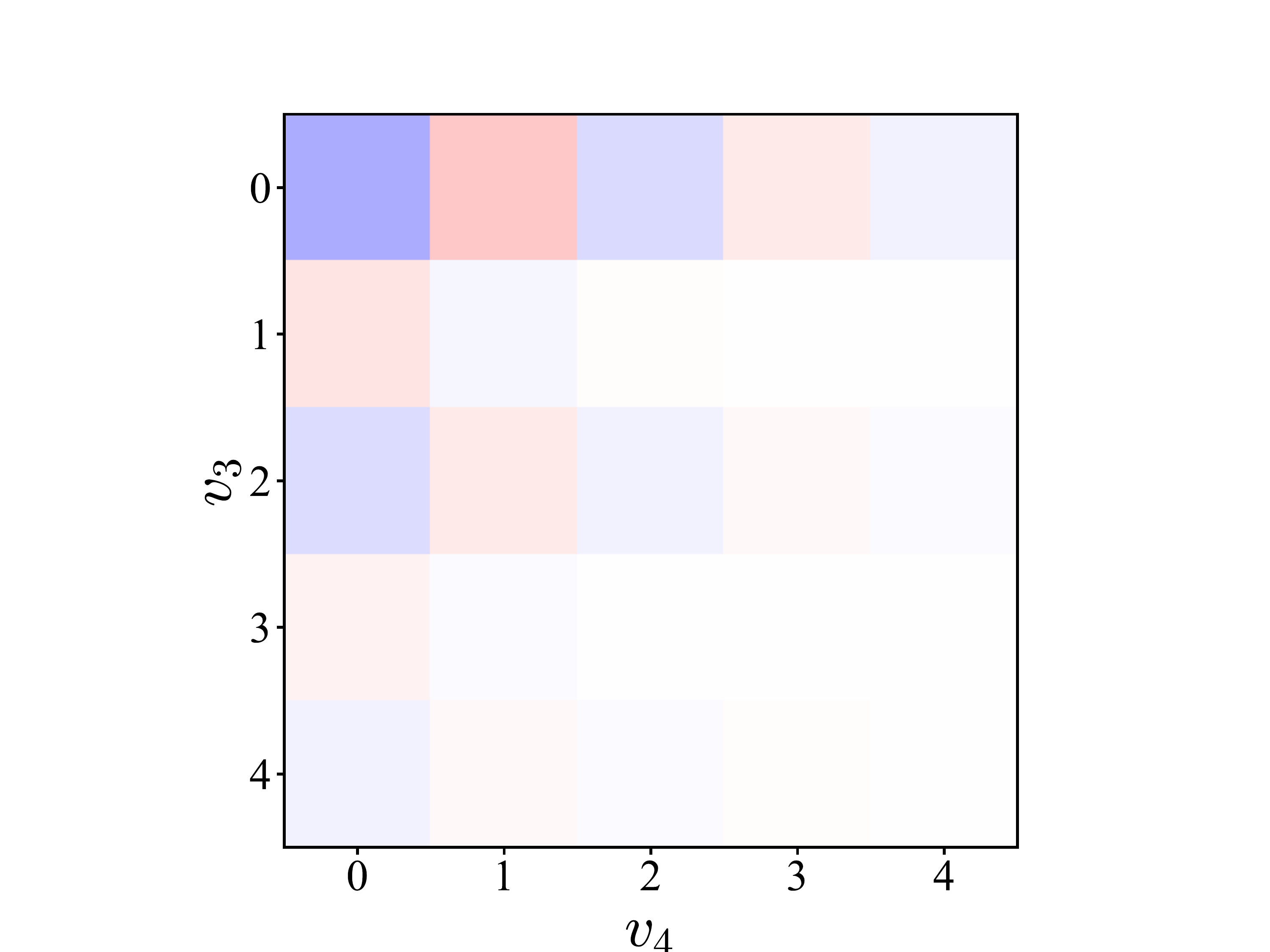}%
    }
\subfloat[]{
\includegraphics[width=0.2\textwidth]{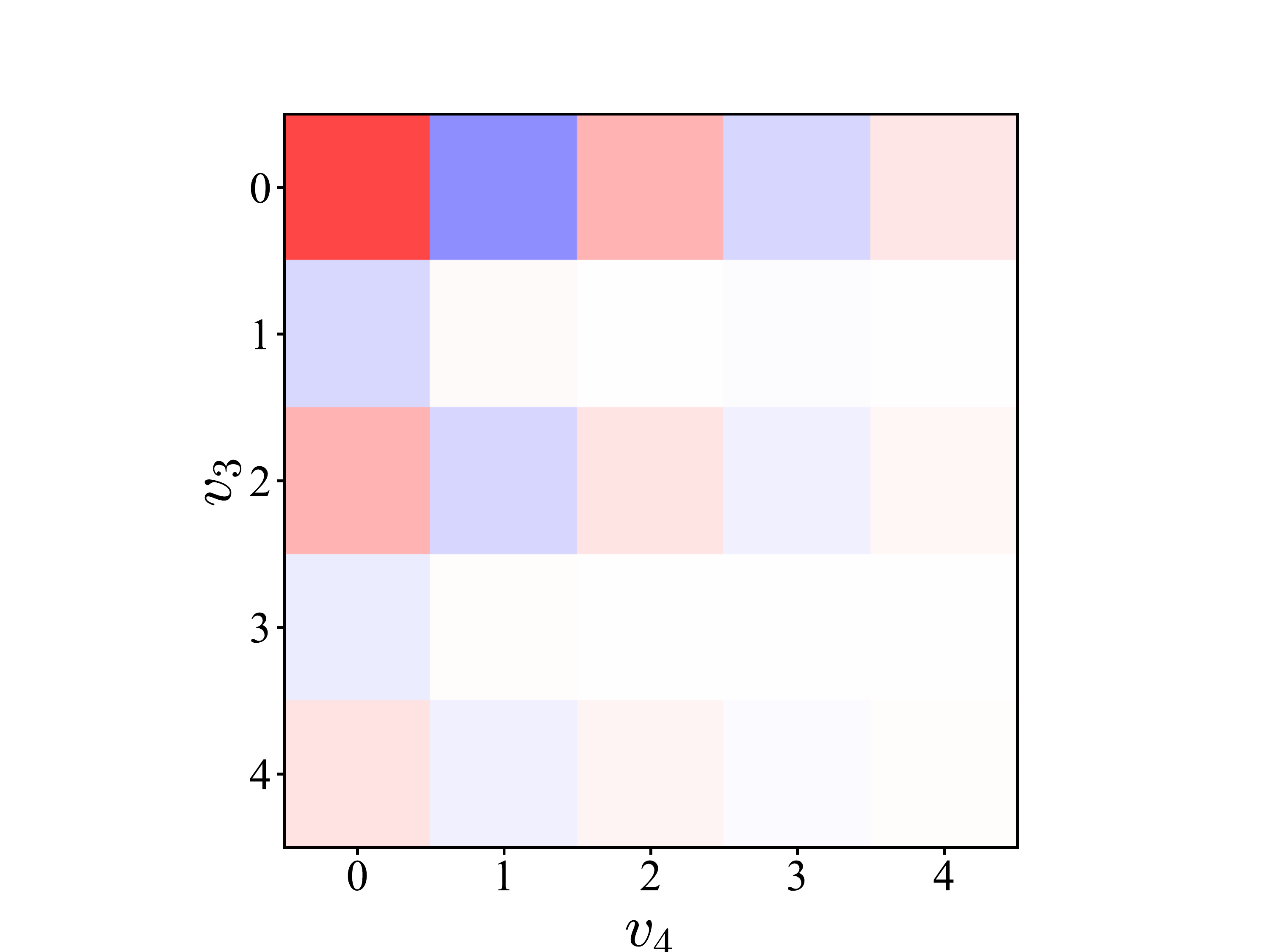}%
    }
\subfloat[]{
\includegraphics[width=0.2\textwidth]{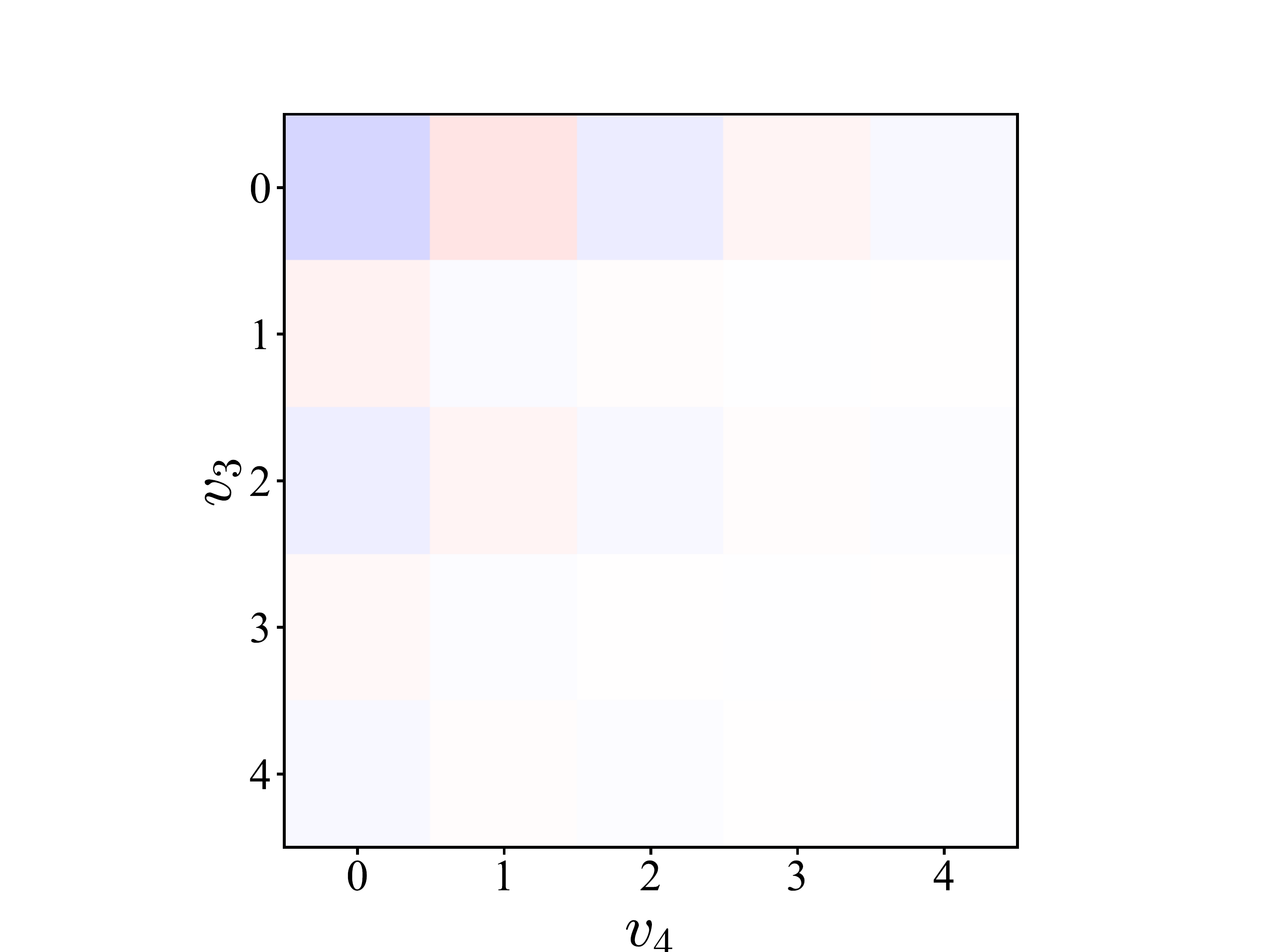}%
    }
\subfloat[]{
\includegraphics[width=0.2\textwidth]{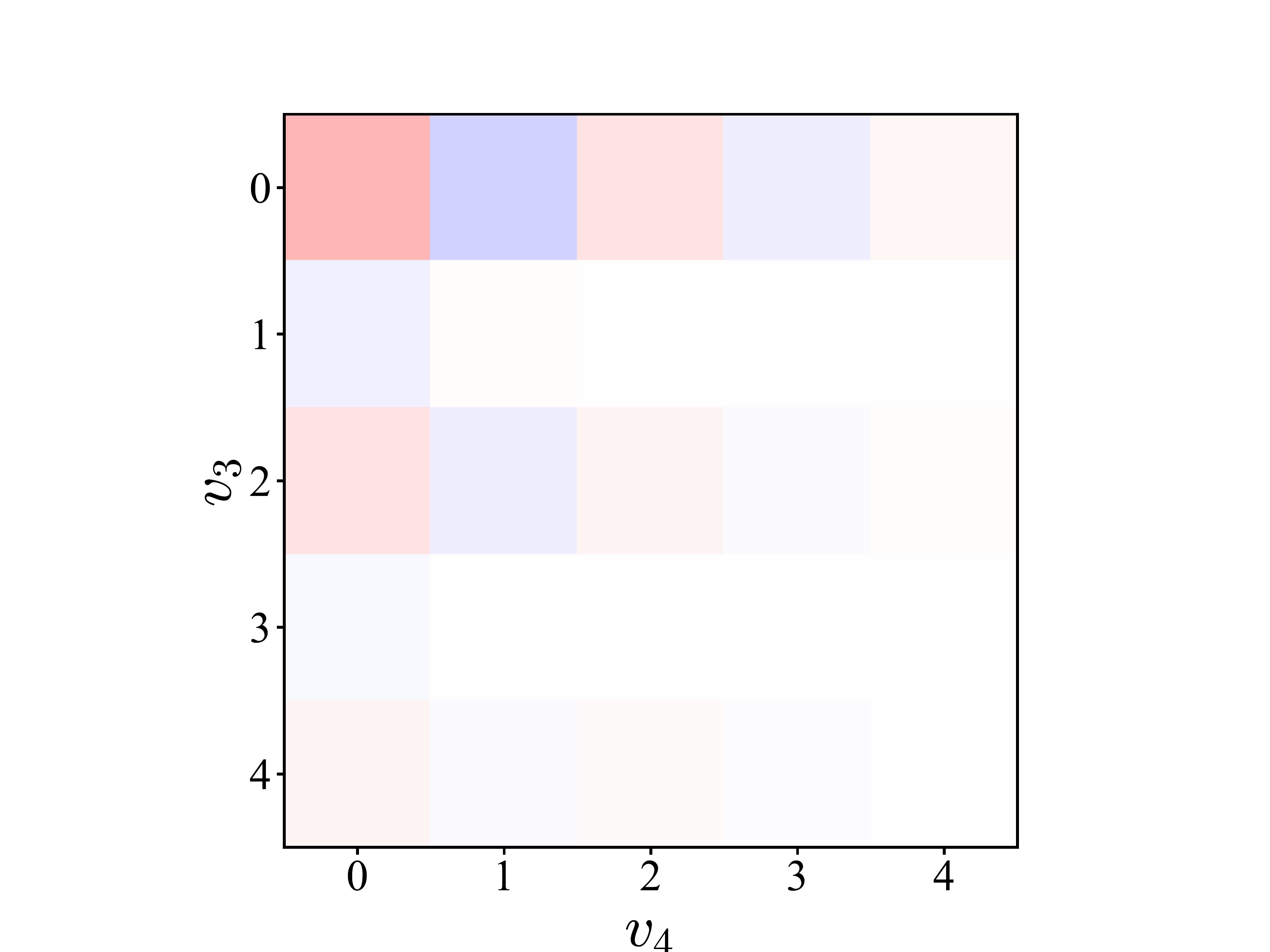}%
    }\\   
    
\caption{\label{fig:four-mode}Benchmark of the sign of overlap $\langle \phi_\mathrm{i}^\mathbf{0} | \phi_\mathrm{f}^{(v_1,v_2,v_3,v_4)} \rangle$ in the four-mode model through exact diagonalization. Red means positive overlap and blue means negative overlap here. Darker color means larger norm of the overlap. Among the $5\times5$ graphic lattice, lines are arranged according to the quanta on mode 1 ((a),(f),(k),(p),(u): $v_1 = 0,1,2,3,4$ and $v_2 = 0$ ) and columns are arranged according to the quanta on mode 2 ((a),(b),(c),(d),(e): $v_1 = 0$ and $v_2 = 0,1,2,3,4$). $v_3$ and $v_4$ are arranged on the y-axis and x-axis on every subplot respectively. The basis for mode $n$ is simple harmonic oscillator with frequency $\omega_{\mathrm{f},n}$ and size as 5.}
\end{figure*}

The reliability in two-mode model may be a special case. For higher dimensions, the total available rotations in $D$ dimension system is $D(D-1)/2$ which can be attributed to each pair of two dimensions. So we also benchmark on a four-mode model to numerically check the reliability of our method , the frequencies of four modes are arranged as $\omega_{\mathrm{f},n} = \omega_0/n$, $n \in \{1,2,3,4\}$ , $\omega_{\mathrm{i},n} = \omega_{\mathrm{f},n}/5$ and also $\omega_0 = 3868 \mathrm{cm^{-1}}$. The rotation matrix among four modes is construct as Eq.~\ref{eq:s-four-mode}, where $\mathbf{S}_{n,m}(\theta)$ means rotation between mode $n$ and mode $m$ with angle $\theta$ according to Eq.~\ref{eq:s-two-mode}. Here we set $\Delta q_{\mathrm{f},1} =\frac{1}{\omega_0}\sqrt{\frac{D}{2}}$ and $\Delta q_{\mathrm{f},2} =\Delta q_{\mathrm{f},3}=\Delta q_{\mathrm{f},4}= -\frac{1}{\omega_0}\sqrt{\frac{D}{2}}$ to show the effect of the sign of displacement on the overlap. The numerical results collected in FIG.~\ref{fig:four-mode} show that quanta increasing/decreasing on a single mode by one causes the sign flipping of the overlap except for $v_1$ on mode one, which matches the sign of displacement and clearly show our sign approximation method still valid at least for configurations with large norm of overlap. ( Those with dark color in the figure) 
\begin{equation}
    \mathbf{S} = \mathbf{S}_{3,4}(\pi/4)\mathbf{S}_{2,4}(\pi/4)\mathbf{S}_{2,3}(\pi/4)\mathbf{S}_{1,4}(\pi/4)\mathbf{S}_{1,3}(\pi/4)\mathbf{S}_{1,2}(\pi/4)
    \label{eq:s-four-mode} 
\end{equation}
\clearpage
\section{Physical parameters of two-mode Morse model}
The two-mode Morse model used in the main article have harmonic $V_\mathrm{i}$ and Morse form $V_\mathrm{f}$. Expressions and parameters are as following.  
\begin{gather}
        V_\mathrm{i} = V_\mathrm{ex}  = \sum_{n=1,2} \frac{1}{2}\omega_{\e,n}^2 q_{\e,n}^2 + V^{(0)}_\mathrm{i} \\
        V_\mathrm{f} = V_\mathrm{gs} = \sum_{n=1,2} D_n(1-e^{- \alpha_n
        q_{\g,n}})^2  + V^{(0)}_\mathrm{f} \label{eq:example}\\ 
        V_{\mathrm{m}} = \sum_{l=1,2} \frac{1}{2}\omega_{\g,n}^2 q_{\g,n}^2 + V^{(0)}_\mathrm{f} \\
        q_{\g,n} =  \sum_{m=1,2} S_{n,l} q_{\e,l} + \Delta q_{\g,n} \\
        V^{(0)}_\mathrm{i} -  V^{(0)}_\mathrm{f} = E_{\ad}
\end{gather}
We have $D_1 = D_2 = 5.52 \, \textrm{eV}$ and $\alpha_1=2 \alpha_2=2.23 \, \textrm{amu}^{-1/2} \textrm{\AA}^{-1}$ ($0.0277 \, \textrm{a.u.}$). Under HA, the harmonic frequency at the equilibrium position for ground state is $\omega_{\g,1} =2\omega_{\g,2} = \sqrt{2 \alpha_1^2 D_1}= 3868 \, \textrm{cm}^{-1}$. The harmonic excited state PES has $\omega_{\e,1} =2\omega_{\e,2} = \omega_{\g,1}/5 =774 \, \textrm{cm}^{-1}$. The adiabatic excitation energy $E_{\ad}=2D$. The displacement used is $\Delta q_{\g,1} = \Delta q_{\g,2} = \frac{1}{\omega_{\g,1}}\sqrt{\frac{D_1}{2}}$ .

\section{Physical parameters of Pyridine}
\begin{figure}
    \centering
    \includegraphics[width=0.8\textwidth]{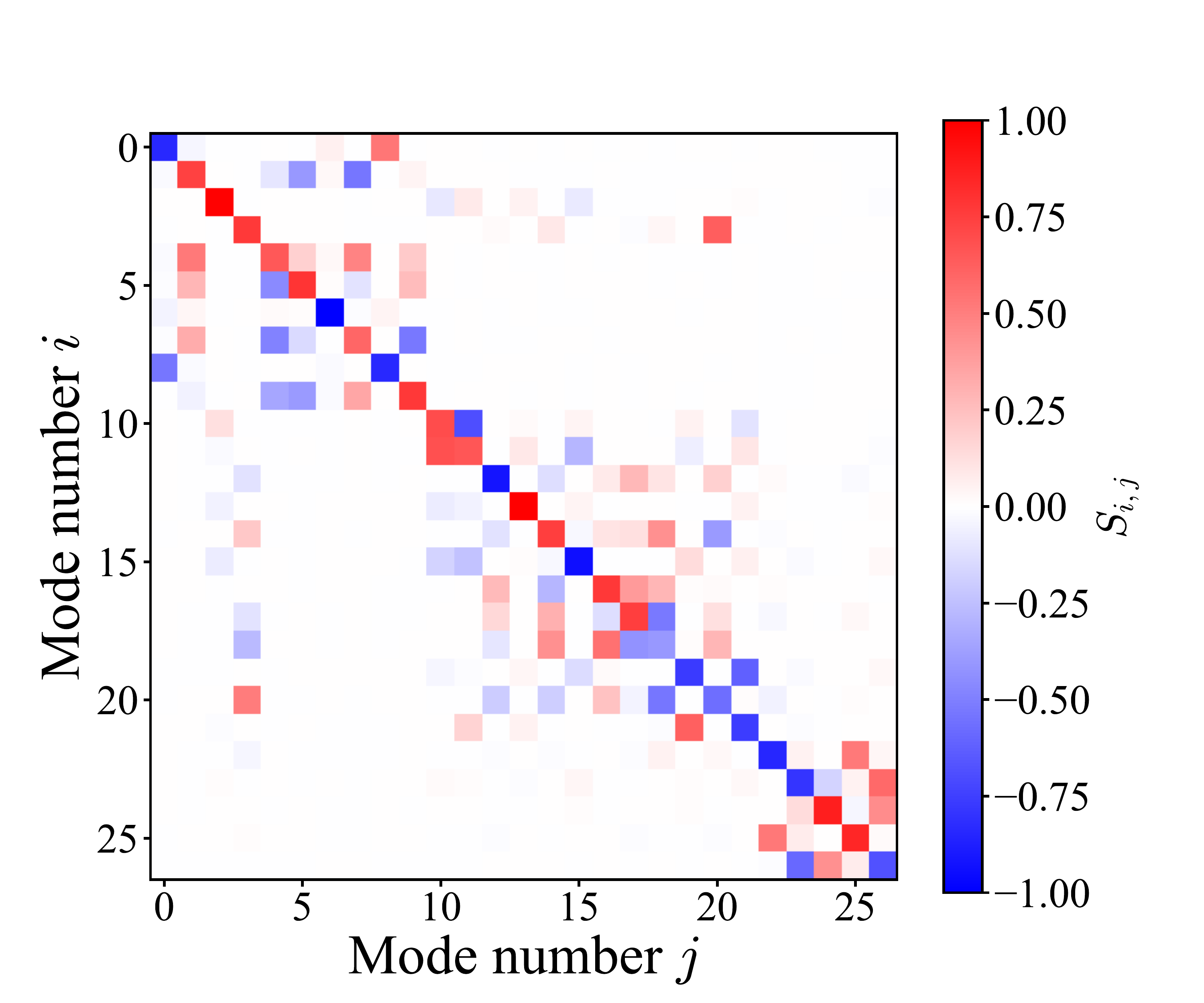}%
    \caption{The duschinsky matrix $\mathbf{S}$ between the ground state ($S_0$) and the excited state ($S_1$) of pyridine.}
    \label{fig:dre matrix}
\end{figure}

Ab-initio physical properties of pyridine used in hybrid method HQCS and the reference method TD-DMRG are calculated through DFT and TD-DFT methods (B3LYP/6-31g(d) in gas phase, Tamm-Dancoff approximation are used for TD-DFT) with Gaussian16\cite{g16}. 
It must be noted that  the ab-initio calculated displacement and duschinsky rotation matrix are  $\mathbf{S}^{-1}$ and $\Delta \mathbf{q}_{\mathrm{gs}} = -\mathbf{S}^{-1} \Delta \mathbf{q}_\mathrm{ex}$ in $\mathbf{q}_\mathrm{gs} = \mathbf{S}^{-1} \mathbf{q}_\mathrm{ex} + \Delta \mathbf{q}_{\mathrm{gs},n}$. (For the emission spectrum we simulated in the main article $\mathrm{'i'}$ is  $\mathrm{'ex'}$ and $\mathrm{'f'}$ is $\mathrm{'gs'}$). In this formalism we screen out 7 important modes with large norm of $\Delta \mathbf{q}_{\mathrm{gs}}$. (Mode 3,11,12,14,16,20 and 22). Displacements and rotation of this 7 modes are used in the main article simulations of vibrationally resolved electronic spectrum for the large computational cost of classical simulation of boson sampling algorithm. The frequency and displacement properties are listed in Table.~\ref{tab:py}, the duschinsky rotation matrix $\mathbf{S}$ are displayed in FIG.~\ref{fig:dre matrix} and the 1-MR anharmonic PES of $S_0$ are shown in FIG.~4,5 and 6.

Though we use $\omega_{\mathrm{m},3} = \mathrm{max}(\omega_{\mathrm{gs},3},\omega_{\mathrm{ex},3}) = \omega_{\mathrm{ex},3} $ in the main article, it should be noted that in fact as $\omega_{\mathrm{ex},3} \approx \omega_{\mathrm{gs},3} $, directly use $\omega_{\mathrm{m},3} = \omega_{\mathrm{gs},3} \textless \omega_{\mathrm{ex},3}$ will not influence the output of sign approximation step. 
From the analytical expression below, we can find when $\omega_{\mathrm{m},n} \ge \omega_{\mathrm{i},n}$, every term of $C_l$ is non-negative so do not influence the sign of the overlap. For another situation when $\omega_{\mathrm{m},n}  \approx \omega_{i,n} $, the $\sum_l C_l $ is mainly contributed by $C_{l=v_\mathrm{m}}$ that is always non-negative, so the sign approximation method usually is still reliable in this looser condition.  We also perform numerical benchmark in the one-mode harmonic model like previously done in two-mode and four-mode model. The difference is frequency parameters is taken from mode 3 with  $\omega_\mathrm{i} = 631.48 \mathrm{cm}^{-1}$ ,$\omega_{\mathrm{m}} = 613.41 \mathrm{cm}^{-1}$ and $ q_\mathrm{m} -q_\mathrm{i} = \Delta q_\mathrm{f} = 45.05618 ~ \mathrm{a.u.}$. The result overlaps $\langle \chi_\mathrm{i}^{0} | \chi_{\mathrm{m}}^{v_\mathrm{m}}\rangle$ are [0.26833, 0.44824, 0.52237, 0.48707, 0.37953, 0.24482,
 0.11396] for $v_\mathrm{m}= 0 \sim 6$. Sign of overlaps in this situation is truly still in agree with those predicted by Eq.~\ref{eq:sign-approximate} from our sign approximation method.
\begin{equation}
\langle \chi_\mathrm{i}^{0} | \chi_{\mathrm{m}}^{v_\mathrm{m}}  \rangle =\Gamma (\Delta q_\mathrm{f})^{v_\mathrm{m}} \times (\sum_l C_l )
\label{eq:one-mode-ana}
\end{equation}
\begin{equation}
\begin{aligned}
    C_l = & \frac{(-1)^{v_\mathrm{m}-l}+1}{2} (\Delta q_\mathrm{f})^{-(v_\mathrm{m}-l)}\beta_\mathrm{m}^{l/2}   \\
     & \frac{2^l}{l!} (\frac{\beta_\mathrm{i}}{\beta_\mathrm{i} +\beta_\mathrm{m}})^{l} (\frac{\beta_\mathrm{m}- \beta_\mathrm{i}}{\beta_\mathrm{i} +\beta_\mathrm{m}})^{\frac{v_\mathrm{m}-l}{2}}  \frac{1}{(\frac{v_\mathrm{m}-l}{2})!}
    \label{eq:cm}
\end{aligned}
\end{equation}
\begin{equation}
\Gamma= (\frac{v_\mathrm{m}!}{2^{v_\mathrm{m}}} \times \frac{2\sqrt{\beta_\mathrm{i} \beta_\mathrm{m} }}{\beta_\mathrm{i} +\beta_\mathrm{m}})^\frac{1}{2} \mathrm{exp}(-\frac{\beta_\mathrm{i} \beta_\mathrm{m} (\Delta q_\mathrm{f})^2}{2(\beta_\mathrm{i} +\beta_\mathrm{m})})
\label{eq:N}
\end{equation}
where $\beta_\mathrm{i/m} = \omega_\mathrm{i/m} / \hbar$, $l$ are non-negative integers meeting the condition $l \le v_\mathrm{m}$. 

\begin{table*}[h]
\caption{\label{tab:py}The harmonic frequency and displacement of the ground state ($S_0$) and the excited state ($S_1$) of pyridine.}

\begin{ruledtabular}
\begin{tabular}{c@{\hspace{2cm}}c@{\hspace{2cm}}c@{\hspace{2cm}}c@{\hspace{2cm}}c}
 Mode number $n$ & $\mathbf{\omega}_{\mathrm{gs},n}$/$\mathrm{cm}^{-1}$& $\mathbf{\omega}_{\mathrm{ex},n}$ /$\mathrm{cm}^{-1}$ & $ \Delta q_{\mathrm{gs},n}$/a.u. \\
 \hline

 1&386.49&324.80 &   0.02030 \\
 2&421.45 & 391.56&   -0.01260  \\ 
 3&613.41 & 631.48&  -45.05618 \\
 4&670.23 & 281.36&  0.06642  \\
 5&718.73 & 523.59&    -0.00225\\
 6&763.10 & 581.22&    0.00173\\
 7& 897.84 & 387.67&   -0.00467\\
 8&956.78 &   532.42& 0.01078 \\
 9& 993.72 &  771.94&  -0.02886  \\
 10&1009.77 &  942.76& -0.00248  \\
 11&1011.76 &  955.73& -24.59055 \\
 12&1050.98 &  987.63& -15.37982\\ 
 13&1086.16 &  941.20&  0.13840 \\
 14&1100.39 & 1084.93&  -7.39151 \\
 15&1181.23& 1176.09& 0.14192 \\
 16&1253.05 & 1159.91& 2.90533 \\
 17&1308.03 & 1514.84&  0.01597  \\
 18&1398.00 & 1321.93&  -0.03015 \\
 19&1487.25& 1304.18&  0.15498 \\
 20&1530.78 & 1480.33& -11.83988 \\
 21&1638.04&  748.22& 0.20669 \\
 22&1643.72 & 1440.97& 15.11741 \\ 
 23&3169.47 & 3231.96& -0.01673 \\
 24&3171.66& 3248.95&  0.06826 \\
 25&3191.21& 3165.74&  -0.33462 \\
 26&3207.48 & 3256.50& 0.03941 \\
 27&3215.41& 3264.19& -1.15443 \\
\end{tabular}
\end{ruledtabular}
\end{table*}

\begin{figure*}[htbp]
\subfloat[]{
\includegraphics[width=0.3\textwidth]{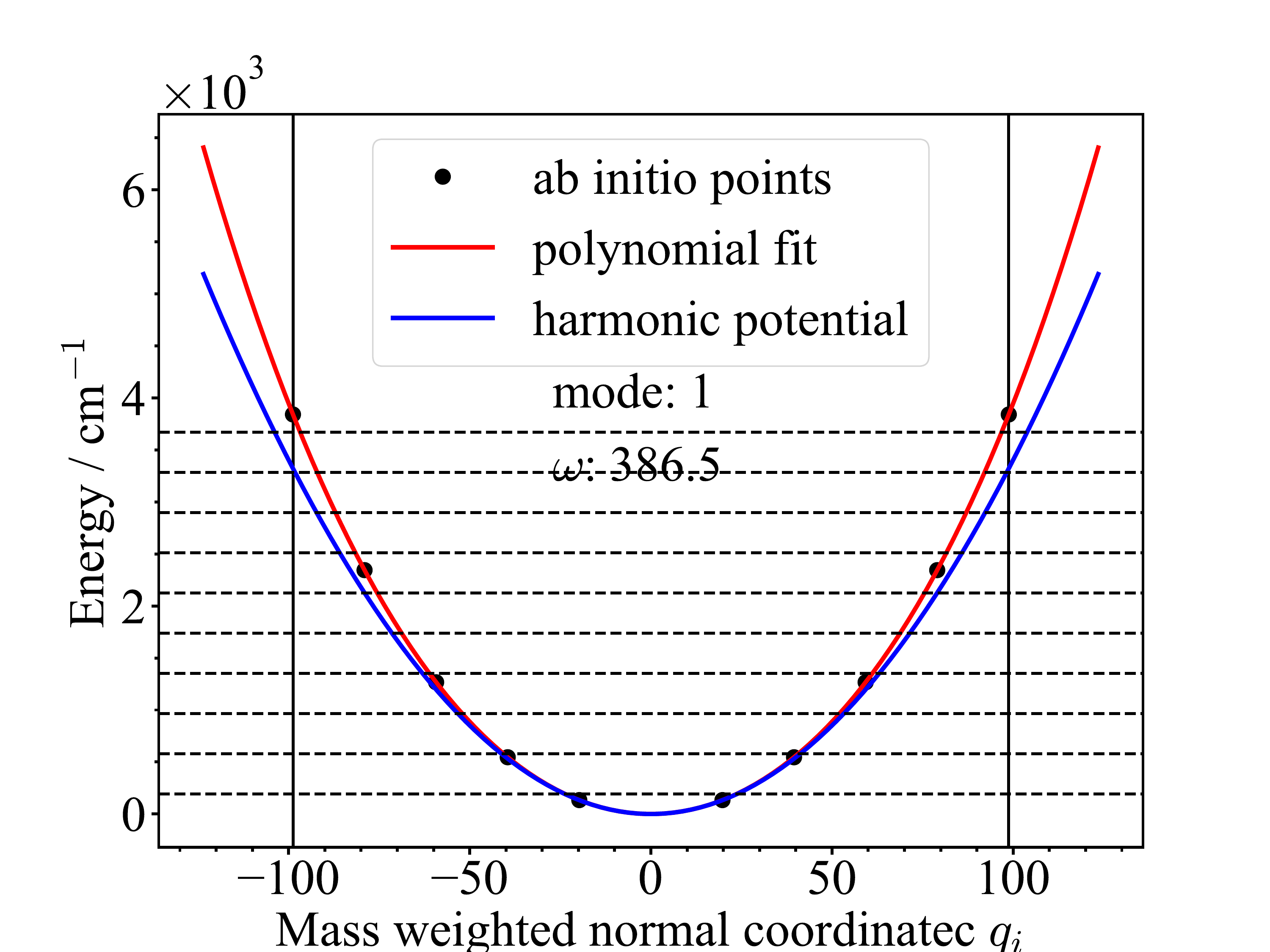}%
    }
\subfloat[]{
\includegraphics[width=0.3\textwidth]{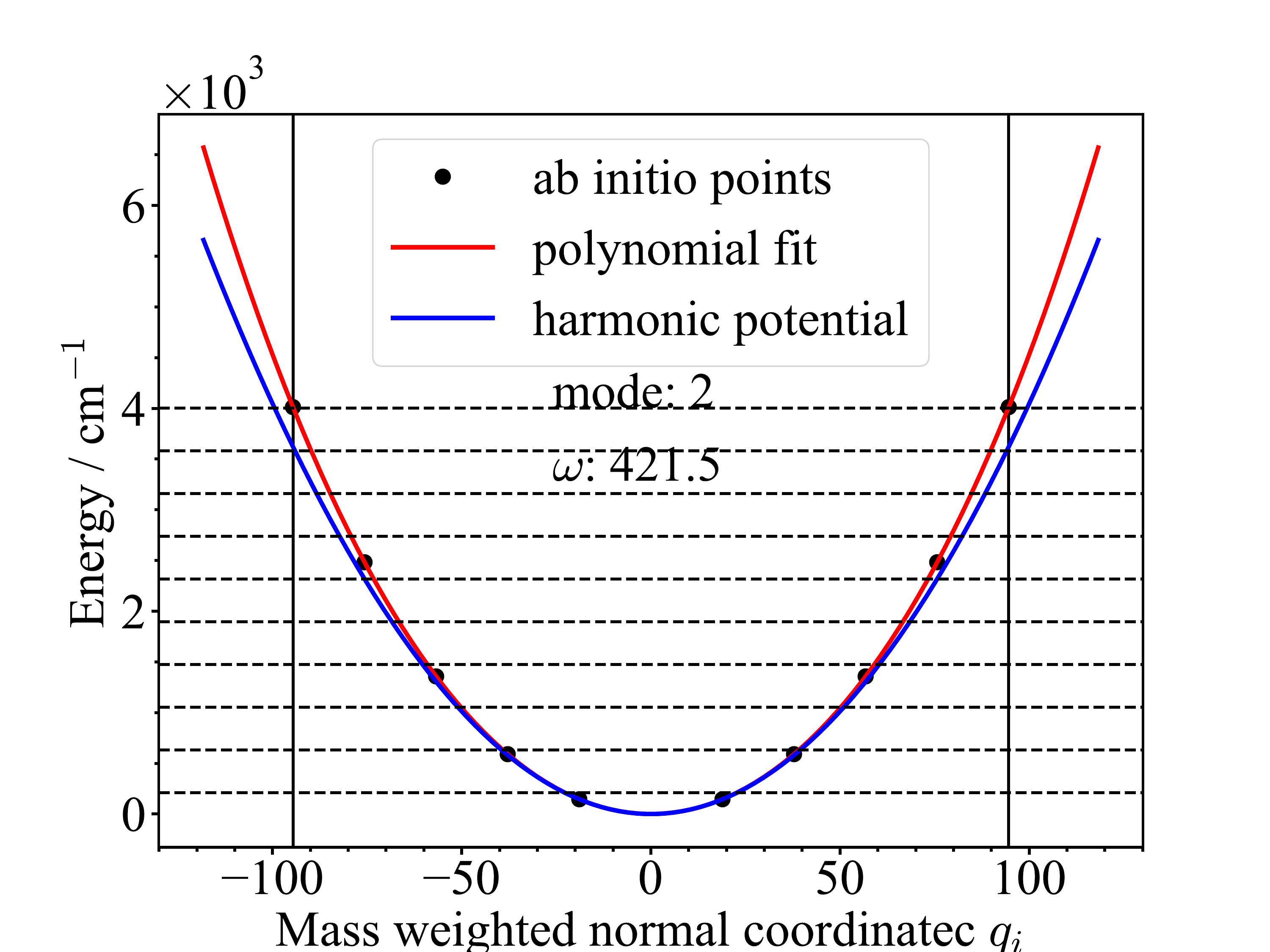}%
    }
\subfloat[]{
\includegraphics[width=0.3\textwidth]{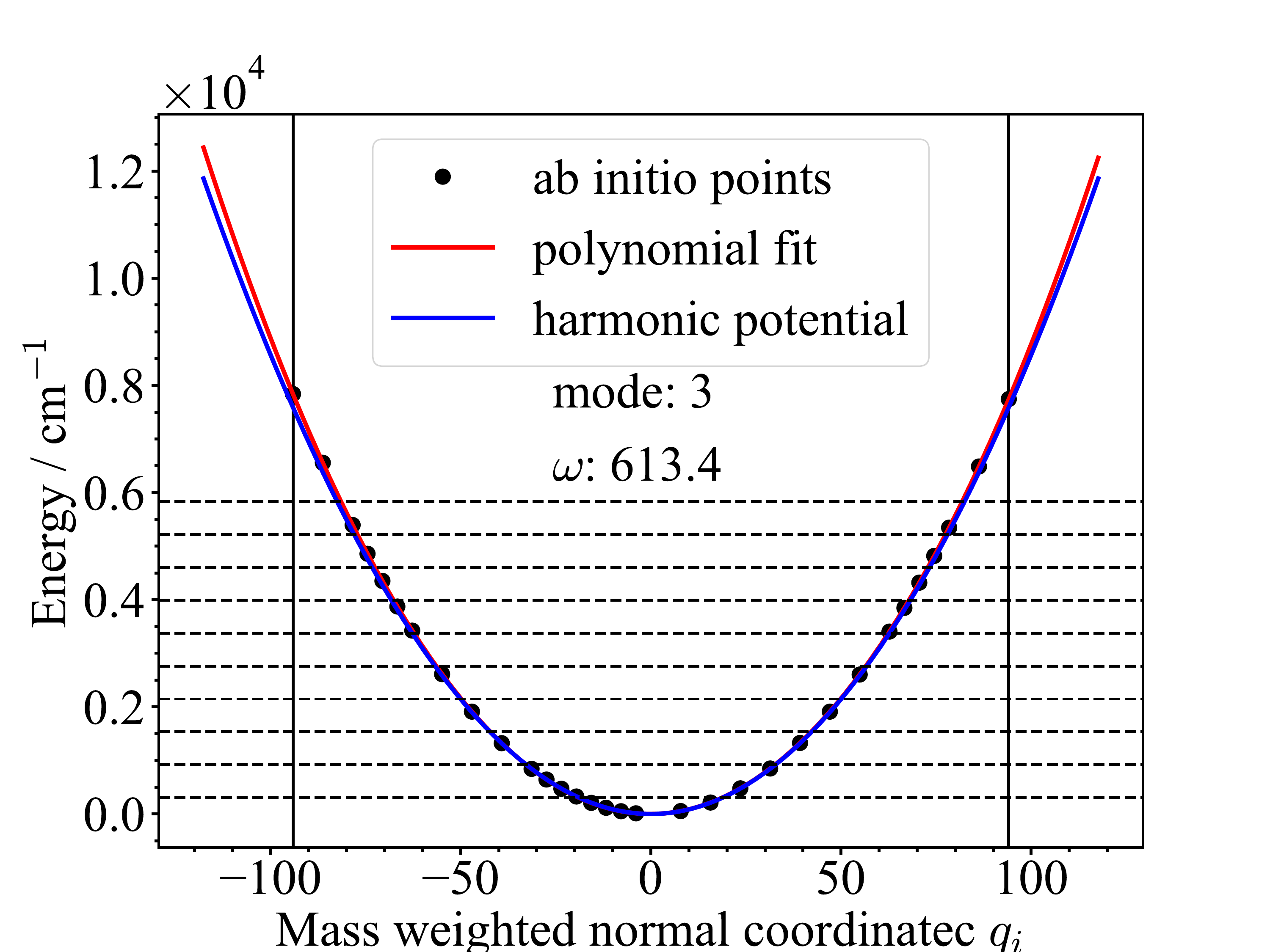}%
    }\\
\subfloat[]{
\includegraphics[width=0.3\textwidth]{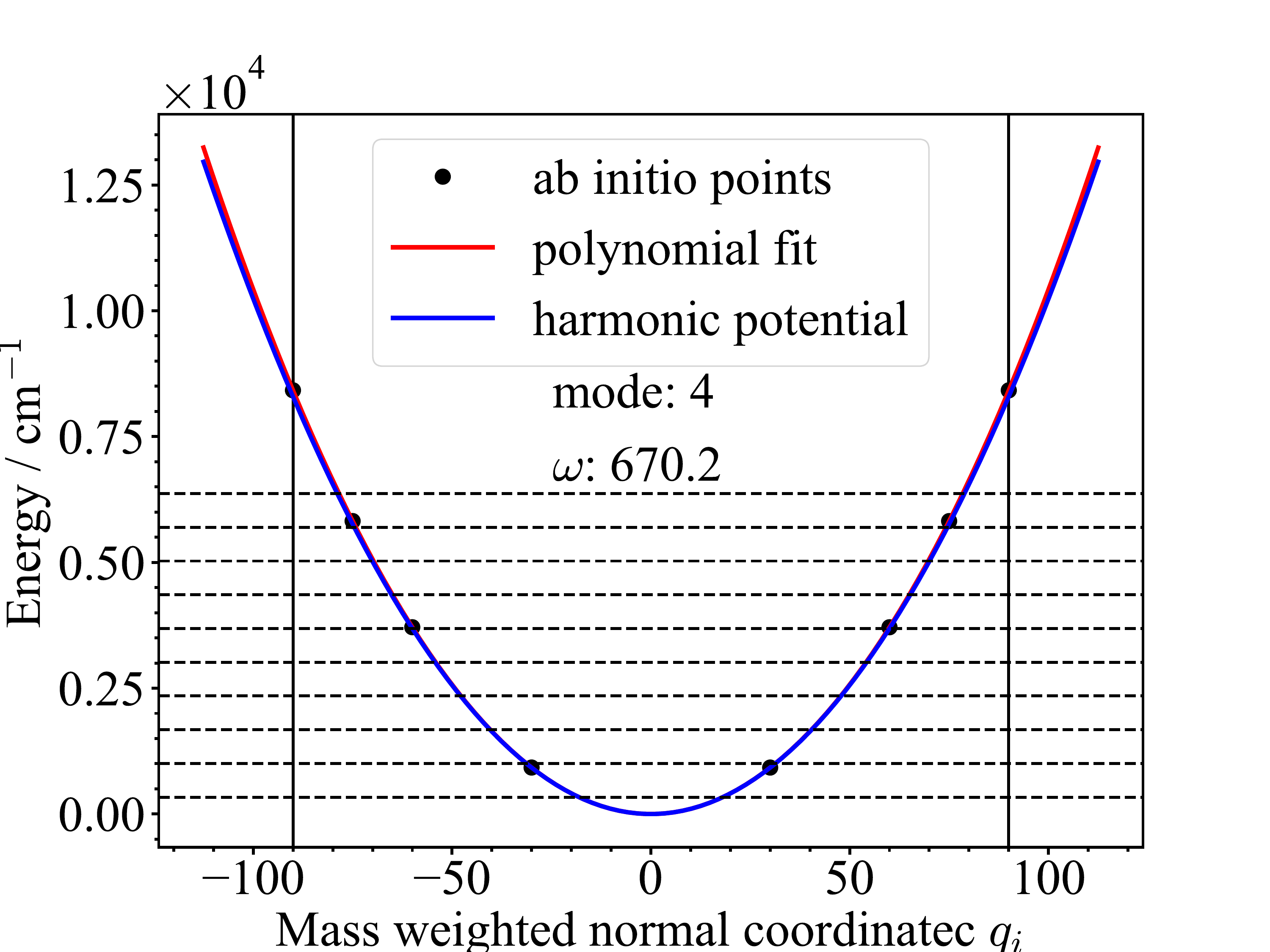}%
    }
\subfloat[]{
\includegraphics[width=0.3\textwidth]{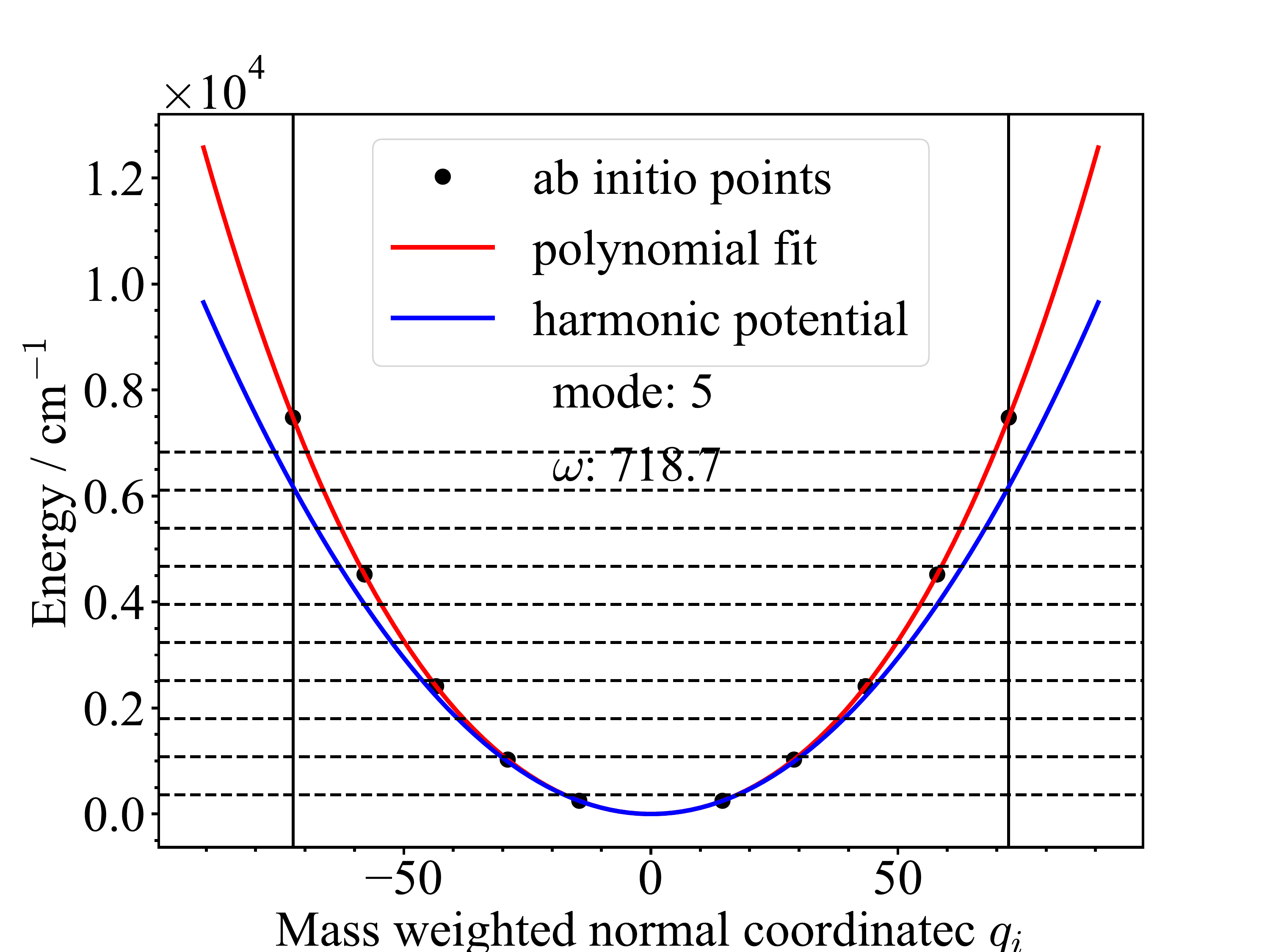}%
    }
\subfloat[]{
\includegraphics[width=0.3\textwidth]{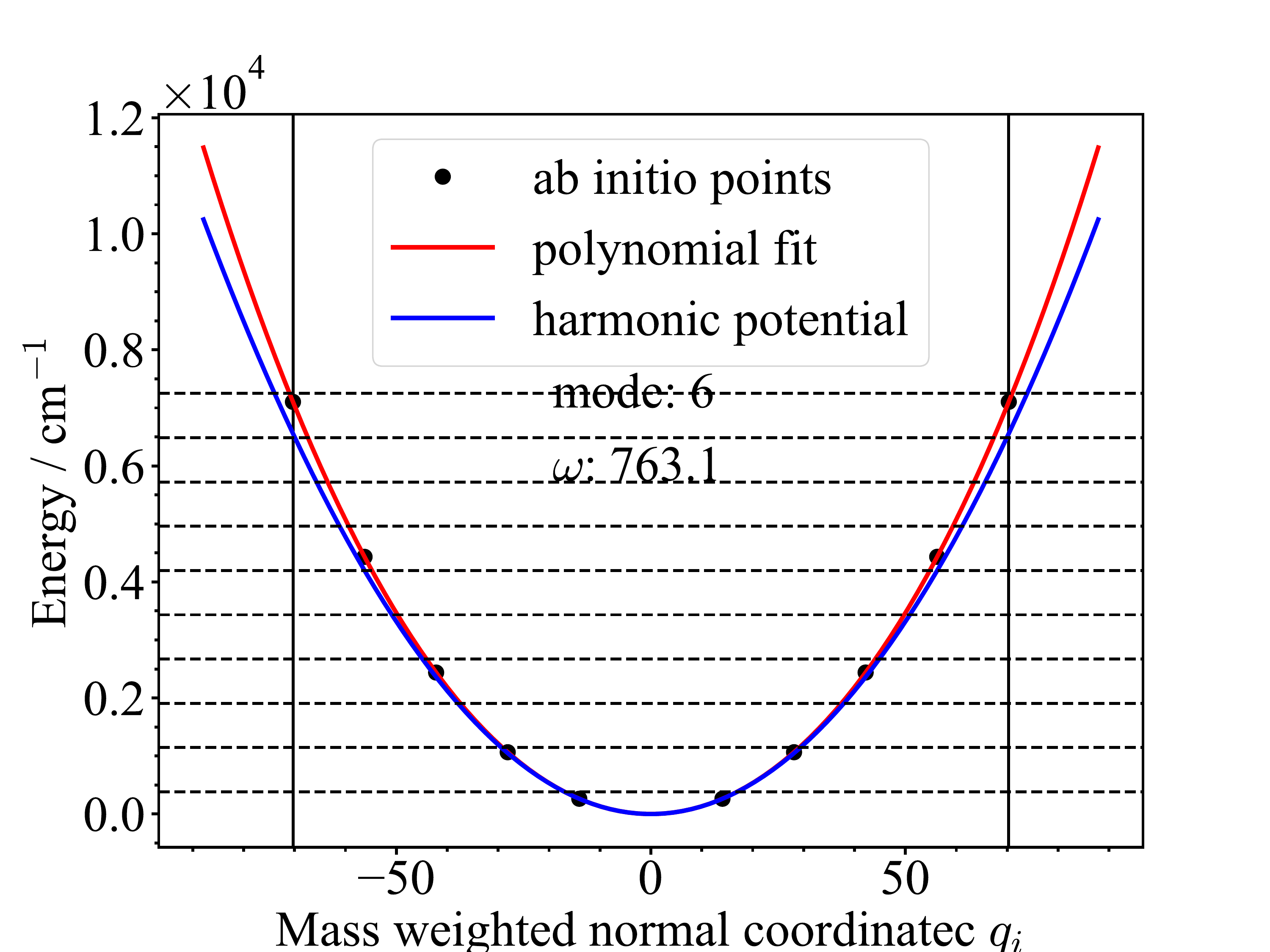}%
    }\\
\subfloat[]{
\includegraphics[width=0.3\textwidth]{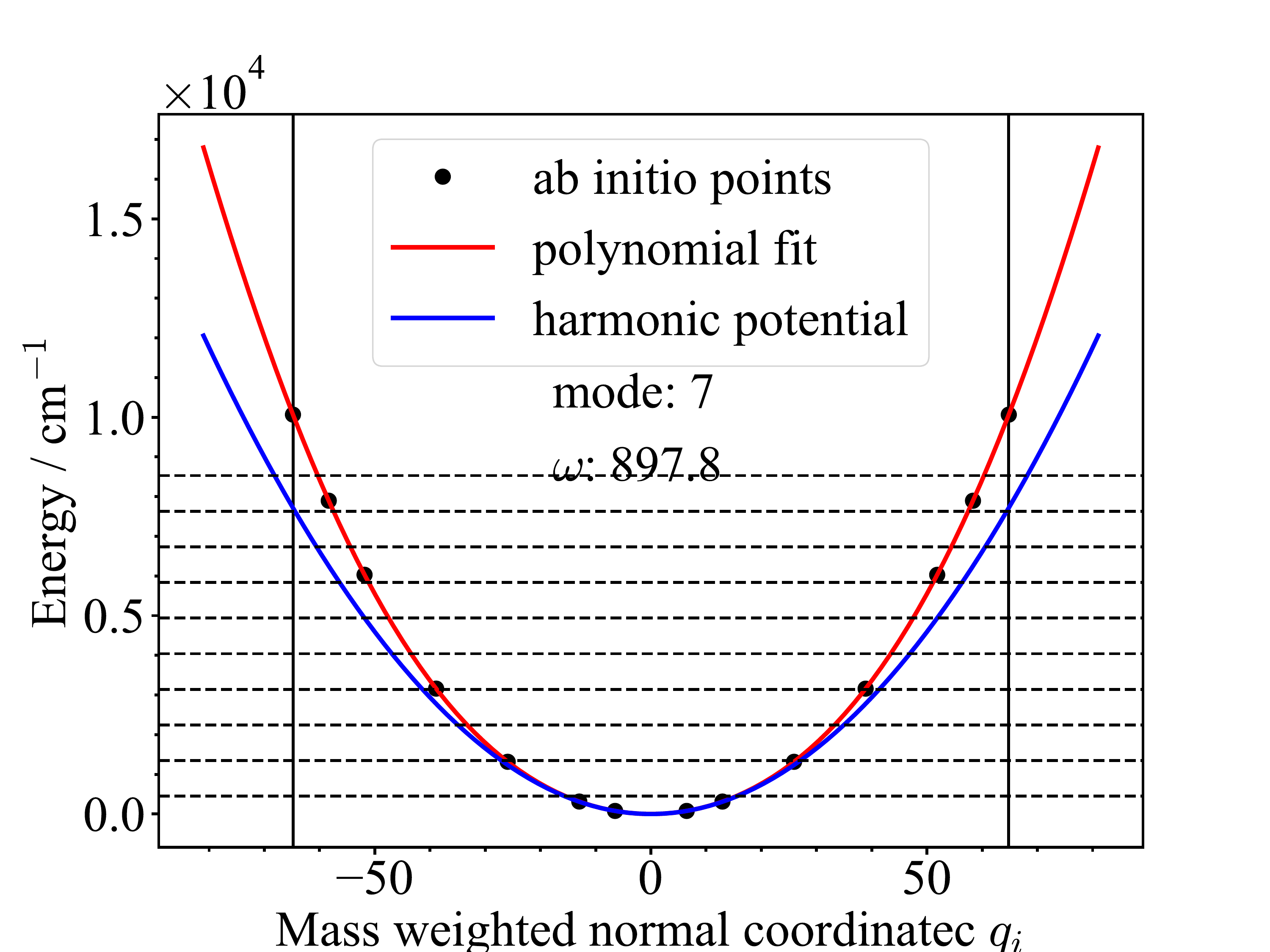}%
    }
\subfloat[]{
\includegraphics[width=0.3\textwidth]{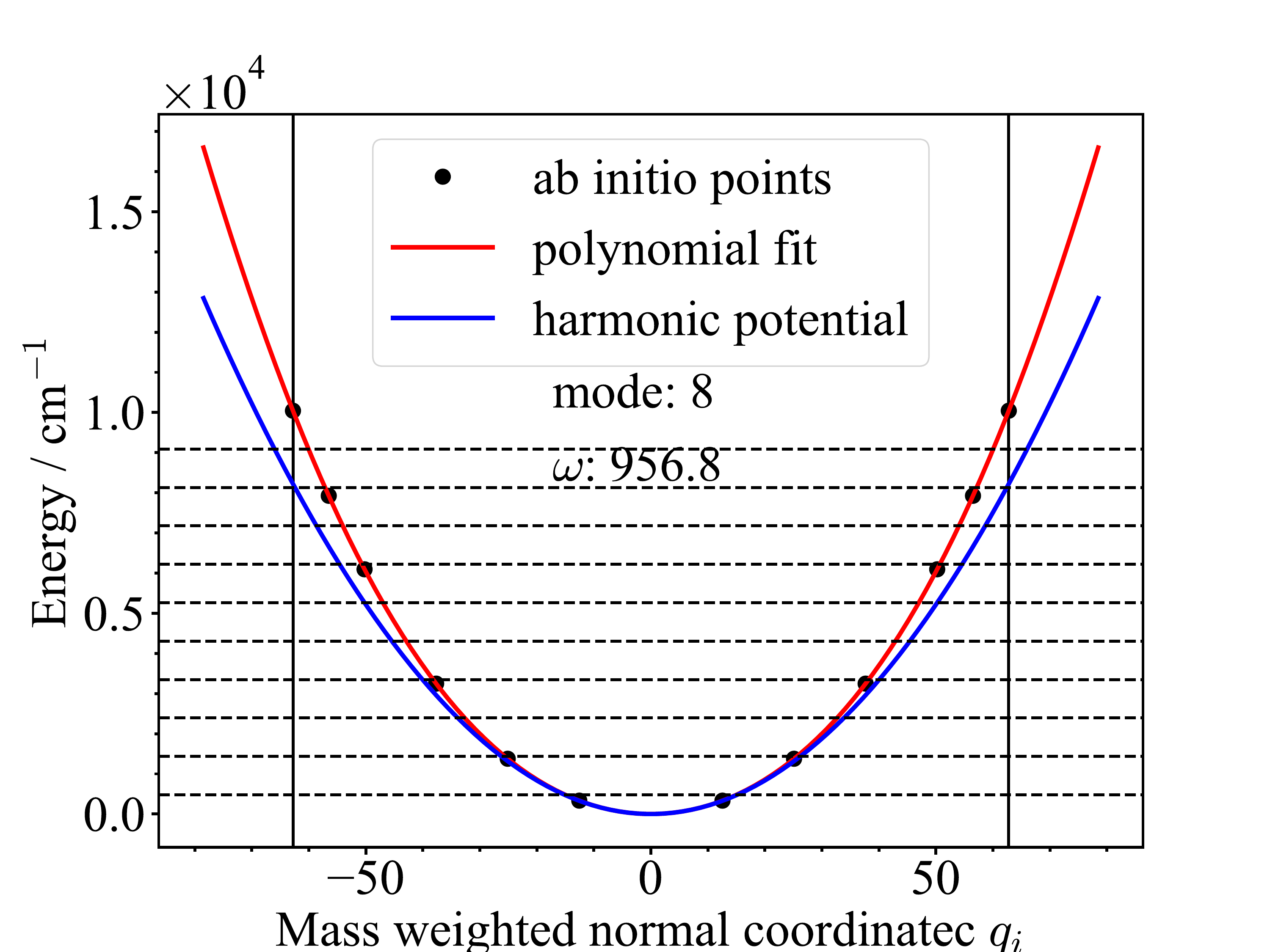}%
    }
\subfloat[]{
\includegraphics[width=0.3\textwidth]{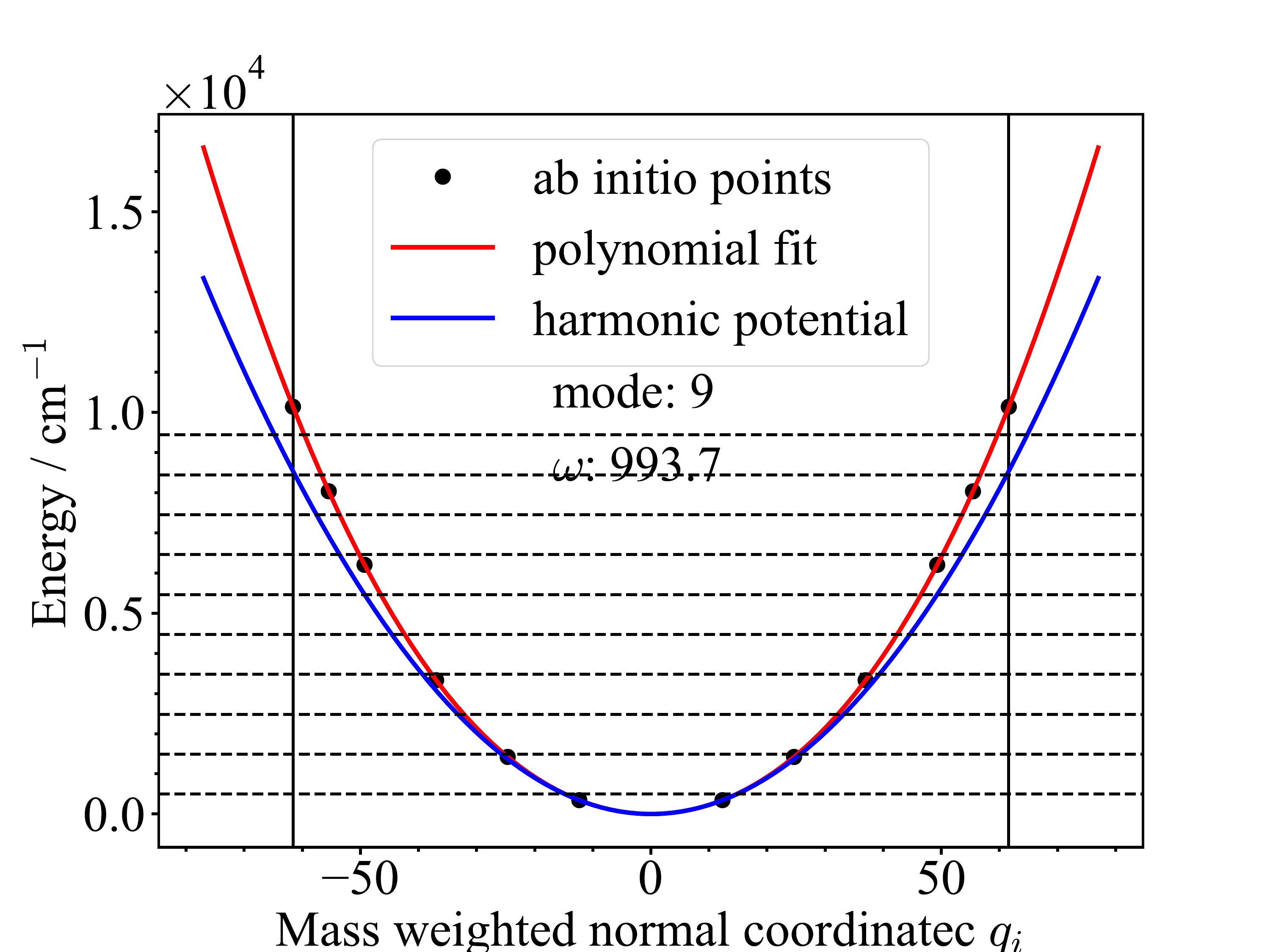}%
    }\\
\subfloat[]{
\includegraphics[width=0.3\textwidth]{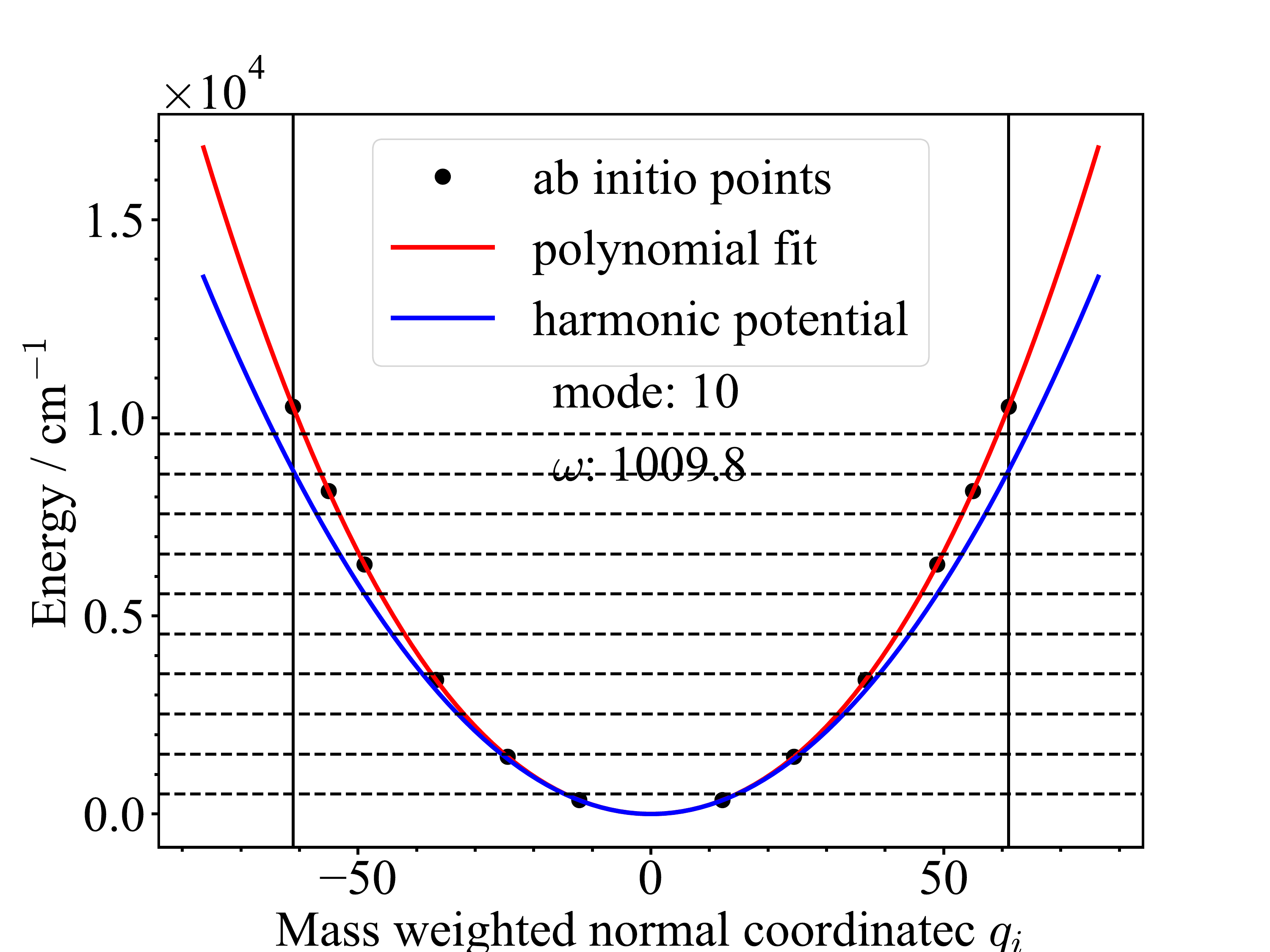}%
    }
\subfloat[]{
\includegraphics[width=0.3\textwidth]{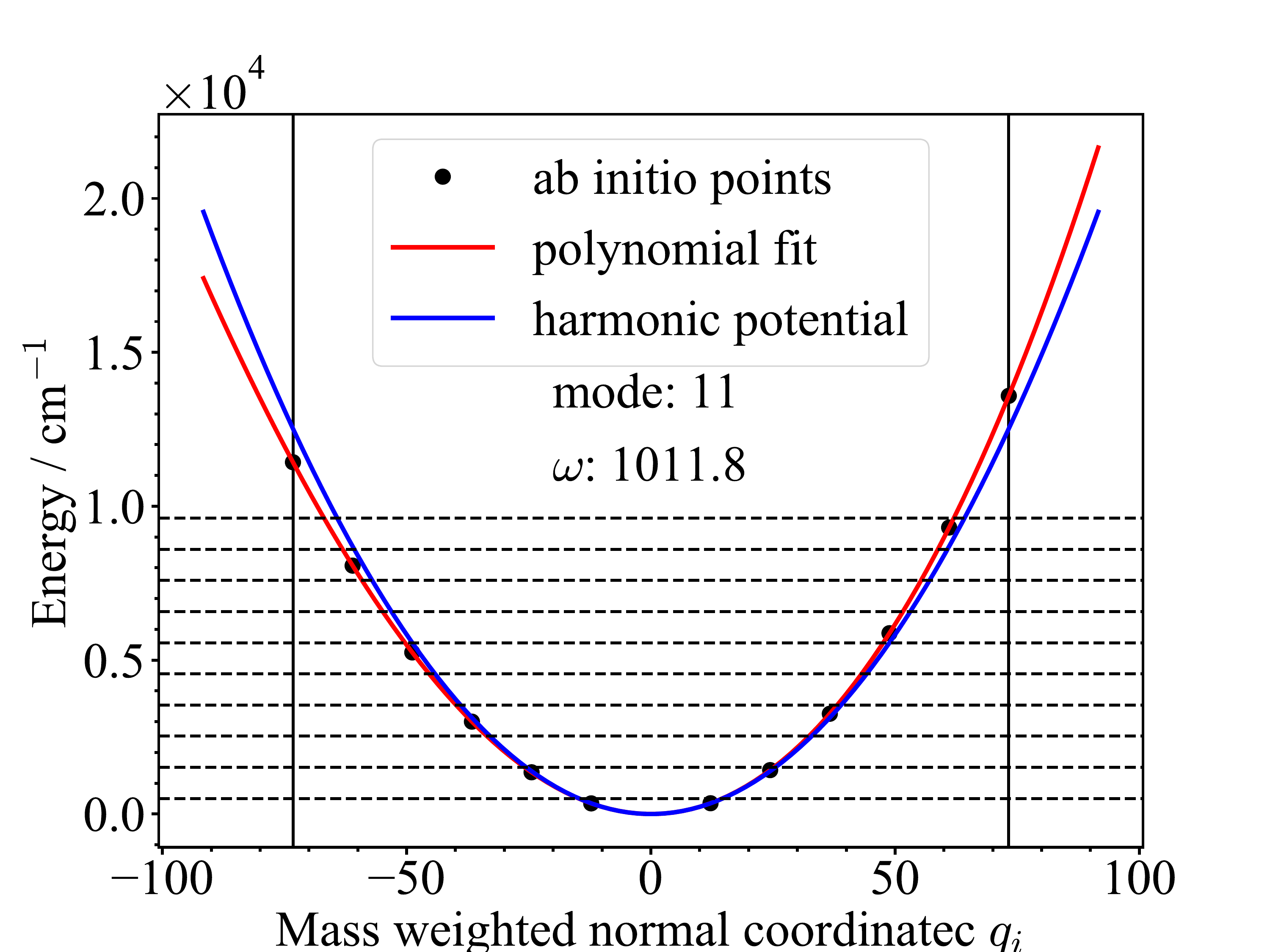}%
    }
\subfloat[]{
\includegraphics[width=0.3\textwidth]{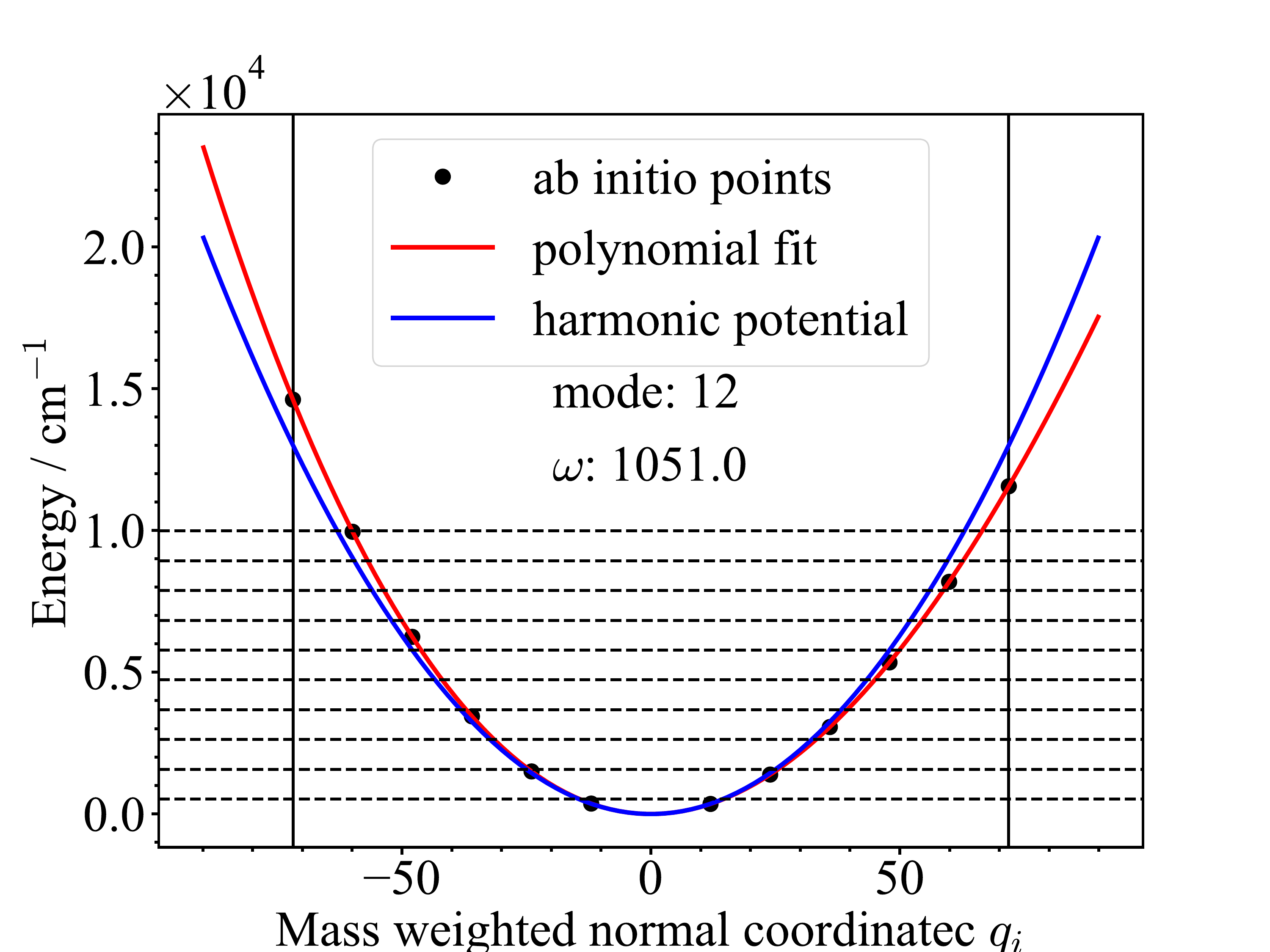}%
    }\\
\caption{\label{fig:anh_1}(a)-(l) The 1-D ground state PES cut of pyridine from mode 1 to mode 12. The unit of mass-weighted normal coordinates is the atomic unit.  The red line is the polynomial fit of ab initio points up to order 12. The blue line is harmonic potential. The vertical solid line is the boundary of the ab-initio points calculated acting as a guide to the eye. The dashed horizon lines are the lowest 10 energy levels of the harmonic potential. $\omega$ is gound state harmonic frequency. The ab-initio calculations are carried out at B3LYP/6-31g(d) level in gas phase.  }
\end{figure*}

\begin{figure*}[htbp]
\subfloat[]{
\includegraphics[width=0.3\textwidth]{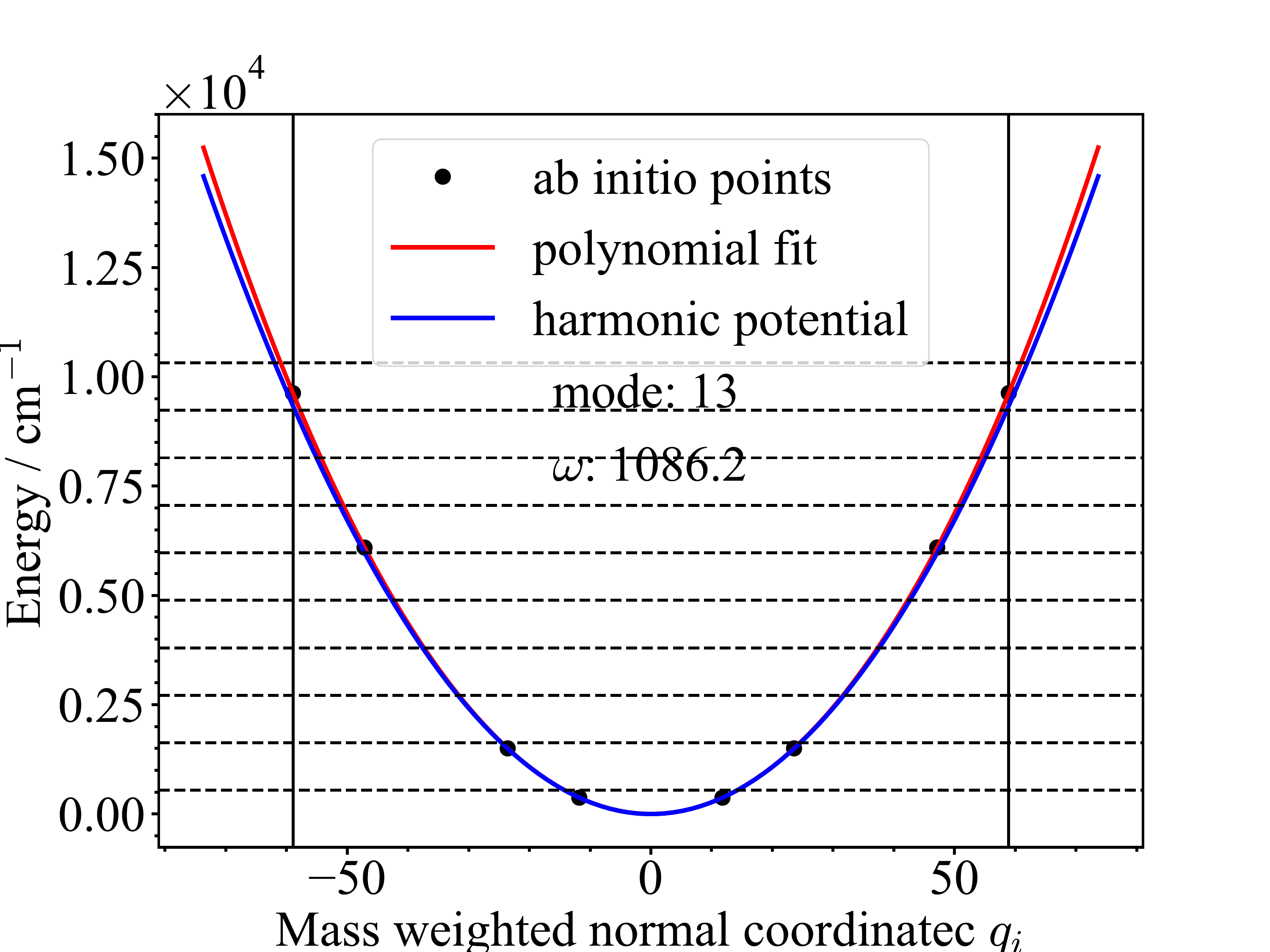}%
    }
\subfloat[]{
\includegraphics[width=0.3\textwidth]{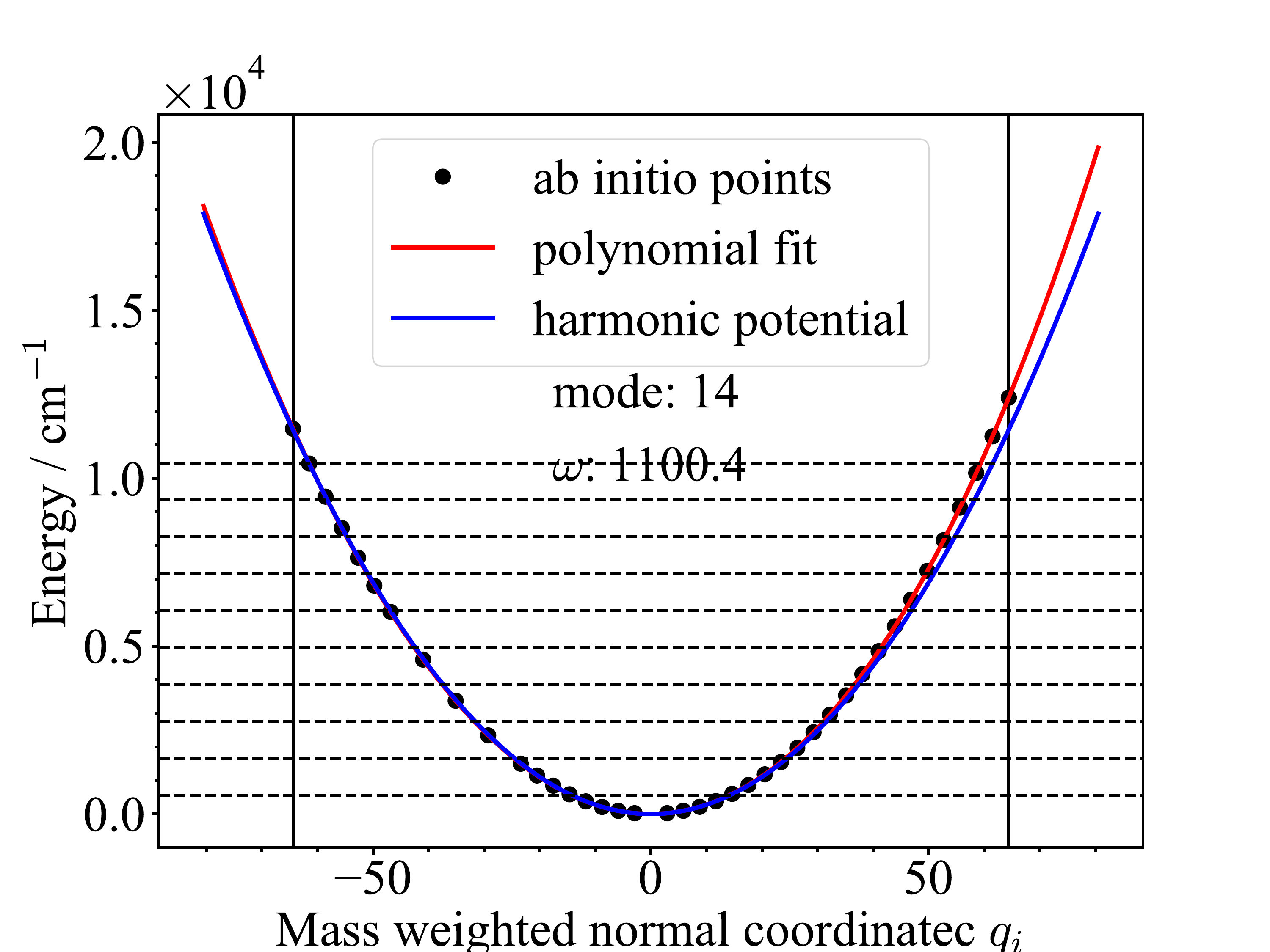}%
    }
\subfloat[]{
\includegraphics[width=0.3\textwidth]{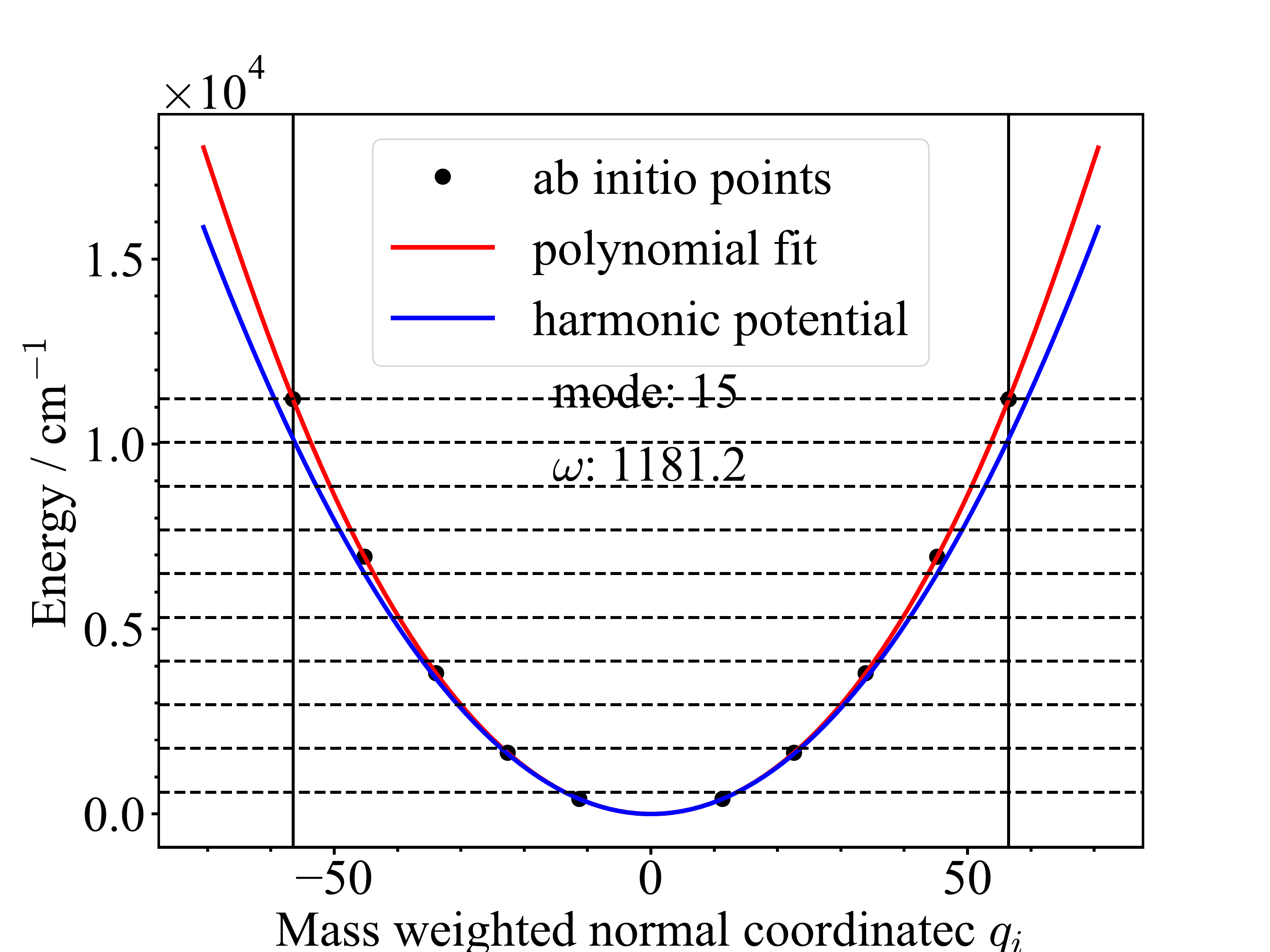}%
    }\\ 
\subfloat[]{
\includegraphics[width=0.3\textwidth]{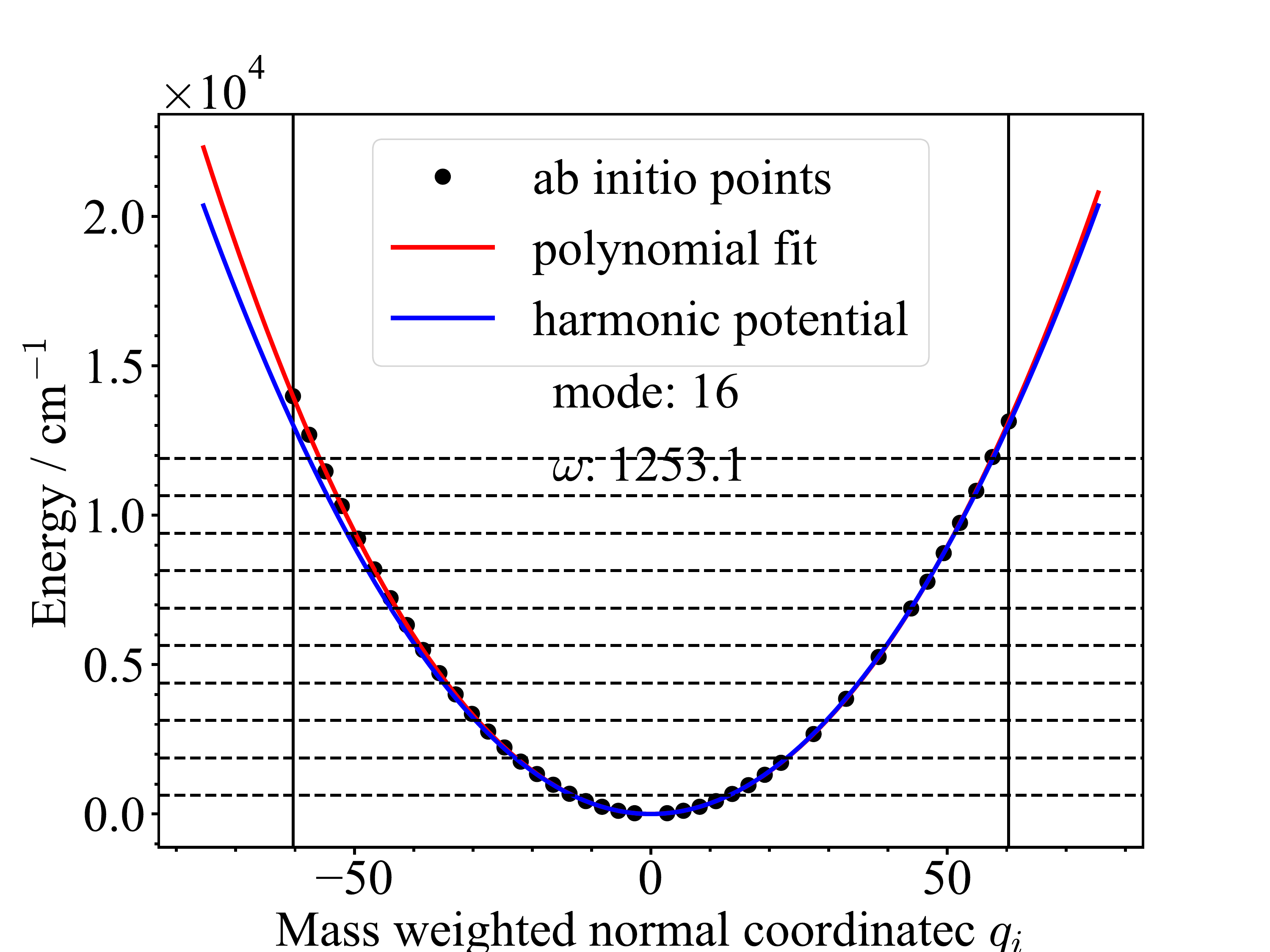}%
    }
\subfloat[]{
\includegraphics[width=0.3\textwidth]{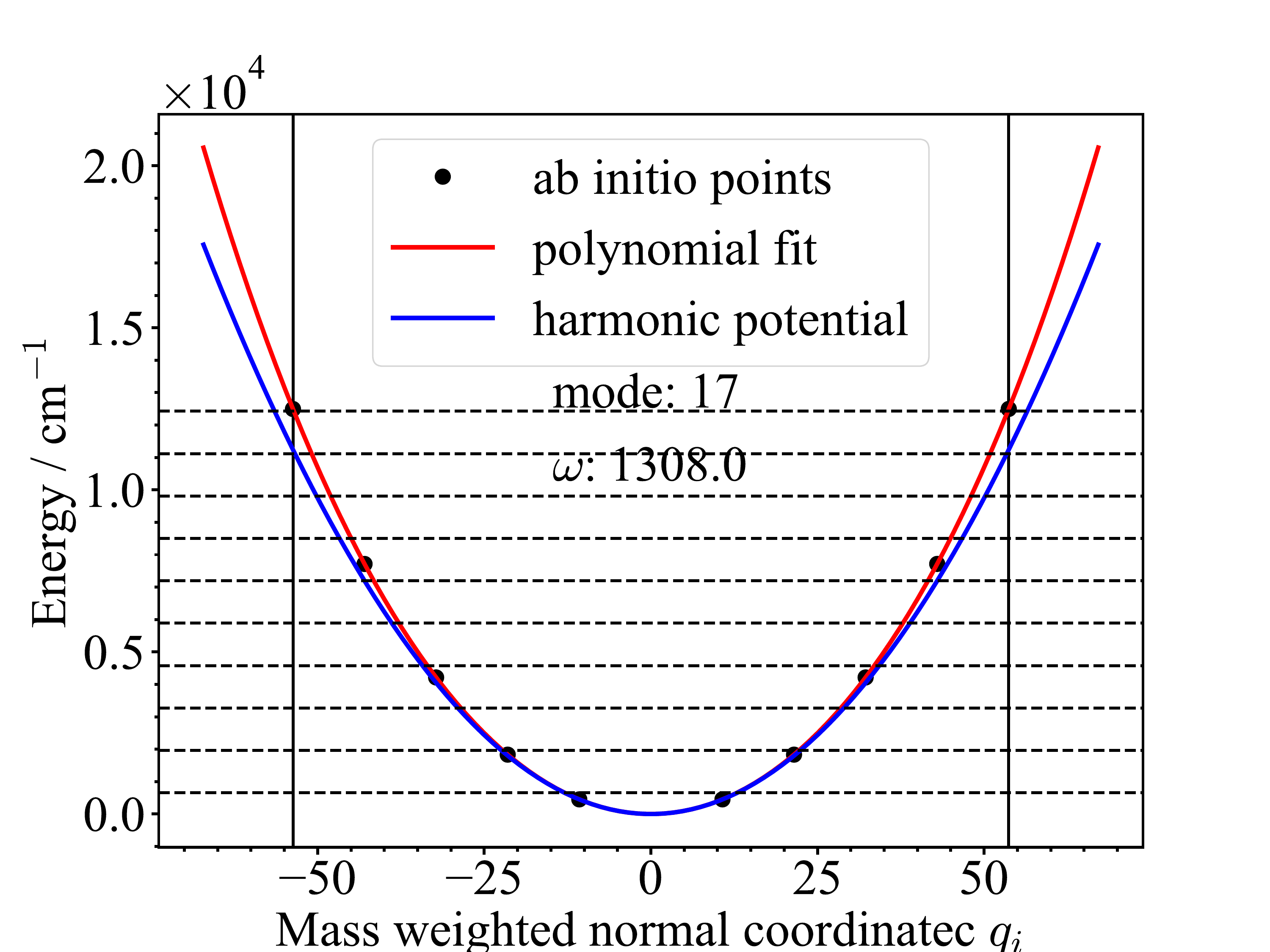}%
    }
\subfloat[]{
\includegraphics[width=0.3\textwidth]{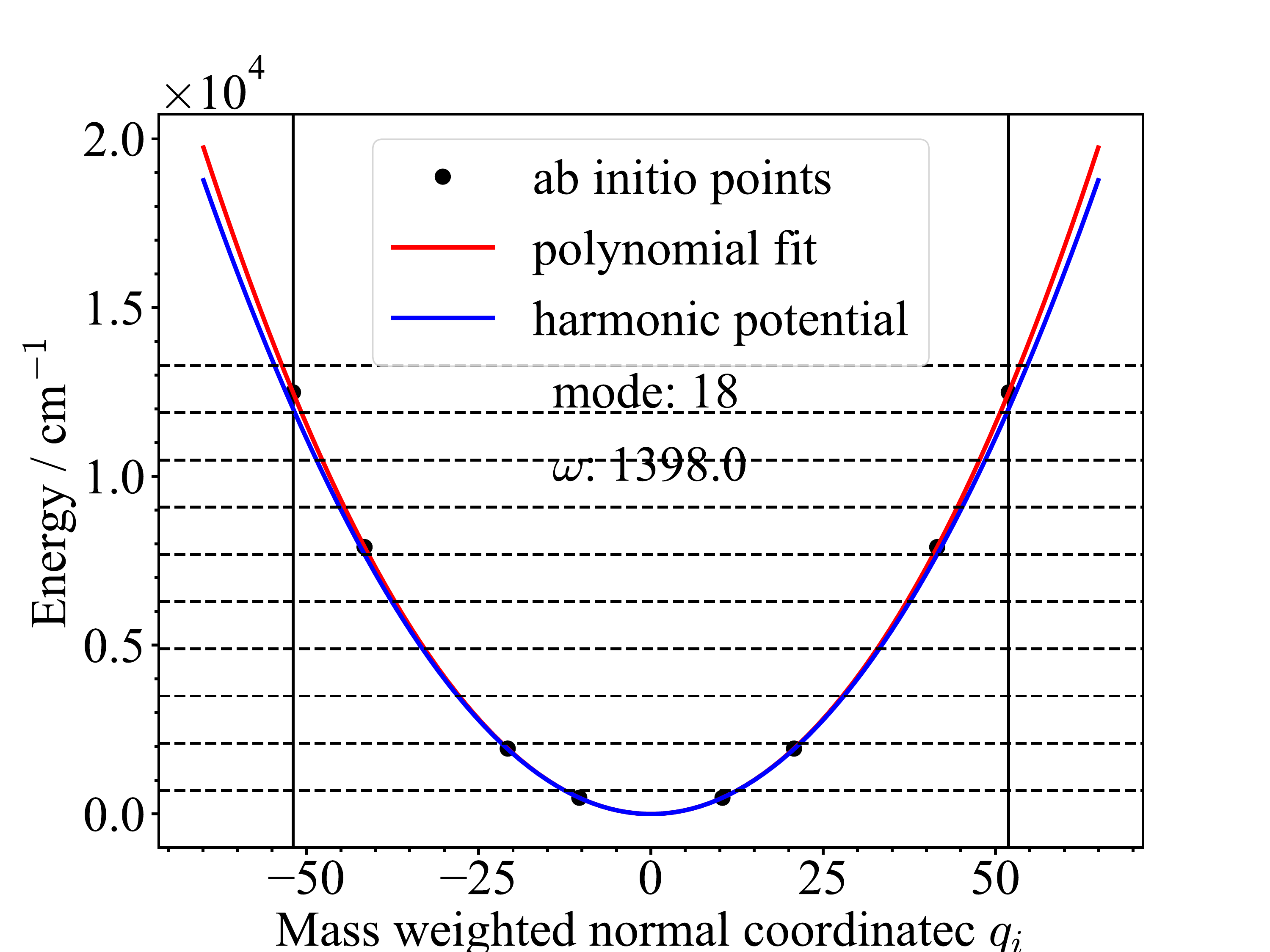}%
    }\\
\subfloat[]{
\includegraphics[width=0.3\textwidth]{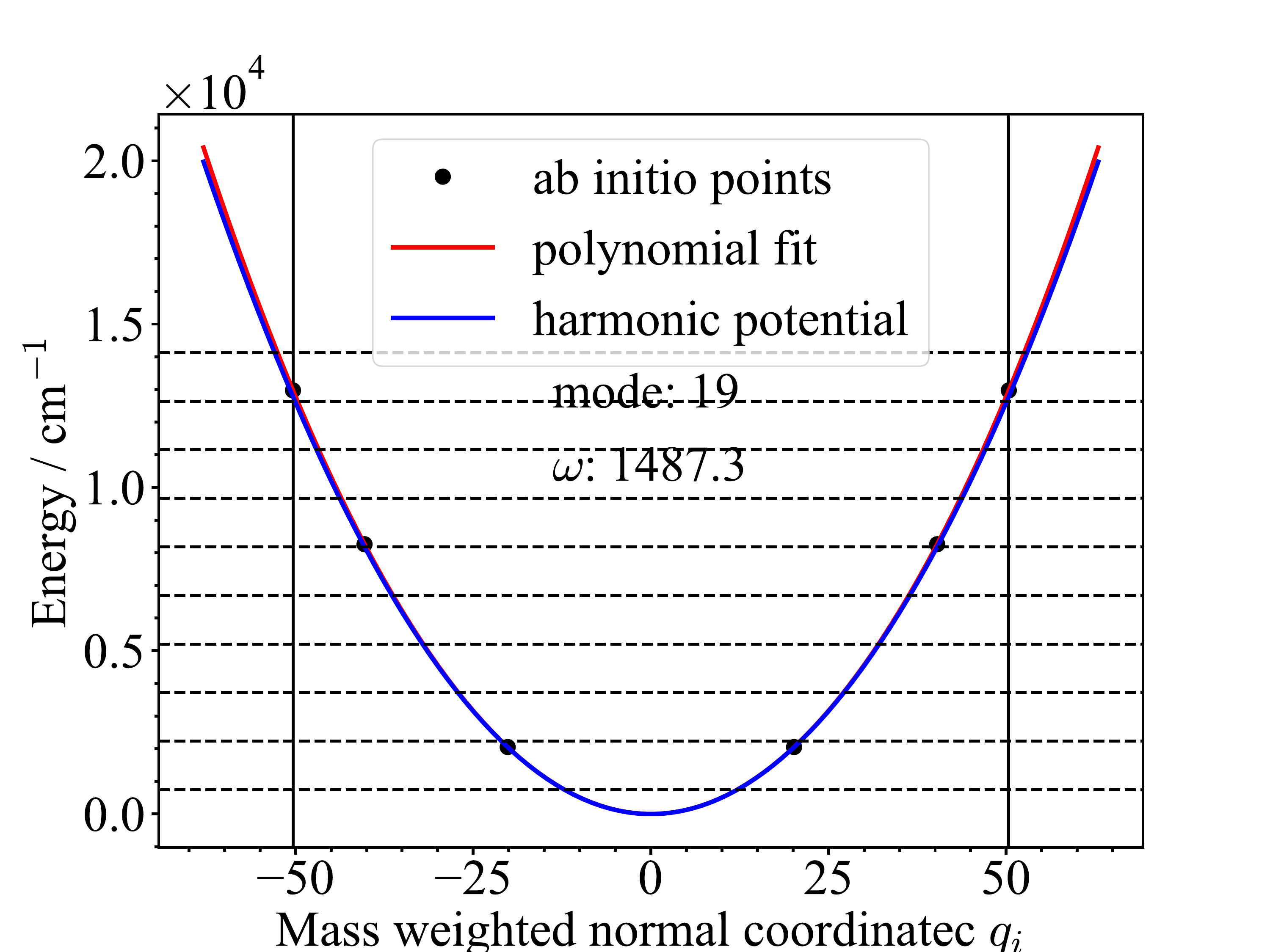}%
    }
\subfloat[]{
\includegraphics[width=0.3\textwidth]{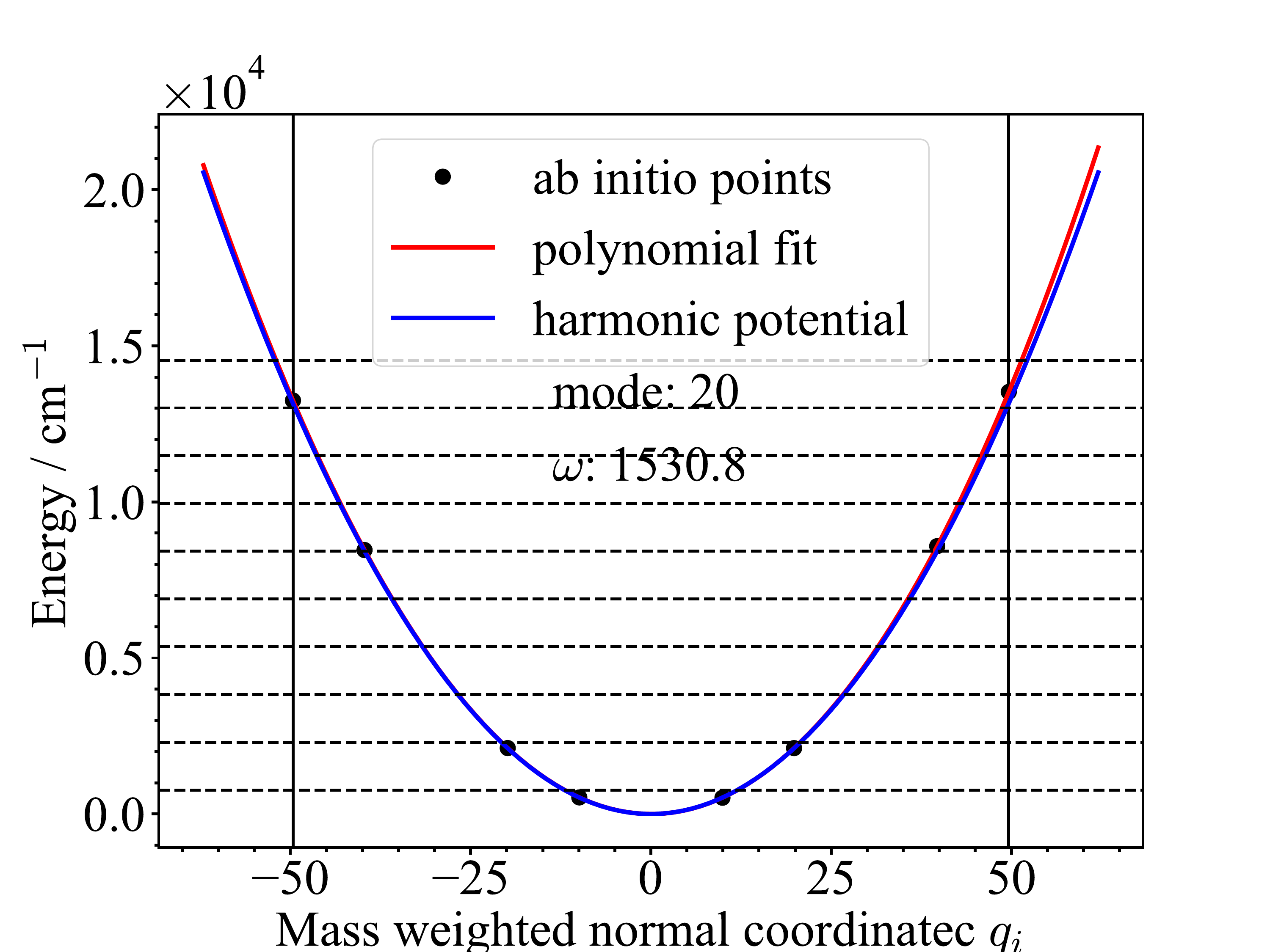}%
    }
\subfloat[]{
\includegraphics[width=0.3\textwidth]{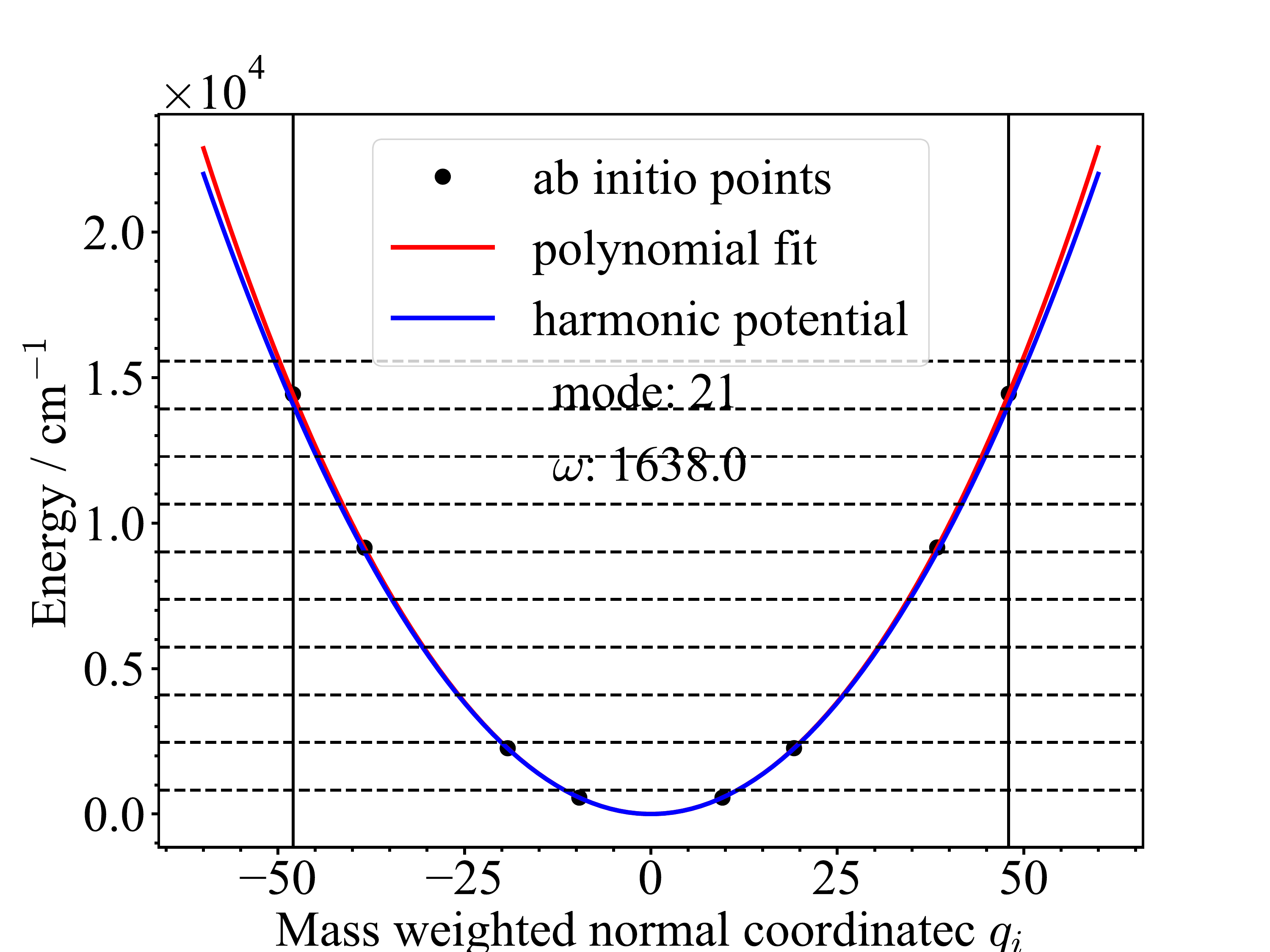}%
    }\\

\caption{\label{fig:anh_2} Similar as Fig.~\ref{fig:anh_1} but for modes 13-21.  }
\end{figure*}

\begin{figure*}
\subfloat[]{
\includegraphics[width=0.3\textwidth]{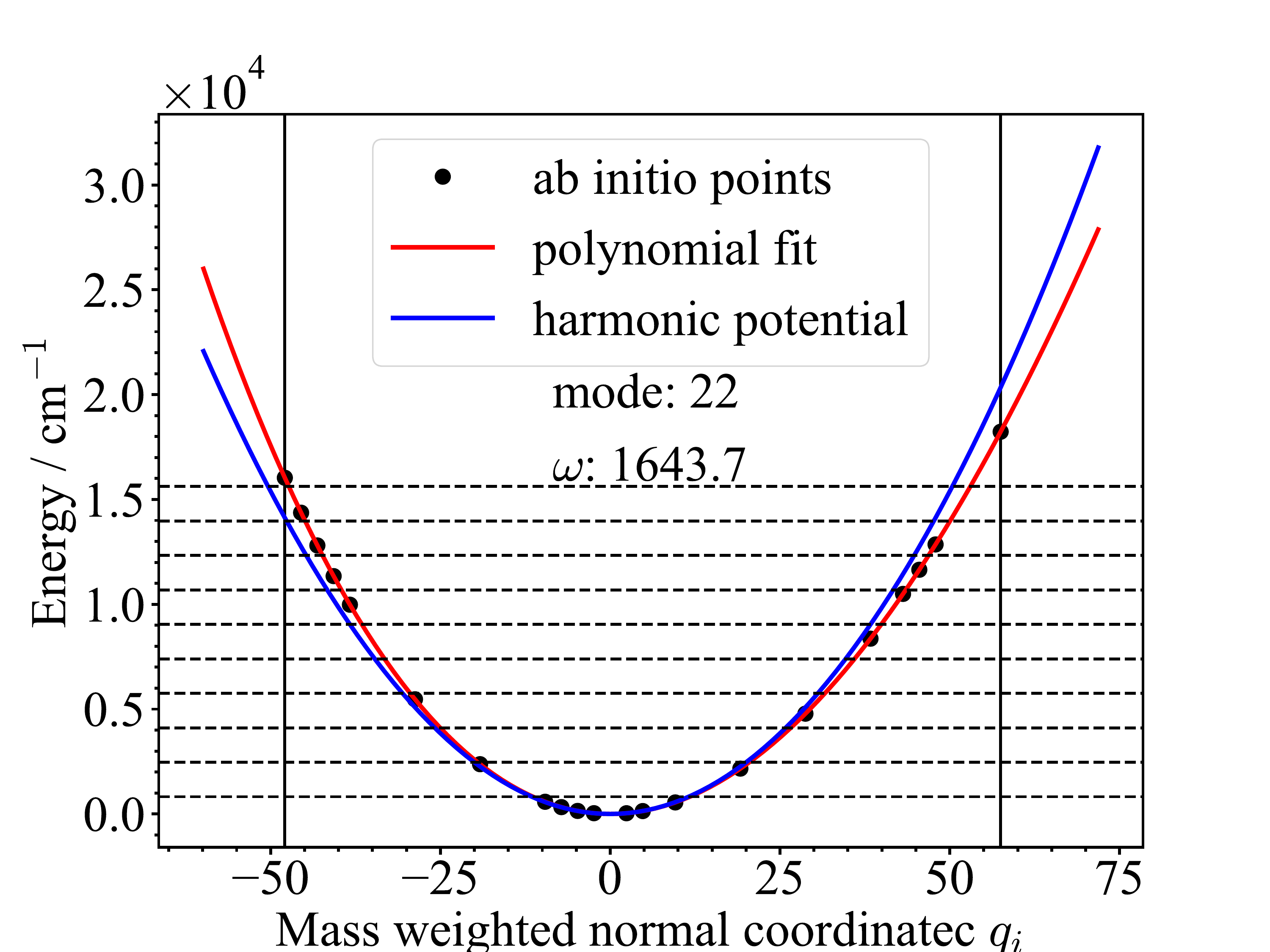}%
    }
\subfloat[]{
\includegraphics[width=0.3\textwidth]{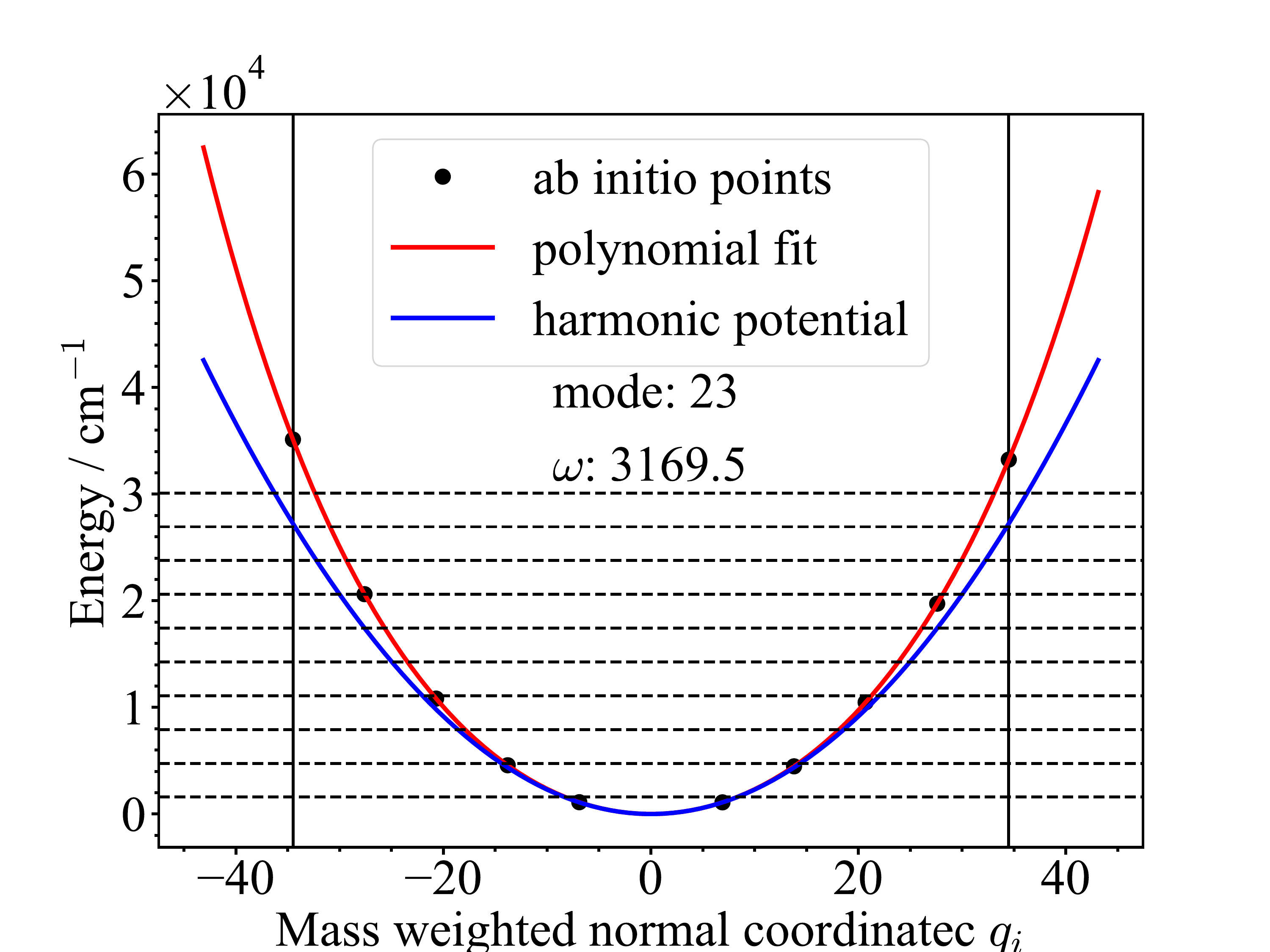}%
    }
\subfloat[]{
\includegraphics[width=0.3\textwidth]{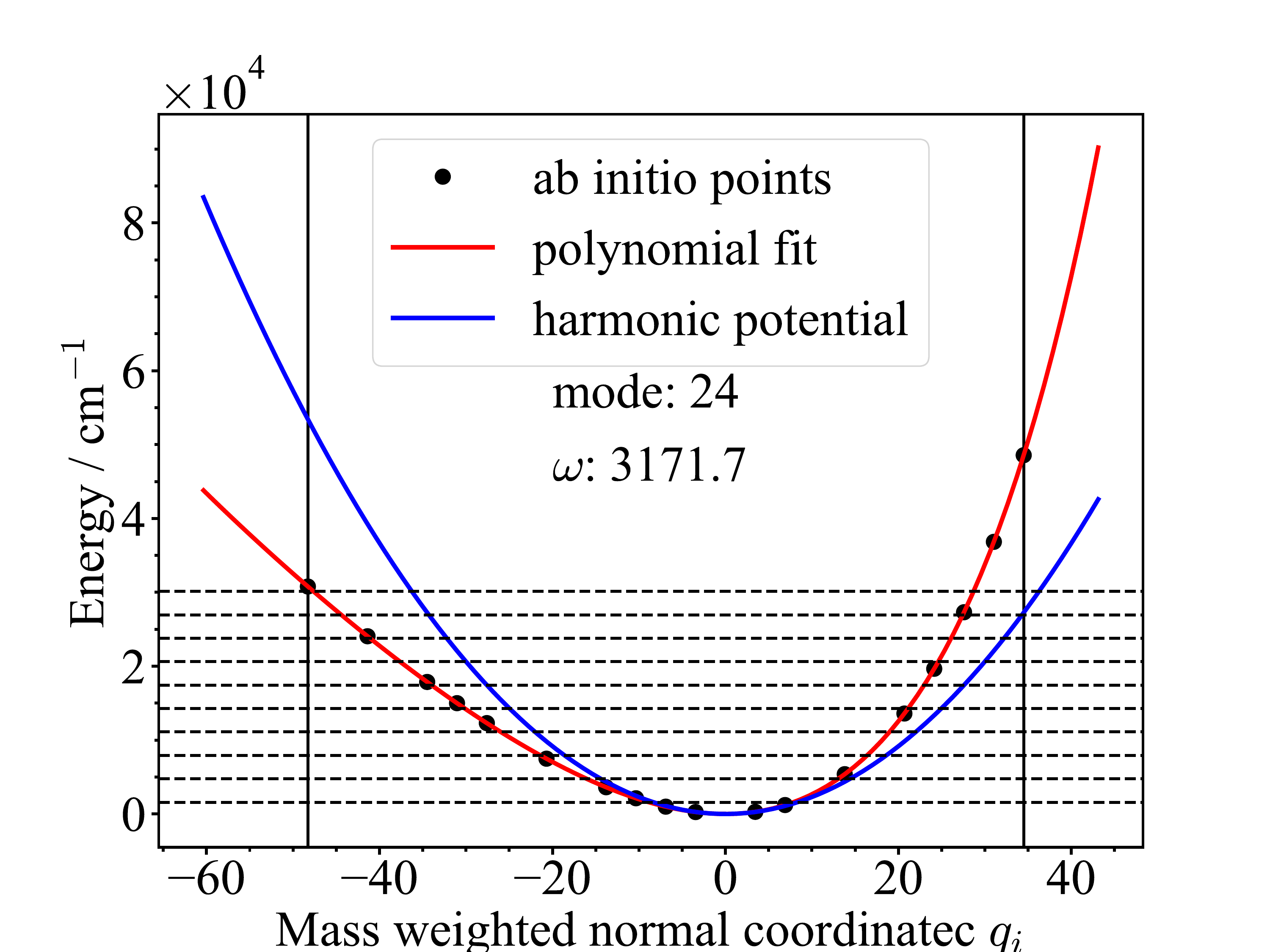}%
    }\\
\subfloat[]{
\includegraphics[width=0.3\textwidth]{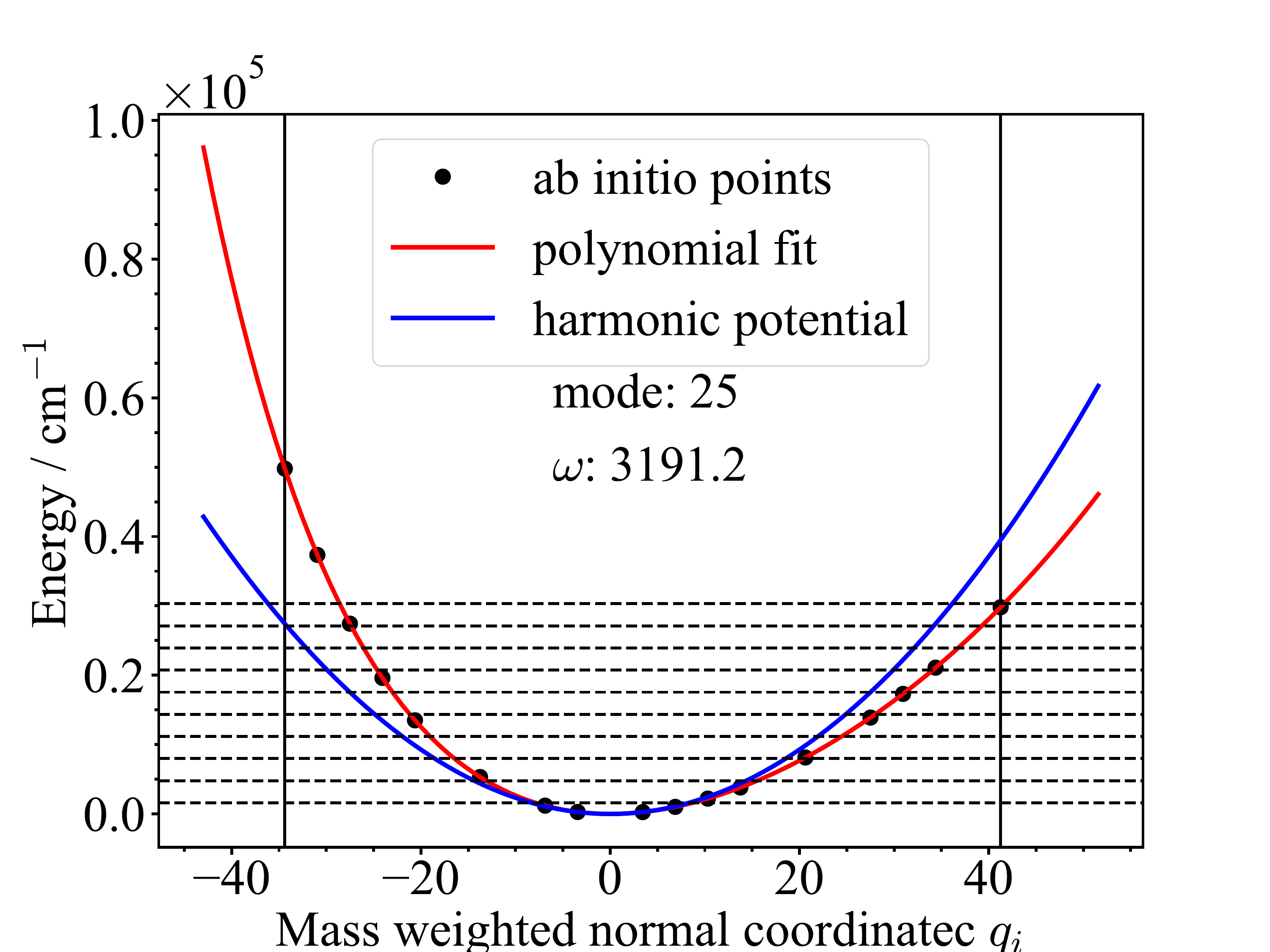}%
    }
\subfloat[]{
\includegraphics[width=0.3\textwidth]{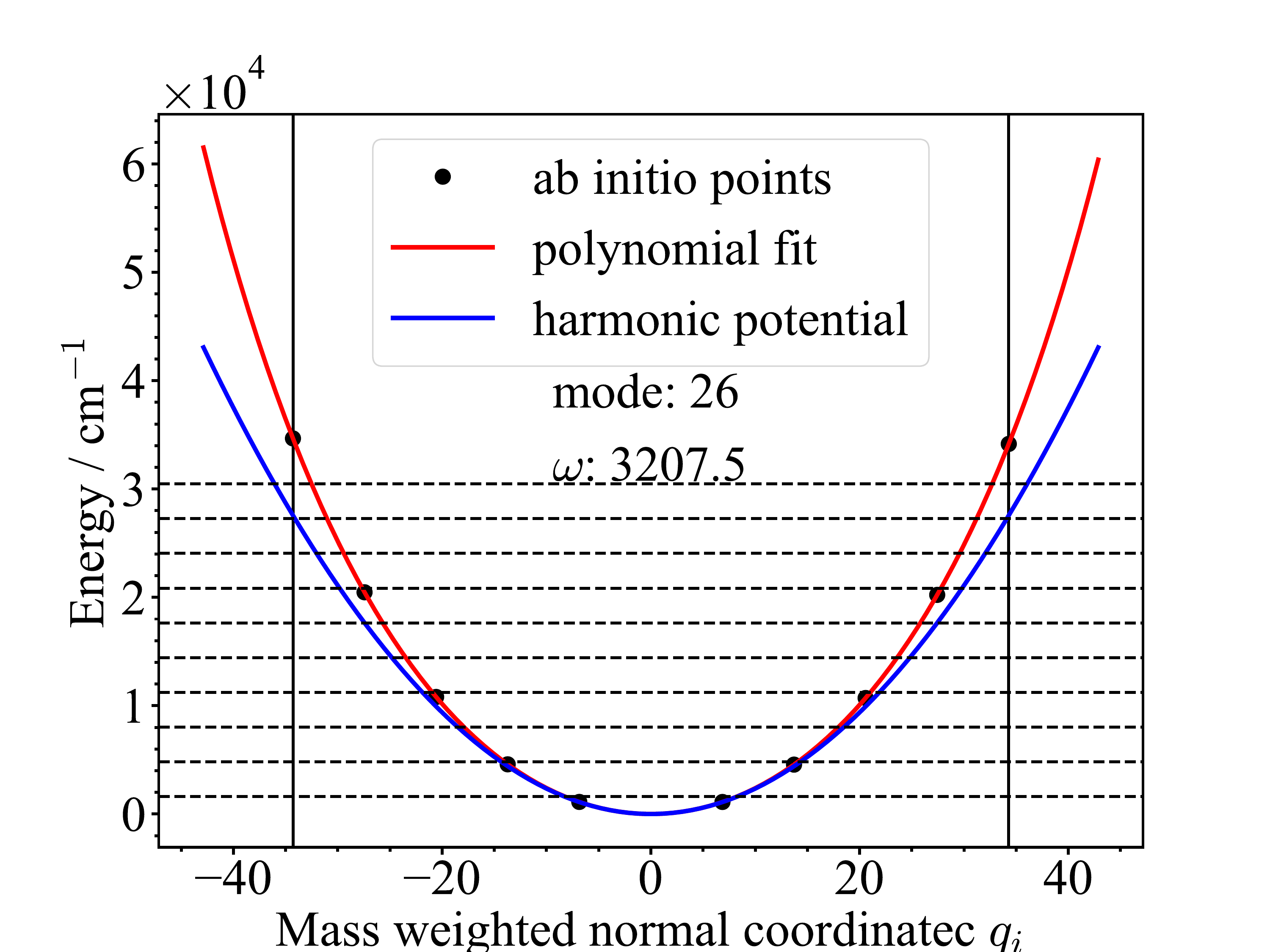}%
    }
\subfloat[]{
\includegraphics[width=0.3\textwidth]{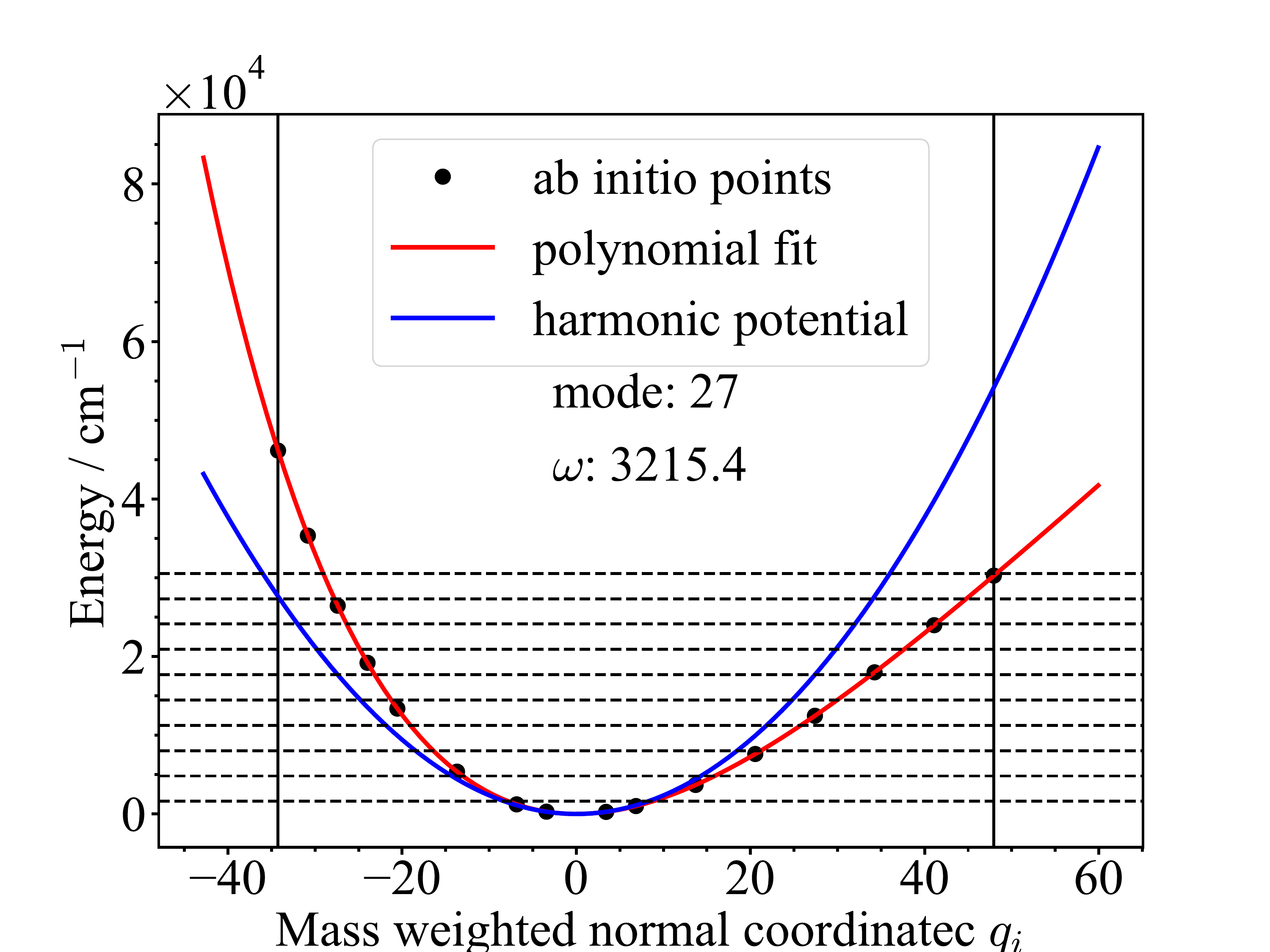}%
    }\\
    
    \caption{\label{fig:anh_3} Similar as Fig.~\ref{fig:anh_1} but for modes 22-27.  }
\end{figure*}

\clearpage
\clearpage
\newpage
\section{Another reference spectrum of pyridine with all 27 vibrational modes }

\begin{figure*}[htbp]
\subfloat[]{
\includegraphics[width=0.75\textwidth]{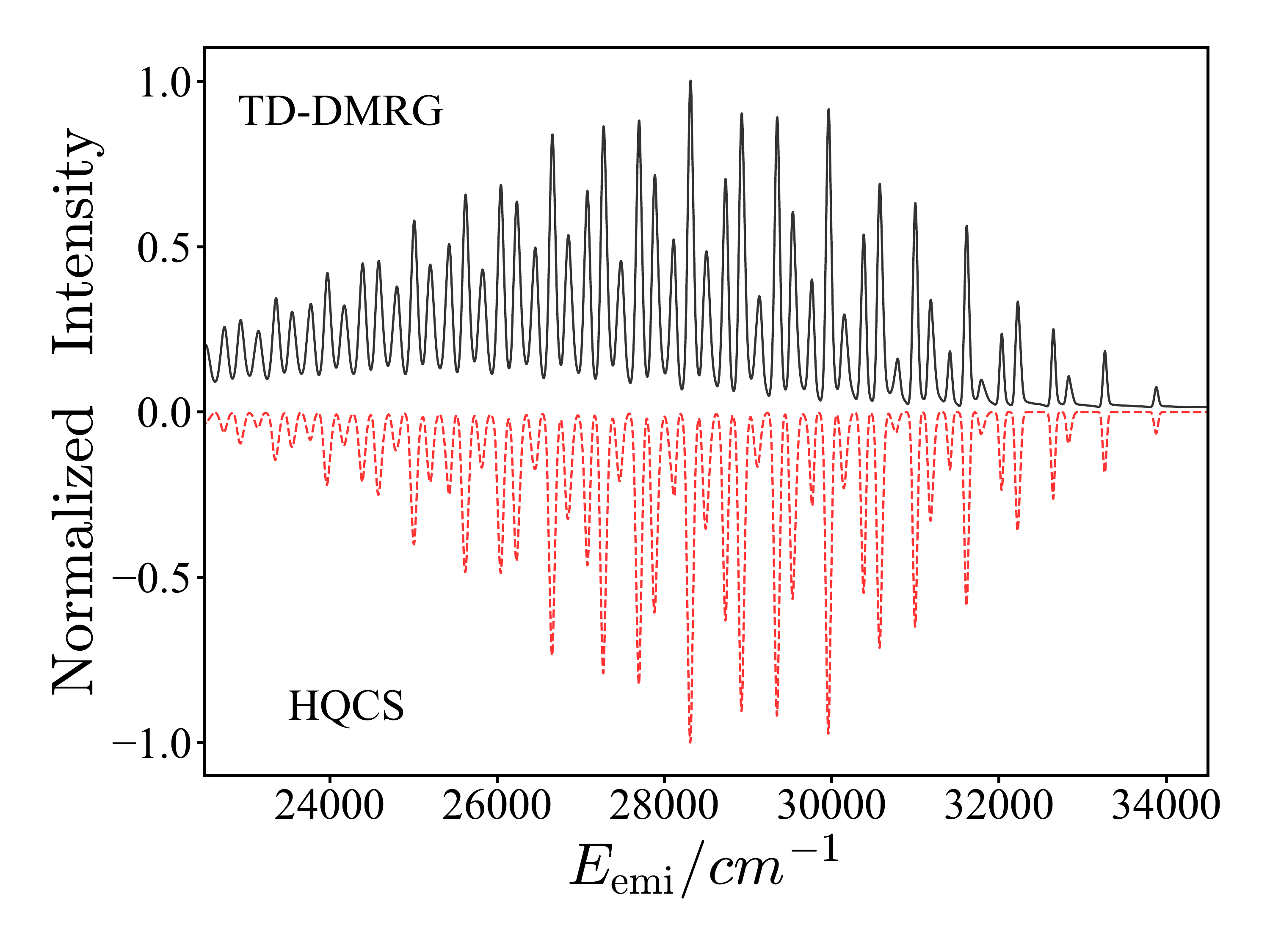}%
    }
    \caption{\label{fig:another reference} With the same condition for pyridine spectrum simulation in the main article. We use TD-DMRG to produce another reference spectrum of pyridine considering the displacement and rotation of all 27 vibration modes (top, black solid line) and compare that with the 7 modes HQCS results (bottom, red dashed line). It can be seen that the chosen 7 modes can reproduce most of vibrational features of the 27 modes spectrum among the showed regime.}
\end{figure*}
\section{Doktorov operator $\mathbf{\hat{U}}_\mathrm{Dok}$}
$\mathbf{\hat{U}}_\mathrm{Dok}$ determined by the harmonic frequency, displacement and rotation of two harmonic PES can be used to calculate the Franck-Condon factors between the harmonic states on the two PES. Here we are going to show its relation with above properties of the two corresponding PES. We need dimensionless normal coordinates $\mathbf{R}=\sqrt{\mathbf{\omega}}\mathbf{q}$. From the relationship between $\mathbf{q}_{\mathrm{m}}$ and $\mathbf{q}_\mathrm{i}$ as Eq.~\ref{eq:mass_r}, we have the relationship between $\mathbf{R}_{\mathrm{m}}$ and $\mathbf{R}_\mathrm{i}$ as Eq.~\ref{eq:freq_r},

\begin{equation}
    \mathbf{q}_{\mathrm{m}} =\mathbf{S}\mathbf{q}_\mathrm{i}+\Delta \mathbf{q}
    \label{eq:mass_r}
\end{equation}
\begin{equation}
\begin{aligned}
\mathbf{R}_{\mathrm{m}} &=\mathbf{A}\mathbf{R}_\mathrm{i}+\mathbf{d} \label{eq:freq_r} \\
\mathbf{A} &= \mathbf{\omega}^\frac{1}{2}_{\mathrm{m}}\mathbf{S}\mathbf{\omega}^{-\frac{1}{2}}_{i} \\
\mathbf{d} &= \mathbf{\omega}^\frac{1}{2}_{\mathrm{m}} \Delta \mathbf{q}
\end{aligned}
\end{equation}
We can write $\mathbf{A}$ in the singular value decomposition (SVD) form as Eq.~\ref{eq:A_SVD} for later usage.
\begin{equation}
\mathbf{A} = \mathbf{O}_\mathrm{L}\mathbf{L}\mathbf{O}^{T}_\mathrm{R} \label{eq:A_SVD}
\end{equation}

where $\mathbf{O}_\mathrm{L/R}$ are orthogonal matrices and $\mathbf{L}=\mathrm{diag}(\mathbf{l})$ is a diagonal matrix.

With Doktorov transformation \cite{doktorov1975dynamical,doktorov1977dynamical}, we can relate the position operators $\mathbf{R}_\mathrm{i}$ and $\mathbf{R}_\mathrm{f}$ with an unitary transformation as Eq.~\ref{eq:doktorv_r}. Here we simply give the result. Detailed information and derivation process is available in former publications~\cite{Quesada2019franck,Huh2015boson,barnett2002methods}.
\begin{equation}
\mathbf{\hat{U}}^{\dagger}_\mathrm{Dok}\mathbf{R}_\mathrm{i}\mathbf{\hat{U}}_\mathrm{Dok}=\mathbf{R}_{\mathrm{m}} = \mathbf{A}\mathbf{R}_\mathrm{i}+\mathbf{d}\label{eq:doktorv_r}
\end{equation}
\begin{equation}
\mathbf{\hat{U}}_\mathrm{Dok} =\mathbf{\hat{U}}(\mathbf{O}^{T}_\mathrm{R})\mathbf{\hat{S}}(\mathbf{ln}(\mathbf{l}))\mathbf{\hat{U}}(\mathbf{O}_\mathrm{L})\mathbf{\hat{D}}(\mathbf{d}/\sqrt{2})
\end{equation}
The Doktorov transformation operator includes rotation operator $\mathbf{\hat{U}}$, 
\begin{equation}
    \mathbf{\hat{U}}^\dagger (\mathbf{O}) \mathbf{R} \mathbf{\hat{U}}(\mathbf{O}) = \mathbf{O} \mathbf{R} 
\end{equation}
squeezing operator $\mathbf{\hat{S}}(\mathbf{\lambda})=\otimes_j \hat{S}_j (\lambda_j)$ and displacement operator $\mathbf{\hat{D}}(\mathbf{\alpha})=\otimes_j \hat{D}_j (\alpha_j)$, where
\begin{equation}
\begin{aligned}
\hat{S}_j (\lambda_j) &= \mathrm{exp}(\frac{\lambda_j}{2}(a_j^{\dagger2}-a_j^2)) \\
\hat{D}_j (\alpha_j) &= \mathrm{exp}(\alpha_j a_j^\dagger  - \alpha_j^* a_j)
\end{aligned}
\end{equation}

\clearpage
\section{references}